\documentclass[english,onecolumn]{IEEEtran}
\usepackage[T1]{fontenc}
\usepackage[latin9]{inputenc}
\usepackage{color}
\usepackage{babel}
\usepackage{mathrsfs}
\usepackage{amsthm}
\usepackage{amsmath}
\usepackage{amssymb}
\usepackage{graphicx}
\PassOptionsToPackage{normalem}{ulem}
\usepackage{ulem}
\usepackage[unicode=true,
 bookmarks=true,bookmarksnumbered=true,bookmarksopen=true,bookmarksopenlevel=1,
 breaklinks=false,pdfborder={0 0 0},backref=false,colorlinks=false]
 {hyperref}
\hypersetup{pdftitle={Your Title},
 pdfauthor={Your Name},
 pdfpagelayout=OneColumn,pdfnewwindow=true,pdfstartview=XYZ,plainpages=false}
\usepackage{breakurl}

\makeatletter

\providecommand{\tabularnewline}{\\}

\theoremstyle{plain}
\newtheorem{thm}{\protect\theoremname}
\theoremstyle{definition}
\newtheorem{example}[thm]{\protect\examplename}
\theoremstyle{plain}
\newtheorem{lem}[thm]{\protect\lemmaname}
\theoremstyle{plain}
\newtheorem{cor}[thm]{\protect\corollaryname}
\theoremstyle{plain}
\newtheorem{prop}[thm]{\protect\propositionname}

\ifCLASSOPTIONcompsoc
\else
\fi

\providecommand{\corollaryname}{Corollary}
\providecommand{\examplename}{Example}
\providecommand{\lemmaname}{Lemma}
\providecommand{\propositionname}{Proposition}
\providecommand{\theoremname}{Theorem}

\makeatother

\providecommand{\corollaryname}{Corollary}
\providecommand{\examplename}{Example}
\providecommand{\lemmaname}{Lemma}
\providecommand{\propositionname}{Proposition}
\providecommand{\theoremname}{Theorem}

\begin{document}

\title{\textcolor{black}{Multiuser Broadcast Erasure Channel with Feedback and Side Information,
and Related Index Coding Results}}

\author{A. Papadopoulos and L. Georgiadis%
\thanks{A. Papadopoulos is with University of California Los Angeles, USA, email:athanasios.papadopoulos@ucla.edu.%
} %
\thanks{L. Georgiadis is with Aristotle University of Thessaloniki and CERTH-ITI, GREECE, email:leonid@auth.gr.%
} %
\thanks{This research has been co-financed by the European Union (European Social Fund -- ESF) and
Greek national funds through the Operational Program ``Education and Lifelong Learning''
of the National Strategic Reference Framework (NSRF) -- Research Funding Program: Thales. Investing
in knowledge society through the European Social Fund.%
}}
\maketitle
\begin{abstract}
\textcolor{black}{We consider the N-user broadcast erasure channel with public feedback and
side information. Before the beginning of transmission, each receiver knows a function of the
messages of some of the other receivers. This situation arises naturally in wireless and in
particular cognitive networks where a node may overhear transmitted messages destined to other
nodes before transmission over a given broadcast channel begins. We provide an upper bound
to the capacity region of this system. Furthermore, when the side information is linear, we
show that the bound is tight for the case of two-user broadcast channels. The special case
where each user knows the whole or nothing of the message of each other node, constitutes a
generalization of the index coding problem. For this instance, and when there are no channel
errors, we show that the bound reduces to the known Maximum Weighted Acyclic Induced Subgraph
bound. We also show how to convert the capacity upper bound to transmission completion rate
(broadcast rate) lower bound and provide examples of codes for certain information graphs for
which the bound is either achieved of closely approximated.}\end{abstract}

\section{\textcolor{black}{\normalsize {Introduction}}}

The multiuser broadcast channel where independent messages must be delivered to each one of
a number $N$ of receivers, has been extensively studied since its introductio\textcolor{black}{n
\cite{cover1972broadcast}. The capacity of this channel under general channel statistics is
not known, although special cases, e.g. ``degraded'' channels \cite{bergmans1973random}
have been solved. The erasure channel has been introduced by Elias \cite{elias1954error} and
received at lot of attention recently because it models well data networks where packets may
be lost due to congestion, excessive delays and buffer overflows \cite{richardson2008modern}.
Also, in data networks it is common for a receiver to send feedback to the transmitter in the
form of Acknowledgments (ACK), if a transmitted packet is received correctly. }

The multiuser broadcast erasure channel with feedback has been studied recently, for the case
of two-receiver channels in \cite{sagduyu2011capacity}\textcolor{magenta}{{} }\textcolor{black}{and
for general number of receivers in \cite{wangr2012capacity}, \cite{gatzianas2013Broadcast}.
In the the latter two works, an upper bound to the capacity of the channel has been developed
and algorithms have been proposed which achieve this bound for $N=3$ users, and, under certain
restrictions on the channels statistics, for an arbitrary number of receivers. }

\textcolor{black}{The problem of determining channel capacity when the transmitter has side
information of messages, has been addressed first by Shannon \cite{shannon1958channels} and
has since been studied under various setups \cite{kuznetsov1974coding}, \cite{heegard1983capacity},
\cite{jafar2006capacity}, \cite{kramer2007capacity}, \cite{timo2013lossy}. The issue of
taking advantage of side information has attracted considerable attention lately in wireless
communications where nodes may overhear transmissions intended for other nodes, either opportunistically
or, as in cooperative cognitive networks, in an organized fashion in order to increase the
overall throughput of the network. }

\textcolor{black}{A related problem addressed in the literature is index coding. In index coding,
a transmitter has messages destined to each one of a set of receivers. Each receiver knows
the messages of some of the receivers and each transmission is received error-free by all receivers.
Several works in this area address the problem of designing algorithms that transmit all messages
in shortest time, or shortest broadcast rate (see Section \ref{sub:All-or-Nothing} for the
definition of broadcast rate), \cite{bar2006index}, \cite{alon2008broadcasting}, \cite{el2010index},
\cite{blasiak2013broadcasting}. In the same setup, the problem of determining the channel
capacity region has been addressed and bounds, or in some cases the exact region, have been
determined \cite{sun2013index}, \cite{arbabjolfaei2013capacity}.}

\noindent \textbf{\textcolor{black}{Contributions of this work}}

\textcolor{black}{In this work we consider the multiuser broadcast erasure channel with feedback
when side information is available. The side information receiver $i$ has about the message
$W_{j}$ of receiver $j$ is of the form $h_{i}^{j}(W_{j})$ where $h_{i}^{j}(\cdot)$ is a
general function. For this channel, we develop an upper bound to its capacity region. We show
that when the side information is in the form of linear equations and for $N=2$ receivers,
this bound is tight. The problem considered in this work can be considered as a generalization
of the index coding problem. When the side information is of the type normally considered in
index coding, i.e., $h_{i}^{j}(W_{j})=W_{j}$ or $h_{i}^{j}(W_{j})=c$ where $c$ is a constant,
the upper bound on the capacity region can be translated into a lower bound on the broadcast
rate. When the channel is errorless, this bound reduces to the Maximum Weighted Acyclic Induced
Subgraph (MWAIS) \cite{arbabjolfaei2013capacity} which is a generalization of the Maximum
Acyclic Induced Subgraph (MAIS) bound derived in \cite{bar2011index} when all messages are
of equal size. Finally, for special cases of information graphs we provide algorithms whose
broadcast rate either achieves the lower bound or gets close to it.}

\section{\textcolor{black}{\normalsize {\label{sec:Notation,-Channel-Model}Notation, channel model
and codes} }}

In this section we present notation and describe the channel model and codes that will be studied
in the current work.

\noindent \textbf{\textcolor{black}{Notation}}

\textcolor{black}{By $\mbox{\ensuremath{[i,j],\ i\leq j\ }we denote the set of integers \ensuremath{\{i,i+1,...j\}}}$;
if $i>j$ we set $\left[i,j\right]\triangleq\emptyset$. We also denote by $[j]\triangleq[1,j]$.}

\textcolor{black}{Vectors are denoted by boldface letters. Let $\boldsymbol{D}=\left(D_{1},D_{2},...,D_{N}\right)$,
$N\geq1$, be an $N\mbox{-dimensional vector.}$ For $\mathcal{B}\subseteq[N],$ we denote
by $\boldsymbol{D}_{\mathcal{B}}\triangleq\left(D_{i}\right)_{i\in\mathcal{B}}$, i.e., the
projection of $\boldsymbol{D}$ onto the coordinates in $\mathcal{B}.$ Whenever empty sets
appear as subscripts of a quantity, e.g. $\boldsymbol{D}_{\emptyset}$, the quantities are
interpreted as ``nonexistent'' in the notation. This convention is adopted in order to avoid
dealing with special cases. }

\noindent \textbf{\textcolor{black}{Channel model}}

\textcolor{black}{We consider the broadcast erasure channel with public feedback. The channel
consists of one transmitter and a set $[N]$ of receivers (referred also as ``nodes'' or
``users''). In the $l$th channel use (time), $l=1,2,...$,} 
\begin{itemize}
\item \textcolor{black}{Symbol $X\left(l\right)$ is transmitted, where $X\left(l\right)$ takes
one of the values in the finite set $\mathcal{X}$. } 
\item \textcolor{black}{Symbol $Y_{i}\left(l\right)$ is received by receiver $i\in[N]$, where $Y_{i}\left(l\right)$
takes values in the finite set $\mathcal{Y=\mathcal{X}\cup\left\{ \varepsilon\right\} },$
and $\varepsilon\notin\mathcal{X}$ . } 
\end{itemize}
\textcolor{black}{The statistical relation between $\boldsymbol{Y}\left(l\right)$ and $X\left(l\right)$
is given by: 
\[
Y_{i}\left(l\right)=\left\{ \begin{array}{cc}
X\left(l\right) & \mbox{if \ensuremath{Z_{i}\left(l\right)=1}}\\*
\varepsilon & \mbox{if }Z_{i}\left(l\right)=0
\end{array}\right.,\ i\in[N],
\]
where $\boldsymbol{Z}\left(l\right),\ l=1,....$ is a sequence of i.i.d. $N$-dimensional vectors,
taking values in the set $\left\{ 0,1\right\} $. $\boldsymbol{Z}\left(l\right),\ l=1,....$
represent channel erasures (1 correct reception, 0 erasure). }

\textcolor{black}{For $\mathcal{B}\subseteq[N],\ \mathcal{B\neq\emptyset}$ we denote $\epsilon_{\mathcal{B}}\triangleq\Pr\left(Z_{i}(l)=0,\ i\in\mathcal{B}\right).$
To avoid trivial cases, we assume for the rest of the paper that $\epsilon_{\{i\}}<1$ for
all $i\in[N]$.}

\textcolor{black}{We denote the set of all possible feedback vectors by $\mathcal{Z}=\left\{ \boldsymbol{z}:\ z_{i}=1\mbox{ or \ensuremath{0,\ i\in[N]}}\right\} $,
and by $\mathcal{Z}_{\mathcal{B}}^{*}\triangleq\left\{ \boldsymbol{z}\in\mathcal{Z}:\:\boldsymbol{z}_{\mathcal{B}}\neq(0,...,0)\right\} $,
$\mathcal{B}\subseteq[N]$ the set of all feedback vectors whose coordinates in set $\mathcal{B}$
are }\textcolor{black}{\emph{not }}\textcolor{black}{all erasures . }

\textcolor{black}{Upon reception of symbol $Y_{i}\left(l\right)$, each receiver $i$ sends
to the transmitter and all other receivers the value $Z_{i}(l)$; hence after the $l$th channel
use the transmitter and all receivers know $\boldsymbol{Z}(l)$ (but only the transmitter and
user $i$ know $Y_{i}(l)$). }

\textcolor{black}{In the following, when referring to a sequence of quantities involving the
time index, e.g., $\boldsymbol{D}\left(l\right)=\boldsymbol{Y}_{\mathcal{B}}(l),$ or $\mathbf{Z}(l),\mbox{ or }X(l),$
we use the notation $\boldsymbol{D}^{l}\triangleq\left(\boldsymbol{D}\left(1\right),...,\boldsymbol{D}\left(l\right)\right),\ l\geq1$.
To avoid special cases, we interpret $\boldsymbol{D}^{0}$ as a fixed constant quantity. }

\noindent \textbf{\textcolor{black}{Message indexes }}

\textcolor{black}{For a given $n=1,2,...$, there are message indexes $\boldsymbol{W}_{[N],n}$
where message index (or simply ``message'') $W_{j,n}$ must be transferred through the channel
from the transmitter to receiver $j$ in $n$ channel uses, and is taking values in the finite
set $\mathcal{W}_{j,n}$ whose size depends on $n$. We make the dependence of message and
message index size on $n$ explicit in order to avoid misconceptions when taking limits involving
quantities that refer to side information. For every $n$, message indexes are selected randomly,
independently and uniformly distributed (u.d.) within their corresponding index set. Also,
message indexes are independent of the channel statistics, i.e., independent of $\boldsymbol{Z}\left(l\right),\ l=1,2,....$}

\noindent \textbf{\textcolor{black}{Side information (SI)}}

\textcolor{black}{Let $\boldsymbol{h}_{i,n}^{j}(W_{j,n})$, $i,j\in[N],$ be functions of the
information messages $W_{j,n},\; j\in[N]$, taking values in some set $\mathcal{F}_{n}^{j}.$
For given $n$ and $i,j\in[N],$ the function $\boldsymbol{h}_{i,n}^{j}(W_{j,n})$ represents
the information node $i$ has about message $W_{j,n}$. The case of a constant function, i.e.,
$\boldsymbol{h}_{i,n}^{j}(W_{j,n})=c$, for all $W_{j,n}\in\mathcal{W}_{j,n}$, is equivalent
to assuming that node $i$ has no information about message $W_{j,n}.$ We denote $\mathcal{H}_{n}\triangleq\left\{ \boldsymbol{h}_{i,n}^{j}(\cdot)\right\} _{i,j\in[N]}$.
The }\textcolor{black}{\emph{form }}\textcolor{black}{(either as a table or through formulas)
of all functions in $\mathcal{H}_{n}$ is known by the transmitter, and all receivers. However,
if $\left(W_{j,n}\right)_{j\in[N]}$ are the messages selected for transmission, only user
$i$ (and the transmitter) knows the }\textcolor{black}{\emph{values }}\textcolor{black}{of
the functions $\boldsymbol{S}_{i,n}^{[N]}\triangleq\left(\boldsymbol{h}_{i}^{j}\left(W_{j,n}\right)\right)_{j\in[N]}$
. For a given sets of node indexes $\mathcal{B}$, ${\cal V}$, and $i\in\mathcal{N},$ we
denote: $\boldsymbol{S}_{{\cal V},n}^{\mathcal{B}}$$\triangleq\left(\boldsymbol{S}_{i,n}^{j}\right)_{j\in\mathcal{B},\ i\in{\cal V}}$. }

\textcolor{black}{In the following to avoid overloading the notation, and whenever there is
no possibility for confusion, we omit the index $n$ when referring to quantities involving
it, such as $W$, $\boldsymbol{S}$, $\boldsymbol{h}$. The dependence on $n$, while important,
will play a role mainly in the final step of the derivations.} 
\begin{example}
\textcolor{black}{\label{ExampleSideInfo} Assume that $N=3,$ receiver 1 knows the messages
of receiver $2$ and 3, receiver 2 the message of receiver 3 and receiver 3 the messages of
receivers 1 and 2. Then, 
\[
h_{1}^{1}\left(W_{1}\right)=c,\ h_{1}^{2}\left(W_{2}\right)=w_{2},\ h_{1}^{3}\left(W_{3}\right)=w_{3}
\]
\[
h_{2}^{1}\left(W_{1}\right)=c,\ h_{2}^{2}\left(W_{2}\right)=c,\ h_{2,n}^{3}\left(W_{3}\right)=w_{3}
\]
\[
h_{3}^{1}\left(W_{1}\right)=w_{1},\ h_{3}^{2}\left(W_{2}\right)=w_{3},\ h_{3}^{3}\left(W_{3}\right)=c.
\]
All receivers know the form of the functions above. If messages $W_{1},\ W_{2},\ W_{3}$ are
selected for transmission, the knowledge each receiver has is, 
\[
\boldsymbol{S}_{1}^{[3]}=\left(c,W_{2},W_{3}\right),\ \boldsymbol{S}_{2}^{[3]}=\left(c,c,W_{3}\right),\ \boldsymbol{S}_{3}^{[3]}=\left(W_{1},W_{2},c\right).
\]
}
\end{example}
\medskip{}

\begin{example}
\textcolor{black}{\label{ExampleLinearSideInfo}Assume also that messages are of the form of
$k_{i}$ ``packets'', $W_{i}=\left(p_{k}^{i}\right)_{k=1}^{k_{i}}$ where each $p_{k}^{i}$
takes values in the same field }$\mathbb{F}_{q}$.\textcolor{black}{{} Then a possible set of
functions is described by the equations, 
\[
h_{i,l}^{j}=\sum_{k=1}^{k_{j}}a_{i,l,k}^{j}p_{k}^{j},\ i,j\in[N],\ l\in[l_{i}^{j}],\ l_{i}^{j}\geq1,
\]
where $a_{i,l,k}^{j}$ are constants taking values in }$\mathbb{F}_{q}$\textcolor{black}{{}
and $l_{i}^{j}$ denotes the number of such equations. In this case, $\boldsymbol{h}_{i}^{j}=\left(h_{i,l}^{j}\right)_{l=1}^{l_{i}^{j}}$
where each $h_{i,l}^{j}\left(\cdot\right)$ takes values in }$\mathbb{F}_{q}$.\textcolor{black}{{}
All receivers know all $a_{i,l,k}^{j}$ and each receiver $i$ knows the values, 
\[
\boldsymbol{S}_{i}^{[N]}=\left(\boldsymbol{h}_{i}^{j}\left(W_{j}\right)\right)_{j\in[N]},
\]
for the messages selected for transmission. }
\end{example}
\noindent \textbf{\textcolor{black}{Channel Codes and Channel Capacity}}

\textcolor{black}{A channel code $C_{n}$ of rate vector $\boldsymbol{R}=\left(R_{i}\right)_{i\in[N]},\ R_{i}\geq0,$
for the broadcast erasure channel with feedback and side information consists of the the following:} 
\begin{itemize}
\item \textcolor{black}{Message index sets $\mathcal{W}_{i,n}$, where $\left|\mathcal{W}_{i,n}\right|=2^{\left\lceil nR_{i}\right\rceil }$,
and messages $\boldsymbol{W}_{[N],n}$ , $W_{i,n}\in{\cal W}_{i,n}.$}
\item \textcolor{black}{Side information functions ${\cal H}_{n}$. }
\item \textcolor{black}{An encoder that in the $l$th channel use transmits symbol $X(l)=f_{n,l}(\boldsymbol{W}_{[N],n},\boldsymbol{Z}^{l-1},{\cal H}_{n})$.} 
\item \textcolor{black}{$N$ decoders $g_{i,n}(Y_{i}^{n},\boldsymbol{Z}^{n},\boldsymbol{S}_{i,n},{\cal H}_{n})$,
one for each receiver $i.$ After $n$ channel uses, receiver $i\in[N]$ calculates the message
index $\hat{W}_{i,n}=g_{i,n}(Y_{i}^{n},\boldsymbol{Z}^{n},\boldsymbol{S}_{i,n},{\cal H}_{n})$.} 
\end{itemize}
\textcolor{black}{The form of functions $f_{n,l}$ and $g_{i,n}\ i\in[N]$ is known by the
transmitter and all receivers. Thus the channel code $C_{n}$ is fully specified by the tuple
$(n,\ 2^{\left\lceil nR_{1}\right\rceil },...,2^{\left\lceil nR_{N}\right\rceil },\ \mathcal{H}_{n},\ (f_{n,l}:l\in[n]),\ (g_{i,n}:i\in[N]))$.
The probability of erroneous decoding of code $C_{n}$ is $\lambda_{n}=\Pr(\underset{i\in[N]}{\cup}\{\hat{W}_{i,n}\neq W_{i,n}\})$.
A vector rate $\boldsymbol{R}=(R_{1},...,R_{N})$ is called achievable under the sequence of
codes $C_{n}$ if $\lim_{n\rightarrow\infty}\lambda_{n}=0$. In this case we also say that
the sequence of code $C_{n}$ achieves rate $\boldsymbol{R}.$ A rate vector $\boldsymbol{R}$
is achievable under a class of codes $\mathscr{C}$ if there is a sequence of codes in $\mathscr{C}$
that achieves $\boldsymbol{R}.$ The closure of the set of rates $\boldsymbol{R}$ that are
achievable under $\mathscr{C}$ constitutes the rate region of $\mathscr{C}$. The capacity
region of the channel, $\mathbb{\mathcal{C}}$, is the closure of the set of all achievable
rates under the class of all codes.}

For easy reference, the notations used thus far are summarized in Table \ref{table:1}.

\begin{table}[tbh]
\begin{centering}
\textcolor{black}{\caption{\textcolor{black}{Summary of notation}}
}
\par\end{centering}

\begin{centering}
\label{table:1} 
\par\end{centering}

\centering{}%
\begin{tabular}{cl}
\hline 
\textcolor{black}{$\boldsymbol{W}_{i}$}  & \textcolor{black}{Message destined for receiver $i$}\tabularnewline
\textcolor{black}{$\boldsymbol{X}^{l}$}  & \textcolor{black}{The vector of symbols sent in the first $l$ time slots}\tabularnewline
\textcolor{black}{$\boldsymbol{Y}_{[i]}^{l}$}  & \textcolor{black}{The vector of symbols received by receivers $1,2,..,i$ in the first $l$
time slots}\tabularnewline
\textcolor{black}{$\boldsymbol{Z}^{l}$}  & \textcolor{black}{The vector of variables representing the erasures in all the channels in
the first $l$ time slots}\tabularnewline
\textcolor{black}{$\boldsymbol{S}_{i}^{j}$}  & \textcolor{black}{The side information receiver $i$ has about the packets of receiver $j$}\tabularnewline
\textcolor{black}{$R_{i}$}  & \multicolumn{1}{l}{\textcolor{black}{The rate of communication to receiver $i$}}\tabularnewline
\textcolor{black}{$\mathcal{X}$}  & \textcolor{black}{The alphabet of $X$}\tabularnewline
\textcolor{black}{$\epsilon_{\mathcal{B}}$}  & \textcolor{black}{Erasure probability to receivers contained in $\mathcal{B}$}\tabularnewline
\hline 
\end{tabular}
\end{table}

\section{\textcolor{black}{\normalsize {\label{sec:Preliminaries}Preliminaries}}}

\textcolor{black}{The following relations, provable by standard information theoretic arguments,
will be used in the proofs that follow. }

\begin{eqnarray}
I(X;\phi(X)\left|W\right.) & = & H(\phi(X)\left|W\right.).\label{eq:5}\\
I\left(X;Z\left|Y\right.\right) & = & 0\ \mbox{if \ensuremath{Z\:}is independent of \ensuremath{\left(X,Y\right)}.}\label{eq:5-1}
\end{eqnarray}

\textcolor{black}{Lemmas \ref{lem:1}-\ref{lem:3} below are generalizations of corresponding
lemmas in \cite{czap2012broadcasting}. Lemma \ref{lem:4} is a corollary of these lemmas.
Their proofs can be found in the Appendix.}

\textcolor{black}{Lemma \ref{lem:1} relates sum-rates to information measures and is based
on Fano's bound.} 
\begin{lem}
\textcolor{black}{\label{lem:1}Assume that the rate vector $\boldsymbol{R}=\left(R_{1},...,R_{N}\right)$
is achievable. Then, 
\[
n\sum_{i=1}^{j}R_{i}\leq I(\boldsymbol{W}_{[j]};\boldsymbol{Y}_{[j]}^{n},\boldsymbol{Z}^{n},\boldsymbol{S}_{[j]}^{[N]})+o\left(n\right).
\]
} 
\end{lem}
\textcolor{black}{For given $l,$ Lemma \ref{lem:2} relates the information that the received
symbol vector $\boldsymbol{Y}_{[1,j]}(l)$ and $\boldsymbol{Z}\left(l\right)$, contains about
a random variable $U$, to the information that $X(l)$ contains about $U$, given that we
already know a related variable $Q$. Lemma \ref{lem:3} makes similar connection between information
carried by vectors $\boldsymbol{Y}_{[1,j]}^{n},\boldsymbol{Z}^{n}$ and each of the elements
of $X^{n}$.} 
\begin{lem}
\textcolor{black}{\label{lem:2}If $(U,Q,X(l))$ is independent of $\boldsymbol{Z}(l)$, it
holds: 
\begin{eqnarray*}
I(U;\boldsymbol{Y}_{[j]}\left(l\right),\boldsymbol{Z}\left(l\right)\mid Q) & = & (1-\epsilon_{[j]})I(U;X\left(l\right)\mid Q).
\end{eqnarray*}
} 
\end{lem}
\textcolor{white}{\tiny {sd}}{\tiny \par}
\begin{lem}
\textcolor{black}{\label{lem:3}If $(U,\boldsymbol{Y}_{[j]}^{l-1},\boldsymbol{Z}^{l-1},Q,X(l))$
are independent of $\boldsymbol{Z}(l)$ for $l\in[N],$ it holds:} 
\end{lem}
\textcolor{black}{
\begin{eqnarray*}
I\left(U;\boldsymbol{Y}_{[j]}^{n},\boldsymbol{Z}^{n}\mid Q\right) & = & \left(1-\epsilon_{[j]}\right)\sum_{l=1}^{n}I\left(U;X\left(l\right)\mid\boldsymbol{Y}_{[j]}^{l-1},\boldsymbol{Z}^{l-1},Q\right).
\end{eqnarray*}
}

\textcolor{black}{The following lemma relates the information that the received symbol vector
$\boldsymbol{Y}_{[j]}^{n}$ and $\boldsymbol{Z}^{n}$, contain about the a random variable
$U$, to the information that $\boldsymbol{Y}_{[j+1]}^{n}$, together with $\boldsymbol{Z}^{n}$,
contains about this variable, given that we already know $Q$. } 
\begin{lem}
\textcolor{black}{\label{lem:4}Let $j\in[N-1].$ If $(U,\boldsymbol{Y}_{[j+1]}^{l-1},\boldsymbol{Z}^{l-1},Q,X(l))$
are independent of $\boldsymbol{Z}(l)$ for $l\in[N],$ it holds, 
\begin{eqnarray*}
\frac{I(U;\boldsymbol{Y}_{[j]}^{n},\boldsymbol{Z}^{n}\mid Q)}{1-\epsilon_{[j]}} & \leq & \frac{I(U;\boldsymbol{Y}_{[j+1]}^{n},\boldsymbol{Z}^{n}\mid Q)}{1-\epsilon_{[j+1]}}+\sum_{l=1}^{n}I(\boldsymbol{Y}_{j+1}^{l-1};X\left(l\right)\mid\boldsymbol{Y}_{[j]}^{l-1},\boldsymbol{Z}^{l-1},Q).
\end{eqnarray*}
} 
\end{lem}

\section{\textcolor{black}{\normalsize {\label{sec:Main-Part}Outer Bound for N receivers}}}

\textcolor{black}{In this section we derive a necessary condition for achievability of a given
vector $\boldsymbol{R}$. First, we need some new definitions.}

\begin{figure}[tbh]
\centering{}\textcolor{black}{\includegraphics[scale=0.6]{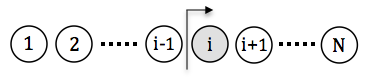}\caption{\label{fig:S-Sybolism1}The $\boldsymbol{S}_{i}^{\Rsh}$ notation: it denotes the information
that $i$ has about and the messages for nodes ``to the right'' of his position, including
him.}
}

\end{figure}

\begin{figure}[tbh]
\centering{}\textcolor{black}{\includegraphics[scale=0.6]{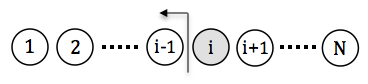}\caption{\label{fig:S-Sybolism2}The $\boldsymbol{S}_{i}^{\Lsh}$ notation: it denotes information that
$i$ has about and messages for nodes ``to the left'' of his position, not including him.}
}
\end{figure}

\textcolor{black}{Recall from Section \ref{sec:Notation,-Channel-Model} the notation, $\boldsymbol{S}_{\mathcal{V}}^{\mathcal{B}}$.
We define $\boldsymbol{S}_{i}^{\Rsh}\triangleq\boldsymbol{S}_{i}^{[i,N]}$ and $\boldsymbol{S}_{i}^{\Lsh}\triangleq\boldsymbol{S}_{i}^{[i-1]}$.
$\boldsymbol{S}_{i}^{\Rsh}$ and $\boldsymbol{S}_{i}^{\Lsh}$ denote the information node $i$
has about the messages for nodes in $[i,N]$ and $[i-1]$ respectively. Figures \ref{fig:S-Sybolism1}
and \ref{fig:S-Sybolism2} illustrate these definitions. We also use the notation: $\boldsymbol{S}_{\mathcal{V}}^{\Rsh}\triangleq\left(\boldsymbol{S}_{i}^{\Rsh}\right)_{i\in\mathcal{V}}$,
$\boldsymbol{S}_{\mathcal{V}}^{\Lsh}\triangleq\left(\boldsymbol{S}_{i}^{\Lsh}\right)_{i\in\mathcal{V}}$. }

\textcolor{black}{The following relations follow directly from the definitions above }

\begin{figure}[tbh]
\centering{}\textcolor{black}{\includegraphics[scale=0.4]{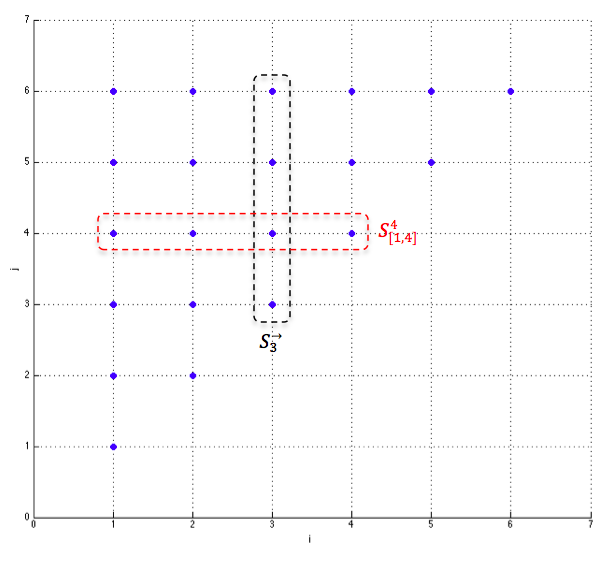}\caption{\label{fig:Reorganization-of-the}Rearrangement of the components in $\boldsymbol{S}^{\Rsh}$.
The dots depict the $S_{i}^{j}$ for different values of $j$ and $i$, with $j>i$, i.e. the
ones that are included in $\boldsymbol{S}^{\Rsh}$. Collecting them in columns we calculate
$\boldsymbol{S}^{\Rsh}\triangleq\left(\boldsymbol{S}_{i}^{\Rsh}\right)_{i\in\mathcal{N}}$.
Collecting them in rows we calculate $\boldsymbol{S}^{\Rsh}=\left(\boldsymbol{S}_{[i]}^{i}\right)_{i\in\mathcal{N}}$.}
}
\end{figure}

\textbf{\textcolor{black}{Side Information Relations (SIR)}} 
\begin{enumerate}
\item \textcolor{black}{\label{enu:4-1}$\boldsymbol{S}_{[j]}^{[N]}=(\boldsymbol{S}_{[j]}^{\Lsh},\boldsymbol{S}_{[j]}^{\Rsh})$.
} 
\item \textcolor{black}{\label{enu:5-1-1}$\boldsymbol{S}_{[N]}^{\Rsh}=\left(\boldsymbol{S}_{[j]}^{\Rsh},\boldsymbol{S}_{[j+1,N]}^{\Rsh}\right)=\left(\boldsymbol{S}_{[i]}^{i}\right)_{i\in[N]}$.
Here the right hand side of the equality is a rearrangement of the elements of the left hand
side (see Figure \ref{fig:Reorganization-of-the}).} 
\item \textcolor{black}{\label{enu:5-1-2}$\ensuremath{\boldsymbol{S}_{[i]}^{i}}$ is a deterministic
function of $W_{i}$.} 
\item \textcolor{black}{\label{enu:5-1}$\ensuremath{\boldsymbol{S}_{[j+1]}^{\Lsh}}$ is a deterministic
function of $\ensuremath{\boldsymbol{W}_{[j]}}.$} 
\end{enumerate}

\textcolor{black}{The following lemma is a calculation of the mutual information between $\boldsymbol{S}_{[N]}^{\Rsh}$
and a subset of the message sets. The result is intuitive. Observing Figure \ref{fig:S-Sybolism1},
we notice that since messages are independent, only messages to the left of $i$ (including
$i$) contain information about $\boldsymbol{W}_{i}$. } 
\begin{lem}
\textcolor{black}{\label{lem:5}It holds for $j\in[N],$ 
\[
I\left(\boldsymbol{W}_{[j]};\boldsymbol{S}_{[N]}^{\Rsh}\right)=\sum_{i=1}^{j}H(\boldsymbol{S}_{[i]}^{i}).
\]
}\end{lem}
\begin{IEEEproof}
\textcolor{black}{Using the chain rule for mutual information we have, 
\begin{eqnarray}
I\left(\boldsymbol{W}_{[j]};\boldsymbol{S}_{[N]}^{\Rsh}\right) & = & I\left(\boldsymbol{W}_{[j]};\left(\boldsymbol{S}_{[i]}^{i}\right)_{i\in[N]}\right)\:\mbox{by SIR \ref{enu:5-1-1}}\nonumber \\
 & = & I\left(\boldsymbol{W}_{[j]};(\boldsymbol{S}_{[1,i]}^{i})_{i\in[j]},(\boldsymbol{S}_{[1,i]}^{i})_{i\in[j+1,N]}\right)\nonumber \\
 & = & I\left(\boldsymbol{W}_{[j]};(\boldsymbol{S}_{[i]}^{i})_{i\in[j]}\right)+I(\boldsymbol{W}_{[j]};(\boldsymbol{S}_{[i]}^{i})_{i\in[j+1,N]}\mid(\boldsymbol{S}_{[i]}^{i})_{i\in[j]})\nonumber \\
 & = & I(\boldsymbol{W}_{[1,j]};(\boldsymbol{S}_{[i]}^{i})_{i\in[j]})\;\textrm{by SIR \ref{enu:5-1-2} and (\ref{eq:5-1})}\nonumber \\
 & = & H((\boldsymbol{S}_{[i]}^{i})_{i\in[j]})\:\textrm{by SIR \ref{enu:5-1-2} and (\ref{eq:5})}.\label{eq:h1}
\end{eqnarray}
By SIR \ref{enu:5-1-2} and the independence of $W_{i},$ $i\in[N],$ it follows that $\boldsymbol{S}_{[i]}^{i},\ i\in[1,j],$
are independent. Hence, 
\begin{equation}
H((\boldsymbol{S}_{[i]}^{i})_{i\in[j]})=\sum_{i=1}^{j}H(\boldsymbol{S}_{[i]}^{i}).\label{eq:h2}
\end{equation}
} 
\end{IEEEproof}
\textcolor{black}{The next two lemmas provide lower and upper partial sum-rate bounds respectively
in terms of relevant information metrics } 
\begin{lem}
\textcolor{black}{\label{lem:6} It holds for $j\in[N-1]$, 
\begin{equation}
n\sum_{i=1}^{j}R_{i}\geq I\left(\boldsymbol{W}_{[j]};\boldsymbol{Y}_{[j+1]}^{n},\boldsymbol{Z}^{n}\mid\boldsymbol{S}_{[N]}^{\Rsh}\right)+\sum_{i=1}^{j}H(\boldsymbol{S}_{[i]}^{i})+H(\boldsymbol{S}_{[j+1]}^{\Lsh}\mid\boldsymbol{Y}_{[j+1]}^{n},\boldsymbol{Z}^{n},\boldsymbol{S}_{[N]}^{\Rsh}).\label{eq:R>}
\end{equation}
}\end{lem}
\begin{IEEEproof}
\textcolor{black}{Write, 
\begin{eqnarray*}
n\sum_{i=1}^{j}R_{i} & = & \sum_{i=1}^{j}H(W_{i})\:\mbox{since \ensuremath{W_{i}}s are u.d.}\\
 & = & H(\boldsymbol{W}_{[j]})\:\textrm{since \ensuremath{W_{i}}s are i.i.d.}\\
 & \geq & I\left(\boldsymbol{W}_{[j]};\boldsymbol{Y}_{[j+1]}^{n},\boldsymbol{Z}^{n},\boldsymbol{S}_{[j+1]}^{\Lsh},\boldsymbol{S}_{[N]}^{\Rsh}\right)\\
 & = & I\left(\boldsymbol{W}_{[j]};\boldsymbol{Y}_{[j+1]}^{n},\boldsymbol{Z}^{n},\boldsymbol{S}_{[N]}^{\Rsh}\right)+I(\boldsymbol{W}_{[j]};\boldsymbol{S}_{[j+1]}^{\Lsh}\mid\boldsymbol{Y}_{[j+1]}^{n},\boldsymbol{Z}^{n},\boldsymbol{S}_{[N]}^{\Rsh})\\
 & = & I\left(\boldsymbol{W}_{[j]};\boldsymbol{Y}_{[j+1]}^{n},\boldsymbol{Z}^{n},\boldsymbol{S}_{[N]}^{\Rsh}\right)+H(\boldsymbol{S}_{[j+1]}^{\Lsh}\mid\boldsymbol{Y}_{[j+1]}^{n},\boldsymbol{Z}^{n},\boldsymbol{S}_{[N]}^{\Rsh})\:\textrm{by SIR \ref{enu:5-1} and (\ref{eq:5})}\\
 & = & I\left(\boldsymbol{W}_{[j]};\boldsymbol{Y}_{[j+1]}^{n},\boldsymbol{Z}^{n}\mid\boldsymbol{S}_{[N]}^{\Rsh}\right)+I(\boldsymbol{W}_{[j]};\boldsymbol{S}_{[N]}^{\Rsh})+H(\boldsymbol{S}_{[j+1]}^{\Lsh}\mid\boldsymbol{Y}_{[j+1]}^{n},\boldsymbol{Z}^{n},\boldsymbol{S}_{[N]}^{\Rsh})\\
 & = & I\left(\boldsymbol{W}_{[j]};\boldsymbol{Y}_{[j+1]}^{n},\boldsymbol{Z}^{n}\mid\boldsymbol{S}_{[N]}^{\Rsh}\right)+\sum_{i=1}^{j}H(\boldsymbol{S}_{[i]}^{i})+H(\boldsymbol{S}_{[j+1]}^{\Lsh}\mid\boldsymbol{Y}_{[j+1]}^{n},\boldsymbol{Z}^{n},\boldsymbol{S}_{[N]}^{\Rsh})\:\textrm{ by Lem. \ref{lem:5}}.
\end{eqnarray*}
}\end{IEEEproof}
\begin{lem}
\textcolor{black}{\label{lem:7}If the vector $\boldsymbol{R}$ is achievable then it holds
for $j\in[N].$ 
\begin{equation}
n\sum_{i=1}^{j}R_{i}\leq I\left(\boldsymbol{W}_{[j]};\boldsymbol{Y}_{[j]}^{n},\boldsymbol{Z}^{n}\mid\boldsymbol{S}_{[N]}^{\Rsh}\right)+\sum_{i=1}^{j}H(\boldsymbol{S}_{[i]}^{i})+H(\boldsymbol{S}_{[j]}^{\Lsh}\mid\boldsymbol{Y}_{[j]}^{n},\boldsymbol{Z}^{n},\boldsymbol{S}_{[N]}^{\Rsh})+o\left(n\right).\label{eq:R<}
\end{equation}
}\end{lem}
\begin{IEEEproof}
\textcolor{black}{Write, 
\begin{eqnarray*}
n\sum_{i=1}^{j}R_{i} & \leq & I\left(\boldsymbol{W}_{[j]};\boldsymbol{Y}_{[j]}^{n},\boldsymbol{Z}^{n},\boldsymbol{S}_{[j]}^{[N]}\right)+o\left(n\right)\:\textrm{by Lem. \ref{lem:1} }\\
 & \leq & I\left(\boldsymbol{W}_{[j]};\boldsymbol{Y}_{[j]}^{n},\boldsymbol{Z}^{n},\boldsymbol{S}_{[j]}^{[N]},\boldsymbol{S}_{[j+1,N]}^{\Rsh}\right)+o\left(n\right)\\
 & = & I\left(\boldsymbol{W}_{[j]};\boldsymbol{Y}_{[j]}^{n},\boldsymbol{Z}^{n},\boldsymbol{S}_{[j]}^{\Lsh},\boldsymbol{S}_{[j]}^{\Rsh},\boldsymbol{S}_{[j+1,N]}^{\Rsh}\right)+o\left(n\right)\ \mbox{by SIR \ref{enu:4-1}}\\
 & = & I\left(\boldsymbol{W}_{[j]};\boldsymbol{Y}_{[j]}^{n},\boldsymbol{Z}^{n},\boldsymbol{S}_{[j]}^{\Lsh},\boldsymbol{S}_{[N]}^{\Rsh}\right)+o\left(n\right)\:\mbox{by SIR \ref{enu:5-1-1}}\\
 & = & I\left(\boldsymbol{W}_{[j]};\boldsymbol{Y}_{[j]}^{n},\boldsymbol{Z}^{n},\boldsymbol{S}_{[N]}^{\Rsh}\right)+I(\boldsymbol{W}_{[j]};\boldsymbol{S}_{[j]}^{\Lsh}\mid\boldsymbol{Y}_{[j]}^{n},\boldsymbol{Z}^{n},\boldsymbol{S}_{[N]}^{\Rsh})+o\left(n\right)\\
 & = & I\left(\boldsymbol{W}_{[j]};\boldsymbol{Y}_{[j]}^{n},\boldsymbol{Z}^{n},\boldsymbol{S}_{[N]}^{\Rsh}\right)+H(\boldsymbol{S}_{[j]}^{\Lsh}\mid\boldsymbol{Y}_{[j]}^{n},\boldsymbol{Z}^{n},\boldsymbol{S}_{[N]}^{\Rsh})+o\left(n\right)\:\textrm{by SIR \ref{enu:5-1} and (\ref{eq:5})}\\
 & = & I\left(\boldsymbol{W}_{[j]};\boldsymbol{Y}_{[j]}^{n},\boldsymbol{Z}^{n}\mid\boldsymbol{S}_{[N]}^{\Rsh}\right)+I(\boldsymbol{W}_{[j]};\boldsymbol{S}_{[N]}^{\Rsh})+H(\boldsymbol{S}_{[j]}^{\Lsh}\mid\boldsymbol{Y}_{[j]}^{n},\boldsymbol{Z}^{n},\boldsymbol{S}_{[N]}^{\Rsh})+o\left(n\right)\\
 & = & I\left(\boldsymbol{W}_{[j]};\boldsymbol{Y}_{[j]}^{n},\boldsymbol{Z}^{n}\mid\boldsymbol{S}_{[N]}^{\Rsh}\right)+\sum_{i=1}^{j}H(\boldsymbol{S}_{[i]}^{i})+H(\boldsymbol{S}_{[j]}^{\Lsh}\mid\boldsymbol{Y}_{[j]}^{n},\boldsymbol{Z}^{n},\boldsymbol{S}_{[N]}^{\Rsh})+o\left(n\right)\:\textrm{by Lem. \ref{lem:5}}.
\end{eqnarray*}
} 
\end{IEEEproof}
\textcolor{black}{Now, we are ready to prove the main result.} 
\begin{thm}
\textcolor{black}{\label{thm:1}If the rate vector $\boldsymbol{R}=\left(R_{1},R_{2},...,R_{N}\right)$
is achievable, it holds for any $n$, 
\begin{equation}
\sum_{i=1}^{N}\frac{nR_{i}}{1-\epsilon_{[i]}}\leq n\log|\mathcal{X}|+\sum_{i=1}^{N}\frac{H\left(\boldsymbol{S}_{[i],n}^{i}\right)}{1-\epsilon_{[i]}}+o\left(n\right).\label{eq:mainTh}
\end{equation}
}

\textcolor{black}{Hence, 
\begin{equation}
\sum_{i=1}^{N}\frac{R_{i}}{1-\epsilon_{[i]}}\leq\log|\mathcal{X}|+\liminf_{n\rightarrow\infty}\frac{1}{n}\sum_{i=1}^{N}\frac{H\left(\boldsymbol{S}_{[i],n}^{i}\right)}{1-\epsilon_{[i]}}.\label{eq:mainTh1}
\end{equation}
In general, for any permutation $\pi_{i}$ of the indexes it holds, 
\begin{equation}
\sum_{i=1}^{N}\frac{R_{\pi_{i}}}{1-\epsilon_{\mathcal{B}_{\pi}(i)}}\leq\log|\mathcal{X}|+\liminf_{n\rightarrow\infty}\frac{1}{n}\sum_{i=1}^{N}\frac{H\left(\boldsymbol{S}_{\mathcal{B}_{\pi}(i),n}^{\pi_{i}}\right)}{1-\epsilon_{\mathcal{B}_{\pi}(i)}},\label{eq:mainTh2}
\end{equation}
where $\mathcal{B}_{\pi}(i)=\left\{ \pi_{1},...,\pi_{i}\right\} .$}\end{thm}
\begin{IEEEproof}
\textcolor{black}{For simplicity in the notation, in the proof we use the identity permutation,
$\pi_{i}=i$. It will be evident that the same arguments hold for any other permutation. }

\textcolor{black}{Using Lemma \ref{lem:7} for $j=N$ and Lemma \ref{lem:3} for $U=\boldsymbol{W}_{[N]}$
and $Q=\boldsymbol{S}_{[N]}^{\Rsh}$ we conclude: 
\begin{eqnarray}
\frac{n\sum_{i=1}^{N}R_{i}}{1-\epsilon_{[N]}} & \leq & \sum_{l=1}^{n}I\left(\boldsymbol{W}_{[N]};X\left(l\right)\mid\boldsymbol{Y}^{l-1},\boldsymbol{Z}^{l-1},\boldsymbol{S}_{[N]}^{\Rsh}\right)\nonumber \\
 &  & +\frac{\sum_{i=1}^{N}H(\boldsymbol{S}_{[i]}^{i})}{1-\epsilon_{[N]}}+\frac{H(\boldsymbol{S}_{[N]}^{\Lsh}\mid\boldsymbol{Y}_{[N]}^{n},\boldsymbol{Z}^{n},\boldsymbol{S}_{[N]}^{\Rsh})}{1-\epsilon_{[N]}}+o\left(n\right).\label{eq:R+R}
\end{eqnarray}
}

\textcolor{black}{Using again Lemma \ref{lem:7} for $U=\boldsymbol{W}_{[j]}$ and $Q=\boldsymbol{S}_{[N]}^{\Rsh}$
we have for $j\in[N-1]$: 
\begin{eqnarray*}
\frac{n\sum_{k=1}^{j}R_{k}}{1-\epsilon_{[j]}} & \leq & \frac{I\left(\boldsymbol{W}_{[j]};\boldsymbol{Y}_{[j]}^{n},\boldsymbol{Z}^{n}\mid\boldsymbol{S}_{[N]}^{\Rsh}\right)}{1-\epsilon_{[j]}}+\frac{\sum_{i=1}^{j}H(\boldsymbol{S}_{[i]}^{i})}{1-\epsilon_{[j]}}+\frac{H(\boldsymbol{S}_{[j]}^{\Lsh}\mid\boldsymbol{Y}_{[j]}^{n},\boldsymbol{Z}^{n},\boldsymbol{S}_{[N]}^{\Rsh})}{1-\epsilon_{[j]}}+o(n)\\
 & \leq & \frac{I\left(\boldsymbol{W}_{[j]};\boldsymbol{Y}_{[j+1]}^{n},\boldsymbol{Z}^{n}\mid\boldsymbol{S}_{[N]}^{\Rsh}\right)}{1-\epsilon_{[j+1]}}+\sum_{l=1}^{n}I\left(\boldsymbol{Y}_{j+1}^{l-1}\left(l\right);X\left(l\right)\left|\boldsymbol{Y}_{[j]}^{l-1},\boldsymbol{Z}^{l-1},\boldsymbol{S}_{[N]}^{\Rsh}\right.\right)+o(n)\\
 &  & +\frac{\sum_{i=1}^{j}H(\boldsymbol{S}_{[i]}^{i})}{1-\epsilon_{[j]}}+\frac{H(\boldsymbol{S}_{[j]}^{\Lsh}\mid\boldsymbol{Y}_{[j]}^{n},\boldsymbol{Z}^{n},\boldsymbol{S}_{[N]}^{\Rsh})}{1-\epsilon_{[j]}}+o(n)\ by\:\mbox{Lem. \ref{lem:4}}\\
 & \leq & \frac{n\sum_{k=1}^{j}R_{k}}{1-\epsilon_{[1,j+1]}}-\frac{\sum_{i=1}^{j}H(\boldsymbol{S}_{[i]}^{i})}{1-\epsilon_{[j+1]}}-\frac{H(\boldsymbol{S}_{[j+1]}^{\Lsh}\mid\boldsymbol{Y}_{[j+1]}^{n},\boldsymbol{Z}^{n},\boldsymbol{S}_{[N]}^{\Rsh})}{1-\epsilon_{[j+1]}}\\
 &  & +\sum_{l=1}^{n}I\left(\boldsymbol{Y}_{j+1}^{l-1}\left(l\right);X\left(l\right)\left|\boldsymbol{Y}_{[j]}^{l-1},\boldsymbol{Z}^{l-1},\boldsymbol{S}_{[N]}^{\Rsh}\right.\right)\\
 &  & +\frac{\sum_{i=1}^{j}H(\boldsymbol{S}_{[i]}^{i})}{1-\epsilon_{[j]}}+\frac{H(\boldsymbol{S}_{[j]}^{\Lsh}\mid\boldsymbol{Y}_{[j]}^{n},\boldsymbol{Z}^{n},\boldsymbol{S}_{[N]}^{\Rsh})}{1-\epsilon_{[j]}}+o(n)\ \mbox{by Lem. \ref{lem:6}}.
\end{eqnarray*}
Rearranging terms in the last inequality we arrive at, 
\begin{eqnarray}
\frac{n\sum_{k=1}^{j}R_{k}}{1-\epsilon_{[1,j]}}-\frac{n\sum_{k=1}^{j}R_{k}}{1-\epsilon_{[1,j+1]}} & \leq & \sum_{l=1}^{n}I\left(\boldsymbol{Y}_{j+1}^{l-1}\left(l\right);X\left(l\right)\left|\boldsymbol{Y}_{[j]}^{l-1},\boldsymbol{Z}^{l-1},\boldsymbol{S}_{[N]}^{\Rsh}\right.\right)+\frac{\sum_{i=1}^{j}H(\boldsymbol{S}_{[i]}^{i})}{1-\epsilon_{[1,j]}}-\frac{\sum_{i=1}^{j}H(\boldsymbol{S}_{[i]}^{i})}{1-\epsilon_{[1,j+1]}}+\nonumber \\
 &  & +\frac{H(\boldsymbol{S}_{[j]}^{\Lsh}\mid\boldsymbol{Y}_{[j]}^{n},\boldsymbol{Z}^{n},\boldsymbol{S}_{[N]}^{\Rsh})}{1-\epsilon_{[j]}}-\frac{H(\boldsymbol{S}_{[j+1]}^{\Lsh}\mid\boldsymbol{Y}_{[j+1]}^{n},\boldsymbol{Z}^{n},\boldsymbol{S}_{[N]}^{\Rsh})}{1-\epsilon_{[j+1]}}+o\left(n\right).\label{eq:R-R}
\end{eqnarray}
Summing $\left(\ref{eq:R+R}\right)$ and inequalities (\ref{eq:R-R}) for $j\in[N-1]$ and
taking into account the following: 
\begin{eqnarray*}
\sum_{j=1}^{N-1}\sum_{k=1}^{j}nR_{k}\left(\frac{1}{1-\epsilon_{[j]}}-\frac{1}{1-\epsilon_{[j+1]}}\right) & = & \sum_{k=1}^{N-1}nR_{k}\sum_{j=k}^{N-1}\left(\frac{1}{1-\epsilon_{[j]}}-\frac{1}{1-\epsilon_{[j+1]}}\right)\\
 & = & \sum_{k=1}^{N-1}nR_{k}\left(\frac{1}{1-\epsilon_{[k]}}-\frac{1}{1-\epsilon_{[N]}}\right)\\
 & = & \sum_{k=1}^{N-1}\frac{nR_{k}}{1-\epsilon_{[k]}}-\frac{n\sum_{k=1}^{N-1}R_{k}}{1-\epsilon_{[N]}},
\end{eqnarray*}
and similarly, }

\textcolor{black}{
\begin{eqnarray*}
\sum_{j=1}^{N-1}\left(\frac{\sum_{i=1}^{j}H(\boldsymbol{S}_{[i]}^{i})}{1-\epsilon_{[j]}}-\frac{\sum_{i=1}^{j}H(\boldsymbol{S}_{[i]}^{i})}{1-\epsilon_{[j+1]}}\right) & = & \sum_{i=1}^{N}\frac{H\left(\boldsymbol{S}_{[i]}^{i}\right)}{1-\epsilon_{[i]}}-\frac{\sum_{i=1}^{N}H(\boldsymbol{S}_{[i]}^{i})}{1-\epsilon_{[N]}},
\end{eqnarray*}
and 
\begin{eqnarray*}
\sum_{j=1}^{N-1}\frac{H(\boldsymbol{S}_{[j+1]}^{\Lsh}\mid\boldsymbol{Y}_{[j+1]}^{n},\boldsymbol{Z}^{n},\boldsymbol{S}_{[N]}^{\Rsh})}{1-\epsilon_{[j+1]}} & = & \sum_{j=2}^{N}\frac{H(\boldsymbol{S}_{[j]}^{\Lsh}\mid\boldsymbol{Y}_{[j]}^{n},\boldsymbol{Z}^{n},\boldsymbol{S}_{[N]}^{\Rsh})}{1-\epsilon_{[j]}}\\
 & = & \sum_{j=2}^{N-1}\frac{H(\boldsymbol{S}_{[j]}^{\Lsh}\mid\boldsymbol{Y}_{[j]}^{n},\boldsymbol{Z}^{n},\boldsymbol{S}_{[N]}^{\Rsh})}{1-\epsilon_{[j]}}+\frac{H(\boldsymbol{S}_{[N]}^{\Lsh}\mid\boldsymbol{Y}^{n},\boldsymbol{Z}^{n},\boldsymbol{S}_{[N]}^{\Rsh})}{1-\epsilon_{[N]}}\\
 & = & \sum_{j=1}^{N-1}\frac{H(\boldsymbol{S}_{[j]}^{\Lsh}\mid\boldsymbol{Y}_{[j]}^{n},\boldsymbol{Z}^{n},\boldsymbol{S}_{[N]}^{\Rsh})}{1-\epsilon_{[j]}}+\frac{H(\boldsymbol{S}_{[N]}^{\Lsh}\mid\boldsymbol{Y}^{n},\boldsymbol{Z}^{n},\boldsymbol{S}_{[N]}^{\Rsh})}{1-\epsilon_{[N]}}\\
 &  & \textrm{\ensuremath{\boldsymbol{S}_{1}^{\Lsh}=c\;}by definition}.
\end{eqnarray*}
we obtain, 
\begin{eqnarray}
\sum_{i=1}^{N}\frac{nR_{i}}{1-\epsilon_{[i]}} & \leq & \sum_{l=1}^{n}I\left(\boldsymbol{W};X(l)\mid\boldsymbol{Y}^{l-1},\boldsymbol{Z}^{l-1},\boldsymbol{S}^{\Rsh}\right)+\sum_{j=1}^{N-1}\sum_{l=1}^{n}I\left(\boldsymbol{Y}_{j+1}^{l-1};X(l)\mid\boldsymbol{Y}_{[j]}^{l-1},\boldsymbol{Z}^{l-1},\boldsymbol{S}_{[N]}^{\Rsh}\right)\nonumber \\
 &  & +\sum_{i=1}^{N}\frac{H\left(\boldsymbol{S}_{[i]}^{i}\right)}{1-\epsilon_{[i]}}+o(n).\label{eq:mp1}
\end{eqnarray}
Finally we write, 
\begin{eqnarray}
n\log|\mathcal{X}| & \geq & \sum_{l=1}^{n}H(X(l))\nonumber \\
 & = & \sum_{l=1}^{n}\left(H\left(X(l)\mid\boldsymbol{Y}_{[N]}^{l-1},\boldsymbol{Z}^{l-1},\boldsymbol{S}_{[N]}^{\Rsh}\right)+I\left(\boldsymbol{Y}_{[N]}^{l-1},\boldsymbol{Z}^{l-1},\boldsymbol{S}_{[N]}^{\Rsh};X(l)\right)\right)\\
 & \geq & \sum_{l=1}^{n}H\left(X(l)\mid\boldsymbol{Y}_{[N]}^{l-1},\boldsymbol{Z}^{l-1},\boldsymbol{S}_{[N]}^{\Rsh}\right)+\sum_{l=1}^{n}I\left(\boldsymbol{Y}_{[N]}^{l-1};X(l)\mid\boldsymbol{Z}^{l-1},\boldsymbol{S}_{[N]}^{\Rsh}\right)\nonumber \\
 & \geq & \sum_{l=1}^{n}I\left(\boldsymbol{W};X(l)\mid\boldsymbol{Y}_{[N]}^{l-1},\boldsymbol{Z}^{l-1},\boldsymbol{S}_{[N]}^{\Rsh}\right)+\sum_{l=1}^{n}I\left(\boldsymbol{Y}_{[N]}^{l-1};X(l)\mid\boldsymbol{Z}^{l-1},\boldsymbol{S}_{[N]}^{\Rsh}\right)\nonumber \\
 & = & \sum_{l=1}^{n}I\left(\boldsymbol{W};X(l)\mid\boldsymbol{Y}_{[N]}^{l-1},\boldsymbol{Z}^{l-1},\boldsymbol{S}^{\Rsh}\right)+\sum_{l=1}^{n}\sum_{j=1}^{N-1}I\left(\boldsymbol{Y}_{j+1}^{l-1};X(l)\mid\boldsymbol{Y}_{[j]}^{l-1},\boldsymbol{Z}^{l-1},\boldsymbol{S}_{[N]}^{\Rsh}\right)\ \nonumber \\
 &  & \mbox{chain rule on subscript indices of \ensuremath{\boldsymbol{Y}_{[N]}^{l-1}}}\\
 & = & \sum_{l=1}^{n}I\left(\boldsymbol{W};X(l)\mid\boldsymbol{Y}_{[N]}^{l-1},\boldsymbol{Z}^{l-1},\boldsymbol{S}^{\Rsh}\right)+\sum_{j=1}^{N-1}\sum_{l=1}^{n}I\left(\boldsymbol{Y}_{j+1}^{l-1};X(l)\mid\boldsymbol{Y}_{[j]}^{l-1},\boldsymbol{Z}^{l-1},\boldsymbol{S}_{[N]}^{\Rsh}\right).\label{eq:mp2}
\end{eqnarray}
From (\ref{eq:mp1}), (\ref{eq:mp2}) the theorem follows. } 
\end{IEEEproof}

\section{\textcolor{black}{\normalsize {\label{sec:Linear-side-information}}Linear side information
and achievability for two-receiver channel }}

In this section we consider that side information is in the form of linear equations. Under
this side information and for the case of two-receiver channel, we show that the bound to the
capacity region implied in Theorem \ref{thm:1} is tight.

\begin{table}[tbh]
\begin{centering}
\textcolor{black}{\caption{\textsf{\textcolor{black}{Summary of notation in Section \ref{sec:Linear-side-information}}}}
}
\par\end{centering}

\begin{centering}
\label{table 2} 
\par\end{centering}

\centering{}%
\begin{tabular}{cl}
\hline 
\textcolor{black}{$r_{i}$}  & \multicolumn{1}{l}{\textcolor{black}{The rate of transmission to receiver $i$ in packets per transmission}}\tabularnewline
\textcolor{black}{$L$}  & \textcolor{black}{The length of packets in bits}\tabularnewline
\textcolor{black}{$\rho(A)$}  & \multicolumn{1}{l}{\textcolor{black}{The rank of matrix $A$}}\tabularnewline
\textcolor{black}{$Sp(A)$}  & \textcolor{black}{The span of the rows of matrix $A$}\tabularnewline
$A_{\mathcal{B}_{\pi}(i)}^{\pi_{i}}$  & \multicolumn{1}{l}{\textcolor{black}{The matrix $\left[A_{\pi(1)}^{\pi_{1}}\cdots A_{\pi(i)}^{\pi_{i}}\right]^{\intercal}$
(corresponds to $S_{\mathcal{B}_{\pi}(i)}^{\pi_{i}}$)}}\tabularnewline
${\cal D}_{i}$ & The space of $k_{i}$-dimensional row vectors with elements from the field $\mathbb{F}_{2^{L}}$\tabularnewline
\textcolor{black}{${\cal P}_{i}^{i}$}  & \textcolor{black}{A set of row vectors of matrix $A_{i}^{i}$ that form a basis for $Sp\left(A_{i}^{i}\right)$}\tabularnewline
\textcolor{black}{${\cal P}_{j}^{i}$} & A set of row vectors of matrix \textcolor{black}{$A_{j}^{i}$ that together with the vectors
in ${\cal P}^{i}$ form a basis for $Sp\left(A_{\{1,2\}}^{i}\right)$}\tabularnewline
${\cal U}^{i}$ & \textcolor{black}{A set of row vectors that together with ${\cal P}^{i}\cup{\cal P}_{j}^{i}$
form a basis for the space ${\cal D}_{i}$}\tabularnewline
$\left\{ \boldsymbol{q}_{j,l}^{i}\right\} _{l\in[K_{j}^{i}]}$  & \textcolor{black}{Packets destined to $i$ and received only by $j$ in Phase 1}\tabularnewline
$\left\{ \boldsymbol{\widetilde{q}}_{j,l}^{i}\right\} _{l\in[K_{j}^{i}+d_{j}^{i}]}$  & \textcolor{black}{Packets destined to $i$ and known only by $j$ at the beginning of Phase
2}\tabularnewline
$T(\boldsymbol{k})$  & \textcolor{black}{The time it takes for the proposed algorithm to deliver $\boldsymbol{k}$
packets to their destinations (random variable)}\tabularnewline
\hline 
\end{tabular}
\end{table}

\noindent \textbf{\textcolor{black}{Linear side information}}

In this section we consider that each message $W_{i},\ i\in[N]$ consists of $k_{i}$ $L-$bit
packets, where bits in all packets are i.i.d. uniformly distributed and packets destined to
all receivers are independent. Hence, at the transmitter there are $k_{i}$ packets destined
to user $i.$ We denote the packets destined to receiver $i$ by $\boldsymbol{p}^{i}=(p_{1}^{i},...,p_{k_{i}}^{i}).$
Packets are considered as elements of the field $\mathbb{F}_{2^{L}}$, hence addition and multiplication
can be performed with these packets in the standard manner. We refer to this message model
as ``Packetized''.

We adopt the same channel model as in Section \ref{sec:Notation,-Channel-Model}, where now
$X(l)$ is a transmitted packet consisting of $L$ bits. We also adopt the same codes as in
Section \ref{sec:Notation,-Channel-Model}. A number of new definitions will be introduced
in this section. For easy reference, these definitions are summarized in Table II. 

Setting $k_{i}=\left\lceil r_{i}n\right\rceil $ (hence $r_{i}$ has units ``number of packets
per transmission'') we have $\left|\mathcal{W}_{i}\right|=2^{\left\lceil r_{i}n\right\rceil L}$
and since $\log\left|\mathcal{X}\right|=L$ Theorem \ref{thm:1} obtains the following form
for the Packetized model. 
\begin{cor}
\label{cor:PacketizedNeccesity}In the packetized model, if the rate vector $\boldsymbol{r}=\left(r_{1},r_{2},...,r_{N}\right)$
is achievable, it holds for any $n$, 
\begin{equation}
\sum_{i=1}^{N}\frac{r_{i}}{1-\epsilon_{[i]}}\leq1+\liminf_{n\rightarrow\infty}\frac{1}{nL}\sum_{i=1}^{N}\frac{H\left(\boldsymbol{S}_{[i],n}^{i}\right)}{1-\epsilon_{[i]}}.\label{eq:mainTh1-1}
\end{equation}
In general, for any permutation $\pi_{i}$ of node indexes it holds, 
\begin{equation}
\sum_{i=1}^{N}\frac{r_{\pi_{i}}}{1-\epsilon_{\mathcal{B}_{\pi}(i)}}\leq1+\liminf_{n\rightarrow\infty}\frac{1}{nL}\sum_{i=1}^{N}\frac{H\left(\boldsymbol{S}_{\mathcal{B}_{\pi}(i),n}^{\pi_{i}}\right)}{1-\epsilon_{\mathcal{B}_{\pi}(i)}},\label{eq:mainTh2-1}
\end{equation}
where $\mathcal{B}_{\pi}(i)=\left\{ \pi_{1},...,\pi_{i}\right\} .$ 
\end{cor}
We assume that receivers have linear side information as described in Example \ref{ExampleLinearSideInfo}.
Specifically, receiver $i$ knows the values of the following linear functions of the packets
destined to receiver $j\in[N]$, 
\[
h_{i,l}^{j}=\sum_{k=1}^{k_{j}}a_{i,l,k}^{j}p_{k}^{j},\ i,j\in[N],\ l\in[l_{i}^{j}],\ l_{i}^{j}\geq1.
\]
Hence receiver $i$ knows $l_{i}^{j}$ linear combinations of receiver $j$ packets (note that
it is possible that $i=j$). The case of no side information, e.g., if receiver $i$ does not
have any side information for the packets of user $j,$ can be represented by assuming that
$l_{i}^{j}=1$ and $a_{i,1,k}^{j}=0$ for all $k\in[k_{j}]$. 

We denote by $A_{i}^{j}$ the $l_{i}^{j}\times k_{j}$ matrix with elements $a_{i,l,k}^{j}$\textcolor{black}{.}
For a permutation of receiver indexes $\pi_{i}$, we denote by $A_{\mathcal{B}_{\pi}(i)}^{\pi_{i}}$
the $(\sum_{m=1}^{i}l_{\pi_{m}}^{\pi_{i}})\times k_{\pi_{i}}$ matrix 
\[
A_{\mathcal{B}_{\pi}(i)}^{\pi_{i}}=\left[\begin{array}{c}
A_{\pi_{1}}^{\pi_{i}}\\
\vdots\\
A_{\pi_{i}}^{\pi_{i}}
\end{array}\right].
\]
 Let $\rho\left(A\right)$ be the rank of matrix $A$ and $Sp\left(A\right)$ be the span of\emph{
row} vectors of $A$, so that $\rho(A)=\dim\left(Sp(A)\right)$. 

To compute $H\left(\boldsymbol{S}_{[i],n}^{i}\right)$ in the current setup we will make use
of the following Lemma (see, e.g. \cite[Appendix A]{gatzianas2013Broadcast})\textcolor{black}{.} 
\begin{lem}
Let $\{\boldsymbol{\upsilon}\}_{m=1}^{k},$ be M-dimensional vectors in $\mathbb{F}_{q}^{M}$.
Denote $\boldsymbol{\Upsilon}=\mbox{span\ensuremath{\left(\{\boldsymbol{\upsilon}_{m}\}_{m=1}^{k}\right)}}$
and let $l=\dim\left(\boldsymbol{\Upsilon}\right)$ with $l\geq1$. Let $\left\{ p_{m}\right\} _{m=1}^{k}$
be independent random variables uniformly distributed in $\mathbb{F}_{q}$ and construct the
random vector $\boldsymbol{\upsilon}=\sum_{m=1}^{k}p_{m}\boldsymbol{\upsilon}_{m}.$ Then $\boldsymbol{\upsilon}$
is uniformly distributed in $\boldsymbol{\Upsilon},$ i.e., 
\[
\Pr\left(\boldsymbol{\upsilon}=\boldsymbol{e}\right)=\frac{1}{q^{l}},\ \mbox{for all, \ensuremath{\boldsymbol{e}\in\mathcal{\boldsymbol{\Upsilon}}}.}
\]

\end{lem}
To apply this Lemma in our case, for matrix $A_{\mathcal{B}_{\pi}(i)}^{\pi_{i}}$, identify
$\boldsymbol{\upsilon}_{m}$ with the $m$-th column of this matrix, and $p_{m}$ with packet
$p_{m}^{\pi_{i}}$. Since packets are assumed to be independent with uniformly random bits
each, and $q=2^{L}$, it follows that the vector $A_{\mathcal{B}_{\pi}(i)}^{\pi_{i}}\boldsymbol{p}^{\pi_{i}}$
is uniformly distributed in $Sp\left(A_{\mathcal{B}_{\pi}(i)}^{\pi_{i}}\right)$, hence for
$\rho\left(A_{\mathcal{B}_{\pi}(i)}^{\pi_{i}}\right)\geq1$,
\begin{equation}
H\left(\cup_{m=1}^{i}\left\{ h_{\pi_{m},l}^{\pi_{i}}\right\} _{l=1}^{l_{\pi_{m}}^{\pi_{i}}}\right)=\rho\left(A_{\mathcal{B}_{\pi}(i)}^{\pi_{i}}\right)L.\label{eq:HtoRank-1}
\end{equation}
If $\rho\left(A_{\mathcal{B}_{\pi}(i)}^{\pi_{i}}\right)=0$, then $(\ref{eq:HtoRank-1}$) still
holds.

Based on the above, and introducing the dependence of the side information on $n,$ Corollary
\ref{cor:PacketizedNeccesity} takes the following form in the current setup. 
\begin{cor}
\label{cor:2ReceiverNecessity-1}In the packetized model, if the rate vector $\boldsymbol{r}=\left(r_{1},...,r_{N}\right)$
is achievable, it holds for permutation $\pi$, 
\begin{eqnarray*}
\sum_{i=1}^{N}\frac{r_{\pi_{i}}}{1-\epsilon_{\mathcal{B}_{\pi}(i)}} & \leq & 1+\liminf_{n\rightarrow\infty}\frac{1}{n}\sum_{i=1}^{N}\frac{\rho\left(A_{\mathcal{B}_{\pi}(i),n}^{\pi_{i}}\right)}{1-\epsilon_{\mathcal{B}_{\pi}(i)}}.
\end{eqnarray*}

\end{cor}
Note that the $\liminf$ involving the quantity $\rho\left(A_{\mathcal{B}_{\pi}(i),n}^{\pi_{i}}\right)$
in Corollary \ref{cor:2ReceiverNecessity-12} depends in general on the rate vector $\boldsymbol{r}.$
For example, if receiver 1 knows all packet destined to receiver 2, then $\rho(A_{1}^{2})=k_{2}=\left\lceil nr_{2}\right\rceil $
and hence $\lim_{n\rightarrow\infty}\left(\rho(A_{1}^{2})/n\right)=r_{2}.$

For $N=2$ Corollary \ref{cor:2ReceiverNecessity-1} specializes to, 
\begin{cor}
\label{cor:2ReceiverNecessity-12}In the packetized model, assume that for the rate vector
$\boldsymbol{r}=\left(r_{1},r_{2}\right)$ the following limits exist for $i,j\in\{1.,2\},$
\[
\hat{\rho}_{i}^{i}\left(\boldsymbol{r}\right)=\lim_{n\rightarrow\infty}\frac{\rho\left(A_{i,n}^{i}\right)}{n},\ \hat{\rho}_{\{i,j\}}^{i}(\boldsymbol{r})=\lim_{n\rightarrow\infty}\frac{\rho\left(A_{\{i,j\},n}^{i}\right)}{n}.
\]
If $\boldsymbol{r}$ is achievable it holds, 
\begin{eqnarray*}
\frac{r_{1}-\hat{\rho}_{1}^{1}\left(\boldsymbol{r}\right)}{1-\epsilon_{\{1\}}}+\frac{r_{2}-\hat{\rho}_{\{1,2\}}^{2}\left(\boldsymbol{r}\right)}{1-\epsilon_{\{1,2\}}} & \leq & 1,\\
\frac{r_{2}-\hat{\rho}_{2}^{2}\left(\boldsymbol{r}\right)}{1-\epsilon_{\{2\}}}+\frac{r_{1}-\hat{\rho}_{\{1,2\}}^{1}\left(\boldsymbol{r}\right)}{1-\epsilon_{\{1,2\}}} & \leq & 1.
\end{eqnarray*}

\end{cor}
For the rest of this section we will present algorithms (one for each $n$) whose rate region
is the same as the region described in Corollary \ref{cor:2ReceiverNecessity-12}. As in previous
sections, for simplicity in notation and whenever there is no possibility for confusion we
omit the time index $n$ from various quantities.

The following ``preprocessing'' is done first.

\noindent \textbf{\textcolor{black}{Preprocessing}} 
\begin{itemize}
\item \uline{Using Gaussian Elimination Construct a basis for} $Sp$$\left(A_{\{1,2\}}^{i}\right)$
as follows: 

\begin{itemize}
\item Select a set $\mathcal{\mathcal{P}}_{i}^{i}$ of $\rho(A_{i}^{i})$ linearly independent rows
from matrix $A_{i}^{i}$. These vectors form a basis for $Sp\left(A_{i}^{i}\right).$
\item Select a set $\mathcal{P}_{j}^{i}$ of $d_{j}^{i}=\rho(A_{\{1,2\}}^{i})-\rho(A_{i}^{i})$ linearly
independent vectors from matrix $A_{j}^{i}$ that together with the vectors in $\mathcal{\mathcal{P}}_{i}^{i}$
form a basis for $Sp\left(A_{\{1,2\}}^{i}\right)$. 
\end{itemize}
\item \uline{Select a set $\mathcal{U}_{i}$ of linearly independent vectors from the space $\mathcal{D}_{i}$
of $k_{i}$-dimensional vectors with elements from the field $\mathbb{F}_{2^{L}}$so that $\mathcal{P}_{i}^{i}\cup\mathcal{P}_{j}^{i}\cup\mathcal{U}^{i}$
forms a basis for ${\cal D}_{i}$ }. Let $\nu^{i}$ be the cardinality of $\mathcal{U}^{i}$,
hence
\begin{equation}
\rho(A_{i}^{i})+d_{j}^{i}+\nu^{i}=k_{i},\label{eq:GlobalSum}
\end{equation}
or 
\begin{equation}
\rho(A_{\{i,j\}}^{i})+\nu^{i}=k_{i}.\label{eq:GlobalSum1}
\end{equation}

\end{itemize}
For $M-$dimensional vectors $\boldsymbol{a},\ \boldsymbol{p},$ denote the inner product,
\[
\left\langle \boldsymbol{a},\boldsymbol{p}\right\rangle =\sum_{m=1}^{M}a_{m}p_{m}.
\]

The following observations follow from this construction.

\noindent \textbf{\textcolor{black}{Observations}} 
\begin{enumerate}
\item If receiver $i$ learns the values of all packets of the form $\left\langle \boldsymbol{a},\boldsymbol{p}^{i}\right\rangle ,\ \boldsymbol{a}\in{\cal P}_{j}^{i}\cup{\cal U}^{i}$,
then since $i$ also knows the values of packets $\left\langle \boldsymbol{a},\boldsymbol{p}^{i}\right\rangle ,\ \boldsymbol{a}\in{\cal P}_{i}^{i}$
and t\textcolor{black}{he set $\mathcal{P}_{i}^{i}\cup\mathcal{P}_{j}^{i}\cup\mathcal{U}^{i}$
is }a basis for the whole space, the receiver can decode packets $\boldsymbol{p}^{i}$. 
\item Receiver $j$ knows the values of all packets $\left\langle \boldsymbol{a},\boldsymbol{p}^{i}\right\rangle ,\ \boldsymbol{a}\in{\cal P}_{j}^{i}$.
\end{enumerate}
Figure \ref{fig:LS1} shows the structure of the various subspaces defined in the Preprocessing
construction. In the figure, ${\cal D}_{i}$ is a $k_{i}$-dimensional space. If receiver $i$
knows the inner product of its packet vector $\boldsymbol{p}^{i}$ with each of $k_{i}$ linearly
independent vectors of ${\cal D}_{i}$, then the receiver can decode $\boldsymbol{p}^{i}.$
The subspace of ${\cal D}_{i}$ spanned by the rows of $A_{i}^{i}$ (rows of $A_{j}^{i}$)
represents the side information receiver $i$ (receiver $j$) has about the packets of receiver
$i$. In the figure we refer to this as ``subspace known by receiver $i$ (receiver $j$ )
through its SI, about the packets of receiver $i$''. As will be seen shortly, upon reception
of a transmitted packet, receiver $i$ may obtain knowledge of the inner product of $\boldsymbol{p}^{j}$
with a vector, say $\boldsymbol{v}$, in ${\cal D}^{j}$ , where $i\in[2]$ . At transmission
time $t$, we refer to the subspace spanned by these vectors ($\boldsymbol{v}$ ) together
with the rows of matrix $A_{i}^{j}$ as the ``subspace known by $i$ about the packets of
receiver $j$ at time $t$''.

\begin{center}
\begin{figure}[tbh]
\begin{centering}
\includegraphics[scale=0.45]{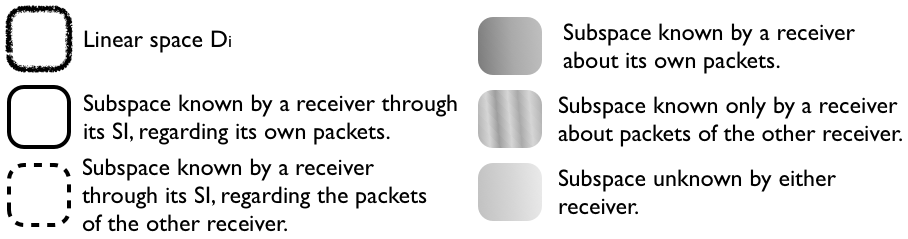} 
\par\end{centering}

\centering{}\textcolor{black}{\includegraphics[scale=0.45]{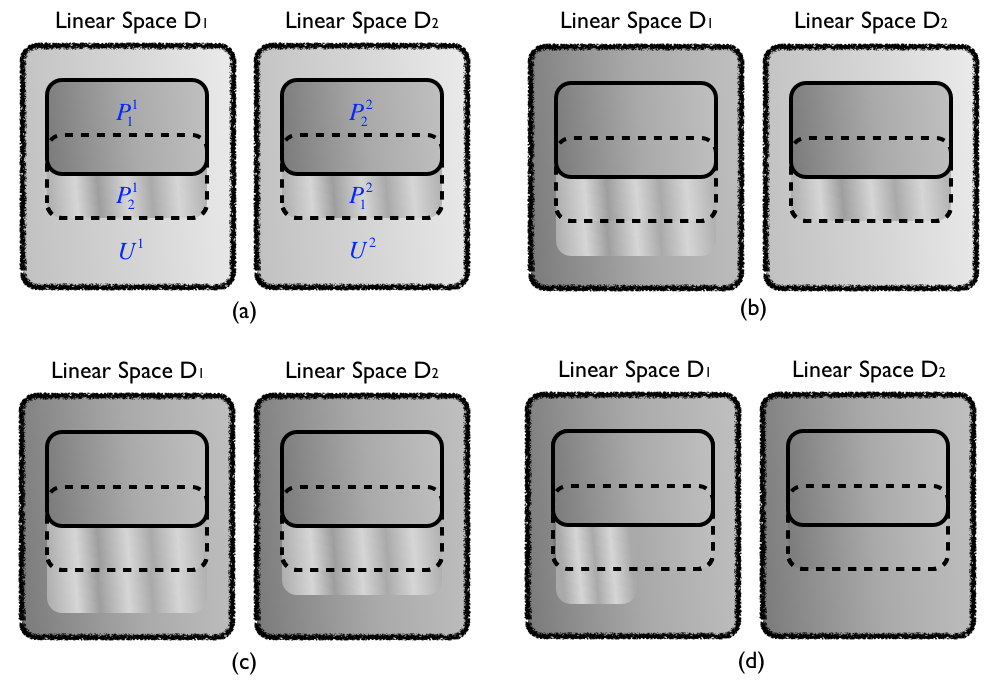}\caption{\label{fig:LS1}Structure of subspaces at preprocessing and during algorithm operations}
}
\end{figure}

\par\end{center}

Based on the above construction and observations, the following algorithm is proposed. The
objective of the algorithm is to ensure that each receiver $i\in[2]$ receives successfully
all packets of the form $\left\langle \boldsymbol{a},\boldsymbol{p}^{i}\right\rangle ,\ \boldsymbol{a}\in{\cal P}_{j}^{i}\cup{\cal U}^{i}$
. In the algorithm, whenever referring to node indexes $i,\ $$j$ it is assumed that $i\neq j$.

\newpage{}

\textbf{\textcolor{black}{Algorithm I }}

\emph{Phase 1} 
\begin{enumerate}
\item \label{enu:Ph1.1}If $\nu^{1}=0$ skip this part. Else,

\begin{enumerate}
\item Form packets $\left\langle \boldsymbol{a},\boldsymbol{p}^{1}\right\rangle ,\ \boldsymbol{a}\in{\cal U}^{1}$
.
\item Transmit each of the formed packets until each one is received by either one of receivers $1,\ 2.$
Let $K_{2}^{1}$ be the number of these packets that are erased at receiver $1$ and received
by receiver $2$. Denote by $\left\{ \boldsymbol{q}_{2,l}^{1}\right\} _{l\in[K_{2}^{1}]}$
the set of theses packets. Figure \ref{fig:LS1}b represents the knowledge space of the receivers
at this point of the algorithm. 
\end{enumerate}

\begin{flushleft}
\label{enu:Ph1.2}Repeat Step \ref{enu:Ph1.1} by replacing $1\leftarrow2$ and $2\leftarrow1$.
Figure \ref{fig:LS1}c represents the knowledge space of the receivers at this point of the
algorithm. 
\par\end{flushleft}

\end{enumerate}
\emph{Note:}

At the end of Phase \ref{enu:Ph1.2} , receiver $i$ has received $\nu^{i}-K_{j}^{i}$ packets
of the form $\left\langle \boldsymbol{a},\boldsymbol{p}^{i}\right\rangle ,\ \boldsymbol{a}\in{\cal U}^{i}$
. Hence, to be able to decode correctly, it must receive the $K_{j}^{i}$ packets $\left\{ \boldsymbol{q}_{j,l}^{i}\right\} _{l\in[K_{j}^{i}]}$
as well as packets the $d_{j}^{i}$ packets $\left\{ \left\langle \boldsymbol{a},\boldsymbol{p}^{i}\right\rangle ,\ \boldsymbol{a}\in{\cal {\cal P}}_{j}^{i}\right\} $;
denote by $\left\{ \boldsymbol{\widetilde{q}}_{j,l}^{i}\right\} _{l\in[K_{j}^{i}+d_{j}^{i}]}$
the set of all these packets and note that they are known by receiver $j.$ Hence one can apply
Network Coding to deliver the remaining packets to their destinations. This technique is based
on the fact that if packet $q=q_{1}^{2}+q_{2}^{1}$ is sent, where packet $q_{j}^{i}$ is a
packet with destination receiver $i$, unknown to $i$ but known to receiver $j$, then any
receiver that receives packet $q$ can decode the packet destined to it. 

\emph{Phase 2 }

\textcolor{black}{If for at least one of the receivers, say receiver $i$ it holds $K_{j}^{i}+d_{j}^{i}=0$
move to Phase 3. }

\textcolor{black}{We denote by $t$ the number of transmissions since the beginning of Phase
2. At the beginning of Phase 2, packet $X_{0}=\widetilde{q}_{1,1}^{2}+\widetilde{q}_{2,1}^{1}$
is transmitted. If at time $t>0$ packet $X_{t}=\widetilde{q}_{1,l_{1}}^{2}\oplus\widetilde{q}_{2,l_{2}}^{1}$
is sent, at time $t+1$ the transmitted packet, depending on the received feedback $\boldsymbol{Z},$
is 
\[
X_{t+1}=\begin{cases}
X_{t} & \mbox{if}\:\boldsymbol{Z}=[0,0]\\
\widetilde{q}_{1,l_{1}}^{2}+\widetilde{q}_{2,l_{2}+1}^{1} & \mbox{if}\:\boldsymbol{Z}=[0,1]\\
\widetilde{q}_{1,l_{1}+1}^{2}+\widetilde{q}_{2,l_{2}}^{1} & \mbox{if}\:\boldsymbol{Z}=[1,0]\\
\widetilde{q}_{1,l_{1}+1}^{2}+\widetilde{q}_{2,l_{2}+1}^{1} & \mbox{if}\:\boldsymbol{Z}=[1,1]
\end{cases},
\]
}This process continues until at least one of the receivers, say receiver $i$, receives all
packets $\left\{ \boldsymbol{\widetilde{q}}_{j,l}^{i}\right\} _{l\in[K_{j}^{i}+d_{j}^{i}]}$.
Figure \ref{fig:LS1}d represents the knowledge space of the receivers at this point of the
algorithm.

\emph{Phase 3}

\textcolor{black}{Transmit the remaining packets (if any) that receiver $j$ needs to receive.}

\noindent \textbf{\textcolor{black}{Performance of the Algorithm}}

Let $\left\lceil n\boldsymbol{r}\right\rceil =\left(\left\lceil nr_{1}\right\rceil ,\ \left\lceil nr_{2}\right\rceil \right)$
be the vector of packets destined to each of the receivers. Let $T$$\left(\left\lceil n\boldsymbol{r}\right\rceil \right)$
be the (random) time it takes for all packets to be delivered to their destinations under Algorithm
I. We then have the following result. 
\begin{prop}
Assume that the following limits exist for $i\in[2],$ 
\[
\hat{\rho}_{i}^{i}(\boldsymbol{r})=\lim_{n\rightarrow\infty}\frac{\rho\left(A_{i,n}^{i}\right)}{n},\ \hat{\rho}_{\{i,j\}}^{i}(\boldsymbol{r})=\lim_{n\rightarrow\infty}\frac{\rho\left(A_{\{i,j\},n}^{i}\right)}{n}.
\]
Then it holds, 
\begin{equation}
\lim_{n\rightarrow\infty}\frac{T\left(\left\lceil n\boldsymbol{r}\right\rceil \right)}{n}=\max\left\{ \frac{r_{1}-\hat{\rho_{1}^{1}(\boldsymbol{r})}}{1-\epsilon_{1}}+\frac{r_{2}-\hat{\rho}_{\{1,2\}}^{2}(\boldsymbol{r})}{1-\epsilon_{\{1,2\}}},\ \frac{r_{2}-\hat{\rho_{2}^{2}}(\boldsymbol{r})}{1-\epsilon_{2}}+\frac{r_{1}-\hat{\rho}_{\{1,2\}}^{1}(\boldsymbol{r})}{1-\epsilon_{\{1,2\}}}\right\} \triangleq\hat{T}\left(\boldsymbol{r}\right).\label{eq:TotaleLength}
\end{equation}
\end{prop}
\begin{IEEEproof}
The proof follows along the lines of corresponding proof in \cite{gatzianas2013Broadcast}
and is based on the strong law of large numbers. We outline the steps here.

Note that the conditions of the Proposition and (\ref{eq:GlobalSum}), (\ref{eq:GlobalSum1})
imply that the following limits exist 
\[
\hat{\nu}^{i}(\boldsymbol{r})=\lim_{n\rightarrow\infty}\frac{\nu^{i}}{n}=r_{i}-\hat{\rho}_{\{i,j\}}^{i}(\boldsymbol{r}),
\]
\[
\hat{d}_{j}^{i}(\boldsymbol{r})=\lim_{n\rightarrow\infty}\frac{d_{j}^{i}}{n}=\hat{\rho}_{\{i,j\}}^{i}(\boldsymbol{r})-\hat{\rho}_{i}^{i}(\boldsymbol{r}).
\]
Henceforth we assume that the limits $v^{i}(\boldsymbol{r})$ are positive. The special cases
of zero values are easily dealt with.

Let $T_{i}^{1}$ be the time it takes for Phase 1 part $i$ to complete. i.e., to deliver the
$\nu^{i}$ packets to either one of the destinations. It follows from the Strong Law of Large
Numbers and the positivity of $\hat{\nu}^{i}(\boldsymbol{r})$ that 
\begin{equation}
\lim_{n\rightarrow\infty}\frac{T_{i}^{1}}{n}=\frac{\hat{\nu}^{i}(\boldsymbol{r})}{1-\epsilon_{\{1,2\}}},\label{eq:lim1}
\end{equation}
\begin{equation}
\lim_{n\rightarrow\infty}\frac{K_{j}^{i}}{n}=\frac{\hat{\nu}^{i}(\boldsymbol{r})\left(\epsilon_{\{1\}}-\epsilon_{\{1,2\}}\right)}{1-\epsilon_{\{1,2\}}}.\label{eq:lim2}
\end{equation}
Let $T_{i}^{2}$ be the time it takes in Phases $2$ and $3$ to deliver the remaining $K_{j}^{i}+d_{j}^{i}$
packets to destination $i$. Applying again the Strong Law of Large Numbers and taking into
account (\ref{eq:lim2}) we have 
\begin{equation}
\lim_{n\rightarrow\infty}\frac{T_{i}^{2}}{n}=\frac{\frac{\hat{\nu}^{i}(\boldsymbol{r})\left(\epsilon_{\{1\}}-\epsilon_{\{1,2\}}\right)}{1-\epsilon_{\{1,2\}}}+\hat{d}_{j}^{i}}{1-\epsilon_{\{i\}}}.\label{eq:lim3}
\end{equation}
Because of the operation of the algorithm in Phases $2$ and $3$, the time $T_{12}^{2}$ it
takes to deliver the remaining packets to both destinations is 
\begin{equation}
T_{12}^{2}=\max\left\{ T_{1}^{2},\ T_{2}^{2}\right\} .\label{eq:maxrel}
\end{equation}
and the total time for the algorithm to complete is, 
\begin{equation}
T\left(\left\lceil n\boldsymbol{r}\right\rceil \right)=T_{1}+T_{2}+T_{12}^{2}.\label{eq:sumtime}
\end{equation}
Dividing (\ref{eq:sumtime}) by $n$, taking limits utilizing (\ref{eq:lim1})-(\ref{eq:maxrel})
the result follows. 
\end{IEEEproof}
Now we are ready to show the partial converse to Corollary \ref{cor:2ReceiverNecessity-12}. 
\begin{cor}
\label{cor:PartialConverse}Assume that the following limits exist for $i,j\in\{1,2\}.$ 
\[
\hat{\rho}_{i}^{i}(\boldsymbol{r})=\lim_{n\rightarrow\infty}\frac{\rho\left(A_{i,n}^{i}\right)}{n},\ \hat{\rho}_{\{i,j\}}^{i}(\boldsymbol{r})=\lim_{n\rightarrow\infty}\frac{\rho\left(A_{\{i,j\},n}^{i}\right)}{n}.
\]
If the rate vector $\boldsymbol{r}=\left(r_{1},r_{2}\right)$ satisfies 
\[
\hat{T}\left(\boldsymbol{r}\right)=\max\left\{ \frac{r_{1}-\hat{\rho}_{1}^{1}(\boldsymbol{r})}{1-\epsilon_{1}}+\frac{r_{2}-\hat{\rho}_{\{1,2\}}^{2}(\boldsymbol{r})}{1-\epsilon_{\{1,2\}}},\ \frac{r_{2}-\hat{\rho}_{2}^{2}(\boldsymbol{r})}{1-\epsilon_{2}}+\frac{r_{1}-\hat{\rho}_{\{1,2\}}^{1}(\boldsymbol{r})}{1-\epsilon_{\{1,2\}}}\right\} <1,
\]
then $\boldsymbol{r}$ is achievable. \end{cor}
\begin{IEEEproof}
The proof is identical to the proof in \cite{gatzianas2010multiuser}. We present it here for
completeness. Consider the following code.

1. Use Algorithm I to transmit $\left\lceil n\boldsymbol{r}\right\rceil $ packets.

2. If $T\left(\left\lceil n\boldsymbol{r}\right\rceil \right)\leq n$ then all receivers receive
correctly their packets.

3. Else declare error.

The probability of error of this code is computed as follows. 
\begin{eqnarray*}
\lim_{n\rightarrow\infty}p_{e}(n) & = & \lim_{n\rightarrow\infty}\Pr\left(T\left(\left\lceil n\boldsymbol{r}\right\rceil \right)>n\right)\\
 & = & \lim_{n\rightarrow\infty}\Pr\left(\frac{T\left(\left\lceil n\boldsymbol{r}\right\rceil \right)}{n}>1\right)\\
 & = & \lim_{n\rightarrow\infty}\Pr\left(\frac{T\left(\left\lceil n\boldsymbol{r}\right\rceil \right)}{n}-\hat{T}\left(\boldsymbol{r}\right)>1-\hat{T}\left(\boldsymbol{r}\right)\right)\\
 & = & 0\ \mbox{by (}\ref{eq:TotaleLength}).
\end{eqnarray*}

\end{IEEEproof}
\emph{Note:} Corollaries \ref{cor:2ReceiverNecessity-12} and \ref{cor:PartialConverse} imply
the following upper and lower bounds for the capacity of the channel under consideration. Let
$\mathcal{C}_{L}=\left\{ \boldsymbol{r}:\ \hat{T}\left(\boldsymbol{r}\right)<1\right\} $ and
$\mathcal{C}_{U}=\left\{ \boldsymbol{r}:\ \hat{T}\left(\boldsymbol{r}\right)\leq1\right\} .$
Then,

\[
\mathcal{C}_{L}\subseteq\overline{\mathcal{C}}_{L}\subseteq\mathcal{C}\subseteq\overline{\mathcal{C}}_{U},
\]
where $\mathcal{\bar{C}}$ denotes the closure of set ${\cal C}$. The next corollary provides
conditions on $\hat{T}\left(\boldsymbol{r}\right)$ under which the capacity can be completely
characterized. Its proof can be found in Appendix B. 
\begin{cor}
\label{cor:CapacityDescription}If for any point $\boldsymbol{r}$ such that $\hat{T}\left(\boldsymbol{r}\right)=1$
there is a sequence $\boldsymbol{r}_{k},\ k=1,2...$ with $\hat{T}\left(\boldsymbol{r}_{k}\right)<1$
and $\lim_{k\rightarrow\infty}\boldsymbol{r}_{k}=\boldsymbol{r},$ then $\mathcal{C}=\mathcal{\bar{C}}_{L}=\mathcal{\bar{C}}_{U}$.

Moreover, if $\hat{T}\left(\boldsymbol{r}\right)$ is continuous function of $\boldsymbol{r}$
then, 
\begin{equation}
\mathcal{C}=\left\{ \boldsymbol{r}:\ \hat{T}\left(\boldsymbol{r}\right)\leq1\right\} .\label{eq:capacity2user}
\end{equation}

\end{cor}

\section{Related Index Coding results}

In Index Coding it is assumed that a) each receiver either knows the message of some other
receiver, or has no knowledge of it - we call this type of side information \emph{All-or-Nothing},
and b) that the channel is errorless. The current work represents a generalization of the Index
Coding problem in the following sense: a) the side information consists of functions of messages
and $b)$ transmission erasures may occur. In the following we examine the implications of
our approach when All-or-Nothing side information is available.

\textcolor{black}{The earlier works on index coding consider the problem of delivering messages
to a number of receivers in shortest time \cite{bar2006index}, \cite{alon2008broadcasting},
\cite{el2010index}, \cite{blasiak2013broadcasting}. Subsequent work considered the problem
of determining the capacity of the channel \cite{sun2013index}, \cite{arbabjolfaei2013capacity}.
As we will see, the two probl}ems are closely related, and conclusions of one can be transformed
into conclusion of the other.

\subsection{\label{sub:All-or-Nothing}All-or-Nothing Side Information}

Assume that node $i$ either knows\textcolor{black}{{} the whole of message $W_{j,n}$ or has
no knowledge of it, i.e., either $h_{i,n}^{j}(W_{j,n})=W_{j,n}$ or $h_{i,n}^{j}(W_{j,n})=c$,
a constant. Construct the ``information graph'' \cite{bar2006index},}\textcolor{red}{{} }$G=([N],\mathcal{E})$
where an edge $(i,j)$ belongs to $\mathcal{E}$ iff node $i$ knows the message of node $j$.
We denote the set of outgoing neighbors of node $i$ in the information graph by $\mathcal{N}_{0}\left(i\right).$
For the rest of the paper we consider that the information graph is independent of $n$.

Since $\boldsymbol{S}_{[i],n}^{i}=W_{i,n}$ if and only if $i\in\mathcal{N}_{0}(j)$ for some
$j\in[i],$ it follows that, 
\[
H\left(\boldsymbol{S}_{[i],n}^{i}\right)=\left\{ \begin{array}{cc}
0 & \mbox{if \ensuremath{i\notin\mathcal{N}_{0}\left(j\right)}for all \ensuremath{j\in[i]}}\\
\left\lceil nR_{i}\right\rceil  & \mbox{otherwise}
\end{array}\right..
\]
Replacing this in (\ref{eq:mainTh1}) we get for any achievable $\boldsymbol{R}$, 
\[
\sum_{i=1}^{N}\frac{\tilde{R}_{i}}{1-\epsilon_{[i]}}\leq\log|\mathcal{X}|,
\]
where 
\[
\tilde{R}_{i}=\left\{ \begin{array}{cc}
R_{i} & \mbox{if }\ensuremath{i\notin\mathcal{N}_{0}\left(j\right)}\mbox{ for all }\ensuremath{j\in[i]}\\
0 & \mbox{otherwise}
\end{array}\right..
\]

Using (\ref{eq:mainTh2}), a similar inequality is derived for any achievable $\boldsymbol{R}$
and any permutation $\pi_{i}$ of node indexes, and we have the following corollary. 
\begin{cor}
If the information is All-or-Nothing, then for any achievable $\boldsymbol{R}$ it holds for
any permutation $\pi_{i}$ of indexes, 
\end{cor}
\[
\sum_{i=1}^{N}\frac{\tilde{R}_{\pi_{i}}}{1-\epsilon_{\mathcal{B}_{\pi}(i)}}\leq\log|\mathcal{X}|,
\]
where $\mathcal{B}_{\pi}(i)=\left\{ \pi_{1},...,\pi_{i}\right\} $ and 
\[
\tilde{R}_{\pi_{i}}=\left\{ \begin{array}{cc}
R_{\pi_{i}} & \mbox{ if \ensuremath{\pi_{i}\notin\mathcal{N}_{0}\left(\pi_{j}\right)}\ensuremath{\mbox{ for all }}}\ensuremath{\pi_{j}\in\mathcal{B}_{\pi}(i)}\\
0 & \mbox{otherwise}
\end{array}\right..
\]
Consider the set 
\[
\hat{\mathcal{C}}=\left\{ \boldsymbol{R}\geq\boldsymbol{0}:\ \sum_{i=1}^{N}\frac{\tilde{R}_{\pi_{i}}}{1-\epsilon_{\mathcal{B}_{\pi}(i)}}\leq\log|\mathcal{X}|,\ \mbox{for all \ensuremath{\pi}}\right\} .
\]
Since $\hat{\mathcal{C}}$ is closed, we have the following corollary. 
\begin{cor}
\label{cor:On-OffSideInfo}If the side information is All-or-Nothing, it holds, 
\[
\mathcal{C}\subseteq\hat{\mathcal{C}}.
\]

\end{cor}
Consider now the Packetized model of Section \ref{sec:Linear-side-information}

Setting $k_{i}=\left\lceil r_{i}n\right\rceil $, we have $\left|\mathcal{W}_{i}\right|=2^{\left\lceil r_{i}n\right\rceil L}$
and since $\log\left|\mathcal{X}\right|=L$, Corollary \ref{cor:On-OffSideInfo} gets the following
form for the Packetized model. 
\begin{cor}
\textcolor{black}{\label{cor:PacketModel} In the packetized model, if side information is
All-or-Nothing and $\boldsymbol{r}$ is achievable it holds under any node index permutation
$\boldsymbol{\pi}=\left\{ \pi_{i}\right\} _{i=1}^{N},$ 
\[
\sum_{i=1}^{N}\frac{\tilde{r}_{\pi_{i}}}{1-\epsilon_{\mathcal{B}_{\pi}(i)}}\leq1.
\]
where $\mathcal{B}_{\pi}(i)=\left\{ \pi_{1},...,\pi_{i}\right\} $ and 
\[
\tilde{r}_{\pi_{i}}=\left\{ \begin{array}{cc}
r_{\pi_{i}} & \mbox{ if \ensuremath{\pi_{i}\notin\mathcal{N}_{0}\left(\pi_{j}\right)}\ensuremath{\mbox{ for all }}}\ensuremath{\pi_{j}\in\mathcal{B}_{\pi}(i)}\\
0 & \mbox{otherwise}
\end{array}\right..
\]
}
\end{cor}
Denote the capacity region of the Packetized system with respect to $\boldsymbol{r}$ by $\mathcal{C}_{p}$
and 
\[
\hat{\mathcal{C}_{p}}=\left\{ \boldsymbol{r}\geq\boldsymbol{0}:\ \sum_{i=1}^{N}\frac{\tilde{r}_{\pi_{i}}}{1-\epsilon_{\mathcal{B}_{\pi}(i)}}\leq1,\ \mbox{for all \ensuremath{\pi}}\right\} .
\]
Then for the Packetized model Corollary \ref{cor:On-OffSideInfo} becomes: 
\begin{cor}
\label{cor:PacketizedOn-Off}In the Packetized model, if the side information is All-or-Nothing,
it holds, 
\[
\mathcal{C}_{p}\subseteq\hat{\mathcal{C}}_{p}.
\]

\end{cor}
For the rest of the paper, we consider the Packetized model.

Instead of stopping after $n$ channel uses we allow transmissions until all receivers decode
\emph{correctly} the packets destined to them. We use the symbol $\tilde{C}$ to distinguish
this type of codes from the codes (denoted by $C$) that operate up to a fixed number $n$
of channel uses. Let $T_{\tilde{C}}\left(\boldsymbol{k}\right)$ be the (random) earliest time
it takes until all receivers decode correctly their $\boldsymbol{k}$ packets under a code
$\tilde{C}$ and define $T_{\tilde{C}}\left(\boldsymbol{0}\right)=0.$ We call $T_{\tilde{C}}\left(\boldsymbol{k}\right)$
``broadcast time'' under code $\tilde{C}$. Denote also: 
\[
\bar{T}_{\tilde{C}}\left(\boldsymbol{k}\right)\triangleq\mathbb{E}\left[T_{\tilde{C}}\left(\boldsymbol{k}\right)\right],
\]
and 
\[
\bar{T}^{*}\left(\boldsymbol{k}\right)\triangleq\inf_{\tilde{C}}\bar{T}_{\tilde{C}}\left(\boldsymbol{k}\right).
\]
We refer to $\bar{T}^{*}\left(\boldsymbol{k}\right)$ as the ``minimum broadcast time''.

Notice that for any $\boldsymbol{k}$ it holds $\bar{T}^{*}\left(\boldsymbol{k}\right)<\infty$
since the code that retransmits each packet until that packet is received by its corresponding
destination has finite expectation. From the fact that a code that transmits successfully $\boldsymbol{l}+\boldsymbol{m},$
$\boldsymbol{l}\geq\boldsymbol{0},\ \boldsymbol{m}\geq\boldsymbol{0},$ packets can be constructed
by using two codes, one for transmitting the $\boldsymbol{k}$ packets first and another one
for transmitting the remaining $\boldsymbol{m}$ packets, it follows that $\bar{T}^{*}\left(\boldsymbol{k}\right)$
is subadditive, i.e., it holds for any $\boldsymbol{l}$, $\boldsymbol{m},$ $\boldsymbol{l}\geq\boldsymbol{0},\ \boldsymbol{m}\geq\boldsymbol{0},$
\[
\bar{T}^{*}\left(\boldsymbol{l}+\boldsymbol{m}\right)\leq\bar{T}^{*}\left(\boldsymbol{l}\right)+\bar{T}^{*}\left(\boldsymbol{m}\right).
\]
For multidimensional subadditive functions the following theorem holds \cite{bingham2008generic},
\cite{georgiadis2012stability}. 
\begin{thm}
\label{thm:subadditiveTheorem}For any $\boldsymbol{r}\geq\boldsymbol{0},$ the limit function,
\[
\hat{T}\left(\boldsymbol{r}\right)=\lim_{n\rightarrow\infty}\frac{\bar{T}^{*}\left(\left\lceil n\boldsymbol{r}\right\rceil \right)}{n},
\]
exists and is finite, conve\textcolor{black}{x, Lipschitz continuous, and positively homogenous,
i.e., for any $\rho\geq0$, $\hat{T}\left(\rho\boldsymbol{r}\right)=\rho\hat{T}\left(\boldsymbol{r}\right).$} 
\end{thm}
\textcolor{black}{We refer to $\hat{T}\left(\boldsymbol{r}\right)$ as the ``broadcast rate''
of $\bar{T}^{*}\left(\boldsymbol{k}\right)$ in the direction $\boldsymbol{r}.$ }

\textcolor{black}{As will be seen, $\hat{T}\left(\boldsymbol{r}\right)$ determines the capacity
region of the channel, and lower (upper) bounds on $\hat{T}\left(\boldsymbol{r}\right)$ can
be translated to upper (lower) bounds }to the capacity region. The next Theorem describes the
capacity region of the system in terms of $\hat{T}\left(\boldsymbol{r}\right)$. 
\begin{thm}
\label{thm:EaualityPacketized}For the packetized system and All-or-Nothing side information
it holds, 
\[
\mathcal{C}_{p}=\left\{ \boldsymbol{r}\geq\boldsymbol{0}:\ \hat{T}\left(\boldsymbol{r}\right)\leq1\right\} \triangleq\mathcal{R}.
\]
\end{thm}
\begin{IEEEproof}
The proof is basically the same as the proof use\textcolor{black}{d in \cite{georgiadis2012stability}
(an ext}ended version can be found at \cite{georgArxiv}). For completeness we provide the
proof in Appendix \ref{sec:TheoremEquality}. 
\end{IEEEproof}
The next corollary provides a lower bound for $\hat{T}\left(\boldsymbol{r}\right).$ 
\begin{cor}
\label{thm:TimeLowerBound}For the packetized system and All-or-Nothing side information it
holds, 
\[
\hat{T}\left(\boldsymbol{r}\right)\geq\max_{\pi\in\Pi}\sum_{i=1}^{N}\frac{\tilde{r}_{\pi_{i}}}{1-\epsilon_{\mathcal{B}_{\pi}(i)}},
\]
where $\Pi$ is the set of permutations $\pi$ of the set $[N],$ $\mathcal{B}_{\pi}(i)=\left\{ \pi_{1},...,\pi_{i}\right\} $
and 
\[
\tilde{r}_{\pi_{i}}=\left\{ \begin{array}{cc}
r_{\pi_{i}} & \mbox{ if \ensuremath{\pi_{i}\notin\mathcal{N}_{0}\left(\pi_{j}\right)}\ensuremath{\mbox{ for all }}}\ensuremath{\pi_{j}\in\mathcal{B}_{\pi}(i)}\\
0 & \mbox{otherwise}
\end{array}\right..
\]
\end{cor}
\begin{IEEEproof}
Assume first that $\hat{T}\left(\boldsymbol{r}\right)=0$. Then by positive homogeneity, $\hat{T}\left(\rho\boldsymbol{r}\right)=\rho\hat{T}\left(\boldsymbol{r}\right)=0<1$
for all $\rho\geq0,$ and by Theorem \ref{thm:EaualityPacketized} and Corollary \ref{cor:PacketizedOn-Off}
we conclude that for all $\rho\geq0$ it holds, 
\[
\sum_{i=1}^{N}\frac{\rho\tilde{r}{}_{\pi_{i}}}{1-\epsilon_{\mathcal{B}_{\pi}(i)}}\leq1,\ \mbox{for all \ensuremath{\pi(\cdot)}},
\]
which in turn implies that $\tilde{r}_{i}=0$ for all $i\in[N],$ hence the corollary holds.

Assume next that $\hat{T}\left(\boldsymbol{r}\right)>0.$ Setting $\rho=1/\hat{T}\left(\boldsymbol{r}\right)$
and using the positive homogeneity of $\hat{T}\left(\boldsymbol{r}\right)$ we have $\hat{T}\left(\rho\boldsymbol{r}\right)=\rho\hat{T}\left(\boldsymbol{r}\right)=1$.
Hence the vector $\rho\boldsymbol{r}$ belongs to the capacity region. By Corollary \ref{cor:PacketizedOn-Off}
then we have 
\begin{eqnarray*}
\max_{\pi\in\Pi}\sum_{i=1}^{N}\frac{\rho\tilde{r}_{\pi(i)}}{1-\epsilon_{\mathcal{B}_{\pi}(i)}} & \leq & 1\\
 & = & \rho\hat{T}\left(\boldsymbol{r}\right).
\end{eqnarray*}
By canceling $\rho$ from both sides of the last inequality, we see that the corollary still
holds. 
\end{IEEEproof}

\subsection{\label{sub:AllAndPerfect}All-or-Nothing Side Information and Errorless Channel}

In this section we consider again All-of-Nothing side information and develop the form of the
lower bound for the case where the channel is errorless, $\epsilon_{\{i\}}=0$ for all $i\in[N]$.
In this case the bound in Corollary \ref{thm:TimeLowerBound} becomes 
\begin{equation}
\hat{T}\left(\boldsymbol{r}\right)\geq\max_{\pi\in\Pi}\sum_{i=1}^{N}\tilde{r}_{\pi_{i}},\label{eq:lowerBoud}
\end{equation}
where 
\[
\tilde{r}_{\pi_{i}}=\left\{ \begin{array}{cc}
r_{\pi_{i}} & \mbox{ if \ensuremath{i\notin\mathcal{N}_{0}\left(j\right)}\ensuremath{\mbox{ for all }}}\ensuremath{j\in\mathcal{B}_{\pi}(i)}\\
0 & \mbox{otherwise}
\end{array}\right.,
\]
$\mathcal{B}_{\pi}(i)=\left\{ \pi_{1},...,\pi_{i}\right\} $.

Consider the information gr\textcolor{black}{aph $G=([N],\mathcal{E})$ and associate with
node $i$ the weight $r_{i}$. For a node set $\mathcal{S}\subseteq[N]$, define its weight
as 
\[
W_{\boldsymbol{r}}\left(\mathcal{S}\right)=\sum_{i\in\mathcal{S}}r_{i}.
\]
}

\textcolor{black}{Let $\mathcal{I}$ be the set of all subsets of nodes in the graph whose
induced subgraph is acyclic. The next theorem shows that in the current setup the lower bound
on $\hat{T}\left(\boldsymbol{r}\right)$ in (\ref{eq:lowerBoud}) is the same as the maximum
weight among the sets in $\mathcal{I}$ , i.e., the Maximum Weighted Acyclic Induced Subgraph
(MWAIS) of $G$. Thus, in this case we get a generalization of the MAIS bound developed in
\cite{bar2006index} for the case where all nodes had the same number of bits to transmit.
The same argument implies the upper bound on channel capacity $\left\{ \boldsymbol{r}:\ \max_{\mathcal{\mathcal{S}}\in\bar{\mathcal{I}}}W_{\boldsymbol{r}}\left(\mathcal{S}\right)\leq1\right\} $
developed in \cite{arbabjolfaei2013capacity}.} 
\begin{lem}
It holds 
\[
\max_{\pi\in\Pi}\sum_{i=1}^{N}\tilde{r}_{\pi_{i}}=\max_{\mathcal{\mathcal{S}}\in\mathcal{I}}W_{\boldsymbol{r}}\left(\mathcal{S}\right)\triangleq W_{\boldsymbol{r}}^{*}.
\]
\end{lem}
\begin{IEEEproof}
We first show that $W_{\boldsymbol{r}}^{*}\leq\max_{\pi\in\Pi}\sum_{i=1}^{N}\tilde{r}_{\pi_{i}}$.
Let $\mathcal{S}^{*}\in\mathcal{I}$ be a set of nodes with maximum weight, and maximum cardinality,
i.e., 
\[
W_{\boldsymbol{r}}(\mathcal{S}^{*})=W_{\boldsymbol{r}}^{*},
\]
and the cardinality of $\mathcal{S}^{*}$ is at least as large as the cardinality of any other
set in $\mathcal{I}$ with maximum weight. Consider the permutation $\tilde{\pi}$ that puts
first the nodes in $\mathcal{S}^{*}$ according to the reverse topological order \cite{leiserson2001introduction}\textcolor{magenta}{{}
}of the subgraph that $\mathcal{S}^{*}$ induces. Here, ``reverse'' means the topological
order obtained by reversing the directions of all links in the induced subgraph of $\mathcal{S}^{*}.$
Hence, it holds for $1\leq i\leq\left|\mathcal{S}^{*}\right|$: $\tilde{\pi_{i}}\notin\mathcal{N}_{0}\left(\tilde{\pi}_{j}\right)\mbox{ \ensuremath{\mbox{ for all }}}\ensuremath{\tilde{\pi_{j}}\in\mathcal{B}_{\tilde{\pi}}(i)}$,
that is, $\tilde{\pi}_{i}$ is not outgoing neighbor of any of the nodes in $\ensuremath{\mathcal{B}_{\tilde{\pi}}(i)},$
which implies that $\tilde{r}_{\tilde{\pi}_{i}}=r_{\tilde{\pi}_{i}}$ for $\tilde{\pi}_{i}\in\mathcal{S}^{*}$.
The rest of the nodes are placed in any order after $\tilde{\pi}_{\left|\mathcal{S}^{*}\right|}$.

Observe that any node $i$ not in $i\notin\mathcal{S}^{*}$ must be an outgoing neighbor for
some of the nodes in $\mathcal{S}^{*}$ and hence $\tilde{r}_{\tilde{\pi}_{i}}=0$ for $\tilde{\pi}_{i}\notin\mathcal{S}^{*}$.
This is so since otherwise the set $\mathcal{S}^{*}\cup\{i\}$ induces an acyclic graph with
weight at least as large as $W_{\boldsymbol{r}}(\mathcal{S}^{*})$ and larger cardinality,
a contradiction.

By construction then it follows, 
\begin{eqnarray*}
\sum_{i=1}^{N}\tilde{r}_{\tilde{\pi}_{i}} & = & \sum_{i=1}^{\left|\mathcal{S}^{*}\right|}\tilde{r}_{\tilde{\pi}_{i}}+\sum_{i=\left|\mathcal{S}^{*}\right|+1}^{N}\tilde{r}_{\tilde{\pi}_{i}}\\
 & = & W_{\boldsymbol{r}}(\mathcal{S}^{*})+0.
\end{eqnarray*}
Hence we conclude that 
\[
W_{\boldsymbol{r}}^{*}\leq\max_{\pi\in\Pi}\sum_{i=1}^{N}\tilde{r}_{\pi_{i}}.
\]
Next we prove the reverse inequality. Consider any permutation $\pi$ and construct the node
set $\mathcal{S}_{\pi}$ as follows: Node $\pi_{i}$ is included in the set $\mathcal{S}_{\pi}$
if and only if $i$ is not an outgoing neighbor of any of the nodes in $\mathcal{B}_{\pi}(i)$.
The induced subgraph of $\mathcal{S}_{\pi}$ is acyclic. To see this, assume that there is
a cycle in this subgraph and consider the node $\pi_{i_{0}}$ in the cycle which is the ``largest''
in the permutation, i.e., $i_{0}>\mbox{\ensuremath{j}}$ for any node $\pi_{j}$ in the cycle.
Then, $i_{0}$ must be an outgoing neighbor of some node in $\mathcal{B}_{\pi}(i_{0})$, which
contradicts the definition of $\mathcal{S}_{\pi}$. Observe also that by construction of $\mathcal{S}_{\pi}$
it holds 
\[
\sum_{i=1}^{N}\tilde{r}_{\pi_{i}}=W_{\boldsymbol{r}}(\mathcal{S}_{\pi}),
\]
hence, 
\[
\max_{\pi\in\Pi}\sum_{i=1}^{N}\tilde{r}_{\pi_{i}}=\max_{\pi\in\Pi}W_{\boldsymbol{r}}(\mathcal{S}_{\pi})\leq\max_{\mathcal{\mathcal{S}}\in\mathcal{I}}W_{\boldsymbol{r}}\left(\mathcal{S}\right)=W_{\boldsymbol{r}}^{*}.
\]

\end{IEEEproof}

\subsection{\label{sub:Tightness-of-bound}Tightness of bound for certain types of information graphs}

For a two-receiver channel since All-or-Nothing side information is linear, we conclude from
Section \ref{sec:Linear-side-information} that the bound in Corollary \ref{thm:TimeLowerBound}
is tight.

Next we examine the tightness of the bound for some types of information graphs assuming errorless
channel. In the discussion below, assuming that $k_{i}$ is the number of packets to be deli\textcolor{black}{vered
to receiver $i$, and under specific conditions on the information graph, we propose codes
and calculate the time $T\left(\boldsymbol{k}\right)$ it takes for all packets to be delivered
to their destinations. To evaluate how close the algorithm performs to the developed bound
we compare the bound with $\lim_{n\rightarrow\infty}\left(T(\left\lceil n\boldsymbol{r}\right\rceil /n\right)$.
We say that a code }\textcolor{black}{\emph{achieves}}\textcolor{black}{{} $W_{\boldsymbol{r}}^{*}$
if $\lim_{n\rightarrow\infty}\left(T(\left\lceil n\boldsymbol{r}\right\rceil /n\right)=W_{\boldsymbol{r}}^{*}$
. Also, by $Q_{i}$ we denote the set or ``queue'' of packets destined to receiver $i$ and
by $p_{i}^{k}$ the $k$-th packet in $Q_{i}.$ All the proofs of this section can be found
in Appendix \ref{sec:Proof-of-Lemmas}.}

\textcolor{black}{The next corollary considers simple cases of $G$. } 
\begin{cor}
\label{cor:CaseDirected}1. If $G$ is acyclic, then for any $\boldsymbol{r}\geq\boldsymbol{0},$
\textup{$W_{\boldsymbol{r}}^{*}=\sum_{i=1}^{N}r_{i}$} and this bound can be achieved by transmitting
each packet separately, hence no benefit can be obtained by coding.

2. If $G$ is a simple (directed) cycle then for any $\boldsymbol{r}\geq\boldsymbol{0},$ 
\begin{equation}
W_{\boldsymbol{r}}^{*}=\sum_{i=1}^{N}r_{i}-\min_{i\in[N]}r_{i}.\label{eq:dircycle}
\end{equation}
This bound can be achieved by pairwise XOR coding operations. 
\end{cor}
In the following, the information graph will be called undirected if whenever link $(i,j)$
belongs to the graph, then link $(j,i)$ belongs as well; hence nodes $i$ and $j$ know each
others' messages. For undirected graphs, $(i,j)$ and $(j,i)$ are considered the same link,
i.e. the order of endpoints does not play a role. For undirected information graphs the set
$\mathcal{I}$ consists of all \emph{independent} nodes sets of the graph and $W_{\boldsymbol{r}}^{*}$
is the\emph{ maximum }weight among the weights of the independent sets of $G$. This is so
since any two neighbors in an undirected graph form a cycle and hence cannot both belong to
an induced acyclic subgraph. We denote by $\mathcal{I}^{*}$ the set of all independent subsets
of $G$ with maximum weight.

For the rest of the paper we consider undirected graphs, which we simply refer to as ``graph''.
We relate $T\left(\boldsymbol{k}\right)$ to $W_{\boldsymbol{k}}^{*}$ with the understanding
that the results can be converted to ``achievability'' through following Corollary \ref{cor:Achieve}. 
\begin{cor}
\label{cor:Achieve}If for some code $T\left(\boldsymbol{k}\right)=W_{\boldsymbol{k}}^{*}$
for all vectors with nonnegative integer components, then for all $\boldsymbol{r}\geq\boldsymbol{0}$,
where $r_{i},i\in[N]$ are real numbers it holds, 
\[
\lim_{n\rightarrow\infty}\frac{T\left(\left\lceil n\boldsymbol{r}\right\rceil \right)}{n}=\lim_{n\rightarrow\infty}\frac{W_{\left\lceil n\boldsymbol{r}\right\rceil }^{*}}{n}=W_{\boldsymbol{r}}^{*},
\]
i.e., the code achieves $W_{\boldsymbol{r}}^{*}.$ 
\end{cor}

\subsubsection*{Trees and Forests }

The next proposition shows achievability when the information graph is a forest. 
\begin{prop}
\label{lem:10-tree}If the information graph is a tree then for any $\boldsymbol{k}\geq\boldsymbol{0}$
there is a code using pairwise XOR operations and $T\left(\boldsymbol{k}\right)=W_{\boldsymbol{k}}^{*}$.
The same holds if the information graph is a forest. 
\end{prop}

\subsubsection*{Cycles}

We now look at cycles. For convenience in the notation, for a cycle $\mathcal{C}=(1,2,...,N,1)$,
node index $N+l,\ 1\leq l\leq N$ is identified with node $l$, and node index $-l,\ 1\leq l\leq N$
is identified with node $N+1-l$. We use the same convention when we refer to indexes of a
subset $\{i_{1},...,i_{m}\}$ of the cycle nodes, by replacing $N$ with $m.$ Also, for $i\neq j$
we denote by $d(i,j)$ the number of links in the path $(i,i+1,....,j)$. Note that in this
notation in general $d(i,j)\neq d(j,i).$

In the discussion below we will need to compare weights between a number of graphs. In such
cases, we use the graph as an index to the quantity related to this graph. For example, $W_{G,\boldsymbol{k}}^{*}$
is the maximum weight among the independent sets of graph $G$ with node weights $\boldsymbol{k}.$

The following proposition describes cases for which codes with completion time $W_{\boldsymbol{k}}^{*}$
can be designed for cycles. 
\begin{prop}
\label{prop:CyclesGeneral}Let the information graph be a cycle $\mathcal{C}=(1,2,...,N,1)$.
If either one of the following conditions holds, 
\begin{enumerate}
\item \label{enu:CyclesGeneralCase1}There is at least one node with zero weight; 
\item \label{enu:CyclesGeneralCase2}$k_{i}>0$ for all $i\in\{1,...,N\}$ and there is an independent
set $\mathcal{S}=\left\{ i_{1},...,i_{\left|\mathcal{S}\right|}\right\} \in\mathcal{I}^{*},$
where $i_{1}<i_{2}<...<i_{\left|\mathcal{S}\right|}$, such that $d(i_{m},i_{m+1})=3$ and
$d(i_{k},i_{k+1})=3$ for some $i_{m},i_{k}\in\mathcal{S},\ i_{k}\neq i_{m};$ 
\end{enumerate}
then there is a code using pairwise XOR operations with $T\left(\boldsymbol{k}\right)=W_{\boldsymbol{k}}^{*}.$ 
\end{prop}
The next proposition examines achievability for even cycles. 
\begin{prop}
\label{lem:11}If the information graph is an even cycle $\mathcal{C}=(1,2,...,N,1)$ with
$N=2g,\ g\geq1$, then for any $\boldsymbol{k}\geq\boldsymbol{0}$ there is a code using pairwise
XOR operations with $T\left(\boldsymbol{k}\right)=W_{\boldsymbol{k}}^{*}.$ 
\end{prop}
The next proposition concerns odd cycles. In this case, the proposed codes are not always achieving
$\hat{T}\left(\boldsymbol{r}\right),$ but the ratio of the deviation from the lower bound
gets close to zero as $N$ increases. Also, for $N=5$ the proposed code achieves channel capacity
as derived in \cite{arbabjolfaei2013capacity}. 
\begin{prop}
\label{lem:OddCycle}Let the information graph be an odd cycle $\mathcal{C}=(1,2,...,N,1)$
with $N=2g+1,\ g\geq2$. Then there is a code using pairwise XOR operations with 
\[
T\left(\boldsymbol{k}\right)\leq W_{\boldsymbol{k}}^{*}+\left\lceil \frac{\min_{i}k_{i}}{2}\right\rceil .
\]
Hence it holds for any $\boldsymbol{r}\geq\boldsymbol{0},$ 
\[
\hat{T}\left(\boldsymbol{r}\right)\leq W_{\boldsymbol{r}}^{*}+\frac{\min_{i}r_{i}}{2}.
\]

\end{prop}
Certain improvements of the algorithm presented in the proof of Proposition \ref{lem:OddCycle}
may be made by taking into account the fact that if case \ref{enu:CyclesGeneralCase2} of Proposition
\ref{prop:CyclesGeneral} holds at some point of the algorithm, then one can employ the algorithm
implied in Proposition \ref{prop:CyclesGeneral} to transmit the remaining packets. In particular,
if case \ref{enu:CyclesGeneralCase2} of Proposition \ref{prop:CyclesGeneral} holds from the
beginning, then $W_{\boldsymbol{k}}^{*}$ can be achieved. However, this improvement complicates
the algorithm while the bound remains the same in the general case.

If $\min_{i}r_{i}>0,$ all sets in ${\cal I}_{\boldsymbol{r}}^{*}$ have size at least $\left\lfloor N/3\right\rfloor $,
hence $W_{\boldsymbol{r}}^{*}\geq\left\lfloor N/3\right\rfloor \min_{i}r_{i}$. Since this
bound also hold when $\min_{i}r_{i}=0$, we have, 
\[
W_{\boldsymbol{r}}^{*}\leq\hat{T}\left(\boldsymbol{r}\right)\leq W_{\boldsymbol{r}}^{*}\left(1+\frac{1}{2\left\lfloor N/3\right\rfloor }\right).
\]
We conclude that for odd cycles there are algorithms whose broadcast rate gets arbitrarily
close to the the lower bound $W_{\boldsymbol{r}}^{*}$ as $N$ increases, \emph{independent}
of node weights.

\subsubsection*{Antiholes}

We now look at antiholes, i.e., the complements of even cycles with $N\geq4$ (the cases with
$N\leq3$ are trivial). We denote an antihole with $N$ nodes by $\mathcal{A}_{N}$ and by
$\mathcal{C}_{N}=\left(1,2,....,N,1\right),\ N\geq4$ the cycle whose complement is the antihole.
According to the definition, for any nodes $i,\ j$ for which $d\left(i,j\right)\geq2$ and
$d\left(j,i\right)\geq2$, there is a link $\left(i,j\right)$ of the antihole. This observation
leads to the following lemma. 
\begin{lem}
\label{lem:EvenHoleLemma}For an antihole, the induced subgraph of any set of nodes $\left\{ i_{1},i_{2},...,i_{m}\right\} $
where $i_{1}<i_{2}<...<i_{m}$ and $d\left(i_{l},i_{l+1}\right)\geq2$ for all $l\in\left\{ 1,...,m\right\} ,$
is a complete graph. 
\end{lem}
The next lemma characterized the maximum weight of independent sets in an antihole. 
\begin{lem}
\label{lem:AntiholeMaxWeight}For an antihole $\mathcal{A}_{N}$, for any $\boldsymbol{k}\geq\boldsymbol{0}$
it holds, 
\[
W_{\boldsymbol{k}}^{*}=\max_{i\in\mathcal{H}_{N}}\left(k_{i}+k_{i+1}\right).
\]

\end{lem}
The next proposition provides a condition under which an algorithm with broadcast completion
time $W_{\boldsymbol{k}}^{*}$ can be designed for antiholes. 
\begin{prop}
\label{prop:AntiholeGeneral}If for an antihole ${\cal A}_{N}$ it holds $k_{i}=0$ for some
node, then there is a code using XOR operations (not necessarily pairwise) with $T(\boldsymbol{k})=W_{\boldsymbol{k}}^{*}.$ 
\end{prop}
We now express the following result concerning even antiholes. 
\begin{prop}
\label{prop:EvenHoleProposition}For an even antihole $\mathcal{A}_{2g}$ and any $\boldsymbol{k}\geq\boldsymbol{0}$
there is a code using XOR operations with $T\left(\boldsymbol{k}\right)=W_{\boldsymbol{k}}^{*}$. 
\end{prop}
The next proposition concerns odd antiholes. 
\begin{prop}
\label{prop:OddAntihole}Let the information graph be an odd antihole ${\cal A}_{2g+1}$. Then
there is a code using XOR operations such that, 
\[
T\left(\boldsymbol{k}\right)\leq W_{\boldsymbol{k}}^{*}+\left\lceil \frac{\min_{i}k_{i}}{\left\lfloor N/2\right\rfloor }\right\rceil .
\]
Hence, 
\[
\hat{T}\left(\boldsymbol{r}\right)\leq W_{\boldsymbol{r}}^{*}+\frac{\min_{i}r_{i}}{\left\lfloor N/2\right\rfloor }.
\]

\end{prop}
Note that since $W_{\boldsymbol{r}}^{*}\geq2\min_{i}r_{i}$ it follows that, 
\[
W_{\boldsymbol{r}}^{*}\leq\hat{T}(\boldsymbol{r})\leq W_{\boldsymbol{r}}^{*}\left(1+\frac{1}{2\left\lfloor N/2\right\rfloor }\right).
\]
As in the case of odd cycles, we conclude that for odd antiholes there are algorithms whose
broadcast rate gets arbitrarily close to the the lower bound $W_{\boldsymbol{r}}^{*}$ as $N$
increases, \emph{independent} of node weights.

\section{Conclusion}

\textcolor{black}{In this work we studied the broadcast erasure channel with feedback and side
information. We provided an upper bound to the capacity region of the system, and showed that
for linear side information and for $N=2$ receivers the bound is achieved. For All-or-Nothing
side information the upper bound on the capacity region is translated to a lower bound of the
broadcast rate of the channel and for the special case of errorless channel, the bound reduces
to MWAIS. Finally, for certain types of information graphs, we provided codes whose broadcast
rate either achieves the lower bound or is close to it and becomes asymptotically tight as
the number of nodes increases. }

The side information considered is of the type $h_{i}^{j}(W_{j})$, where $h_{i}^{j}\left(\cdot\right)$
represents the information receiver $i$ has about the message $W_{j}$ of receiver $j$. The
approach used in this paper can be generalized to include side information of the type $h_{i}\left(W_{1},....,W_{N}\right)$,
i.e., the side information each receiver has is a function of messages of all receivers. This
type of side information may occur naturally in wireless networks where receivers may overhear
transmitted packets which are combinations of messages of others receivers. Forthcoming work
will involve results concerning this type of side information.

For the case of errorless channels, bounds on the capacity region tighter than MWAIS are known
\cite{arbabjolfaei2013capacity}.\textcolor{magenta}{{} }It will be interesting to examine
whether bounds tighter than the one developed in the current paper can be developed when the
channels have erasures.

\textbf{Acknowledgment. }We thank Nestoras Chatzidiamantis for many helpful discussions regarding
this work.

\bibliographystyle{plain}
\bibliography{BECwithSI}

\appendices{}

\section{Proofs of Lemmas in Section \ref{sec:Preliminaries}}

\textcolor{black}{Here we prove Lemmas \ref{lem:1}-\ref{lem:4}. }

\textcolor{black}{For easy reference, we first state the following important facts, which are
direct consequences of the definitions, and will be used extensively in the proofs that follow. }

\textbf{\textcolor{black}{Facts}} 
\begin{enumerate}
\item \textcolor{black}{\label{enu:1-1}$Y_{i}\left(l\right)$ is a deterministic function of $X\left(l\right),Z_{i}\left(l\right)$,
hence of $X\left(l\right),\boldsymbol{Z}\left(l\right)$.} 
\item \textcolor{black}{\label{enu:3-1} If $\boldsymbol{Z}_{[j]}\left(l\right)=(0,...,0)$ then
$\boldsymbol{Y}_{[j]}\left(l\right)=(\varepsilon,...,\varepsilon)$.} 
\item \textcolor{black}{\label{enu:2-1} If $\boldsymbol{Z}_{[j]}\left(l\right)\neq(0,...,0)$ then
$X\mbox{\ensuremath{(l)=}}Y_{i}\left(l\right)$, for some $i\in[j],$ that is, given $\boldsymbol{Z}_{[j]}\left(l\right)=\boldsymbol{z}\neq(0,...,0)$
, $X\left(l\right)$ is a deterministic function of $\boldsymbol{Y}_{[j]}\left(l\right).$} 
\item \textcolor{black}{\label{enu:6-1}$\hat{\boldsymbol{W}}_{[j]}$ is a deterministic function
of $(\boldsymbol{Y}_{[j]}^{n},\boldsymbol{Z}^{n},\boldsymbol{S}_{[j]}).$} 
\end{enumerate}
\textcolor{black}{The following relations, provable by standard information theoretic arguments,
will be used in the proofs that follow. By $\phi\left(\cdot\right)$ we denote a general deterministic
function. }

\textbf{\textcolor{black}{Information Theoretic relations}}

\textcolor{black}{
\begin{eqnarray}
I\left(X;Y,Z\left|W\right.\right) & = & I\left(X;Y\left|W,Z\right.\right)\:\mbox{if \ensuremath{Z}\ensuremath{\mbox{is independent of \ensuremath{\left(X,W\right)}}}}\label{eq:1}\\
H\left(X\left|W,Z=z\right.\right) & = & H\left(X\left|W\right.\right)\:\mbox{if}\: Z\:\mbox{is}\:\mbox{\mbox{independen}t}\:\mbox{of}\:(X,W)\label{eq:2}\\
I\left(X;Y\left|W,Z,\phi\left(Z\right)\right.\right) & = & I\left(X;Y\left|W,Z\right.\right)\label{eq:3}\\
I\left(X;Y\left|W\right.\right) & \leq & I\left(X;Y\left|W,Z\right.\right)+I\left(X;Z\left|W\right.\right)\label{eq:4}
\end{eqnarray}
}\textbf{\textcolor{black}{Lemma \ref{lem:1}.}}\textcolor{black}{\emph{ Assume that the rate
vector $\boldsymbol{R}=\left(R_{1},...,R_{N}\right)$ is achievable. Then, 
\[
n\sum_{i=1}^{j}R_{i}\leq I(\boldsymbol{W}_{[j]};\boldsymbol{Y}_{[j]}^{n},\boldsymbol{Z}^{n},\boldsymbol{S}_{[j]}^{[N]})+o\left(n\right).
\]
}} 
\begin{IEEEproof}
\textcolor{black}{Fix a sequence of codes that achieves $\boldsymbol{R}$ and write, 
\begin{eqnarray*}
n\sum_{i=1}^{j}R_{i} & = & \sum_{i=1}^{j}H\left(W_{i}\right)\;\mbox{since \ensuremath{W_{i}s}are u.d. }\\
 & = & H(\boldsymbol{W}_{[j]})\;\textrm{since \ensuremath{W_{i}}s are i.i.d}\\
 & = & I(\boldsymbol{W}_{[j]};\boldsymbol{Y}_{[j]}^{n},\boldsymbol{Z}^{n},\boldsymbol{S}_{[j]}^{[N]})+H(\boldsymbol{W}_{[j]}\mid\boldsymbol{Y}_{[j]}^{n},\boldsymbol{Z}^{n},\boldsymbol{S}_{[j]}^{[N]})\\
 & = & I(\boldsymbol{W}_{[j]};\boldsymbol{Y}_{[j]}^{n},\boldsymbol{Z}^{n},\boldsymbol{S}_{[j]}^{[N]})+H(\boldsymbol{W}_{[j]}\mid\boldsymbol{Y}_{[j]}^{n},\boldsymbol{Z}^{n},\boldsymbol{S}_{[j]}^{[N]},\hat{\boldsymbol{W}}_{[j]})\mbox{ by Fact \ref{eq:3} and (\ref{eq:3})}\\
 & \leq & I(\boldsymbol{W}_{[j]};\boldsymbol{Y}_{[j]}^{n},\boldsymbol{Z}^{n},\boldsymbol{S}_{[j]}^{[N]})+H(\boldsymbol{W}_{[j]}\mid\hat{\boldsymbol{W}}_{[j]})\:\mbox{since cond. decreases entropy}\\
 & \leq & I(\boldsymbol{W}_{[j]};\boldsymbol{Y}_{[j]}^{n},\boldsymbol{Z}^{n},\boldsymbol{S}_{[j]}^{[N]})+1+\lambda_{n}n\sum_{i=1}^{j}R_{i}\;\mbox{by Fano ineq.}\\
 & = & I(\boldsymbol{W}_{[j]};\boldsymbol{Y}_{[j]}^{n},\boldsymbol{Z}^{n},\boldsymbol{S}_{[j]}^{[N]})+o\left(n\right)\ \mbox{since \ensuremath{\boldsymbol{R}}\ achievable.}
\end{eqnarray*}
} 
\end{IEEEproof}
\textbf{\textcolor{black}{Lemma \ref{lem:2}. }}\textcolor{black}{\emph{If $(U,Q,X(l))$ is
independent of $\boldsymbol{Z}(l)$, it holds: 
\begin{eqnarray*}
I(U;\boldsymbol{Y}_{[j]}\left(l\right),\boldsymbol{Z}\left(l\right)\mid Q) & = & (1-\epsilon_{[j]})I(U;X\left(l\right)\mid Q).
\end{eqnarray*}
}} 
\begin{IEEEproof}
\textcolor{black}{Write}

\textcolor{black}{
\begin{eqnarray}
I\left(U;\boldsymbol{Y}_{[j]}\left(l\right),\boldsymbol{Z}\left(l\right)\left|Q\right.\right) & = & I\left(U;\boldsymbol{Y}_{[j]}\left(l\right)\left|\boldsymbol{Z}\left(l\right),Q\right.\right)\mbox{ by (\ref{eq:1})}\nonumber \\
 & = & \sum_{z\in\mathcal{Z}}I\left(U;\boldsymbol{Y}_{[j]}\left(l\right)\left|\boldsymbol{Z}\left(l\right)=\boldsymbol{z},Q\right.\right)\Pr\left(\boldsymbol{Z}\left(l\right)=\boldsymbol{z}\right)\nonumber \\
 & = & \sum_{z\in\mathcal{Z}_{[j]}^{*}}I\left(U;\boldsymbol{Y}_{[j]}\left(l\right)\left|\boldsymbol{Z}\left(l\right)=\boldsymbol{z},Q\right.\right)\Pr\left(\boldsymbol{Z}\left(l\right)=\boldsymbol{z}\right)\ \mbox{by Fact \ref{enu:3-1}}.\label{eq:l1-1}
\end{eqnarray}
Now, for $\boldsymbol{z}\in\mathcal{Z}_{[j]}^{*}$: 
\begin{eqnarray}
I\left(U;\boldsymbol{Y}_{[j]}\left(l\right)\mid\boldsymbol{Z}\left(l\right)=\boldsymbol{z},Q\right) & = & H\left(U\mid\boldsymbol{Z}\left(l\right)=\boldsymbol{z},Q\right)-H\left(U\mid\boldsymbol{Z}\left(l\right)=\boldsymbol{z},\boldsymbol{Y}_{[j]}\left(l\right),Q\right)\nonumber \\
 & = & H\left(U\mid Q\right)-H\left(U\mid\boldsymbol{Z}\left(l\right)=\boldsymbol{z},\boldsymbol{Y}_{[j]}\left(l\right),Q\right)\;\mbox{by (\ref{eq:2})}\nonumber \\
 & = & H\left(U\mid Q\right)-H\left(U\mid\boldsymbol{Z}\left(l\right)=\boldsymbol{z},\boldsymbol{Y}_{[j]}\left(l\right),X\left(l\right),Q\right)\;\mbox{by Fact \ref{enu:2-1} and (\ref{eq:3})}\nonumber \\
 & = & H\left(U\mid Q\right)-H\left(U\mid\boldsymbol{Z}\left(l\right)=\boldsymbol{z},X\left(l\right),Q\right)\;\mbox{by Fact \ref{enu:1-1} and (\ref{eq:3})}\nonumber \\
 & = & H\left(U\mid Q\right)-H\left(U\mid X\left(l\right),Q\right)\;\mbox{ \mbox{by (\ref{eq:2})}}\nonumber \\
 & = & I\left(U;X\left(l\right)\mid Q\right).\label{eq:l2-1}
\end{eqnarray}
Replacing (\ref{eq:l2-1}) to (\ref{eq:l1-1}) we have, 
\begin{eqnarray*}
I(U;\boldsymbol{Y}_{[j]}\left(l\right),\boldsymbol{Z}\left(l\right)\mid Q) & = & I(U;X\left(l\right)\mid Q)\sum_{z\in\mathcal{Z}_{[j]}^{*}}\Pr\left(\boldsymbol{Z}\left(l\right)=\boldsymbol{z}\right)\\
 & = & (1-\epsilon_{[j]})I(U;X\left(l\right)\mid Q).
\end{eqnarray*}
} 
\end{IEEEproof}
\textbf{\textcolor{black}{Lemma \ref{lem:3}.}}\textcolor{black}{{} }\textcolor{black}{\emph{If
$(U,\boldsymbol{Y}_{[j]}^{l-1},\boldsymbol{Z}^{l-1},Q,X(l))$ are independent of $\boldsymbol{Z}(l)$
for $l\in[N],$ it holds,}}

\textcolor{black}{\emph{
\begin{eqnarray*}
I\left(U;\boldsymbol{Y}_{[j]}^{n},\boldsymbol{Z}^{n}\mid Q\right) & = & \left(1-\epsilon_{[j]}\right)\sum_{l=1}^{n}I\left(U;X\left(l\right)\mid\boldsymbol{Y}_{[j]}^{l-1},\boldsymbol{Z}^{l-1},Q\right).
\end{eqnarray*}
}} 
\begin{IEEEproof}
\textcolor{black}{Write using the chain rule, 
\begin{eqnarray*}
I\left(U;\boldsymbol{Y}_{[j]}^{n},\boldsymbol{Z}^{n}\mid Q\right) & = & \sum_{l=1}^{n}I\left(U;\boldsymbol{Y}_{[j]}(l),\boldsymbol{Z}(l)\mid\boldsymbol{Y}_{[j]}^{l-1},\boldsymbol{Z}^{l-1},Q\right)\\
 & = & \left(1-\epsilon_{[j]}\right)\sum_{l=1}^{n}I\left(U;X\left(l\right)\mid\boldsymbol{Y}_{[j]}^{l-1},\boldsymbol{Z}^{l-1},Q\right)\ \mbox{by Lem. \ref{lem:2}}.
\end{eqnarray*}
} 
\end{IEEEproof}
\textbf{\textcolor{black}{Lemma \ref{lem:4}. }}\textcolor{black}{\emph{Let $j\in[N-1].$ If
$(U,\boldsymbol{Y}_{[j+1]}^{l-1},\boldsymbol{Z}^{l-1},Q,X(l))$ are independent of $\boldsymbol{Z}(l)$
for $l\in[N],$ it holds: 
\begin{eqnarray*}
\frac{I(U;\boldsymbol{Y}_{[j]}^{n},\boldsymbol{Z}^{n}\mid Q)}{1-\epsilon_{[j]}} & \leq & \frac{I(U;\boldsymbol{Y}_{[j+1]}^{n},\boldsymbol{Z}^{n}\mid Q)}{1-\epsilon_{[j+1]}}+\sum_{l=1}^{n}I(\boldsymbol{Y}_{j+1}^{l-1};X\left(l\right)\mid\boldsymbol{Y}_{[j]}^{l-1},\boldsymbol{Z}^{l-1},Q).
\end{eqnarray*}
}} 
\begin{IEEEproof}
\textcolor{black}{Write,}

\textcolor{black}{
\begin{eqnarray*}
\frac{I(U;\boldsymbol{Y}_{[j]}^{n},\boldsymbol{Z}^{n}\mid Q)}{1-\epsilon_{[j]}} & = & \sum_{l=1}^{n}I(U;X\left(l\right)\mid\boldsymbol{Y}_{[j]}^{l-1},\boldsymbol{Z}^{l-1},Q)\:\textrm{by Lem. \ref{lem:3}}\\
 & \leq & \sum_{l=1}^{n}\left(I(U;X\left(l\right)\mid\boldsymbol{Y}_{[j+1]}^{l-1},\boldsymbol{Z}^{l-1},Q)+I(\boldsymbol{Y}_{j+1}^{l-1};X\left(l\right)\mid\boldsymbol{Y}_{[j]}^{l-1},\boldsymbol{Z}^{l-1},Q)\right)\textrm{by (\ref{eq:4})}\\
 & = & \frac{I(U;\boldsymbol{Y}_{[j+1]}^{n},\boldsymbol{Z}^{n}\mid Q)}{1-\epsilon_{[j+1]}}+\sum_{l=1}^{n}I(\boldsymbol{Y}_{j+1}^{l-1};X\left(l\right)\mid\boldsymbol{Y}_{[j]}^{l-1},\boldsymbol{Z}^{l-1},Q)\:\textrm{ by Lem. \ref{lem:3}}.
\end{eqnarray*}
}

\textcolor{black}{Relations (\ref{eq:h1}) and (\ref{eq:h2}) imply the lemma.} 
\end{IEEEproof}

\section{Proof of Corollary \ref{cor:CapacityDescription}}

\textbf{Corollary \ref{cor:CapacityDescription}.}\emph{ If for any point $\boldsymbol{r}$
such that $\hat{T}\left(\boldsymbol{r}\right)=1$ there is a sequence $\boldsymbol{r}_{k},\ k=1,2...$
with $\hat{T}\left(\boldsymbol{r}_{k}\right)<1$ and $\lim_{k\rightarrow\infty}\boldsymbol{r}_{k}=\boldsymbol{r},$
then $\mathcal{C}=\mathcal{\bar{C}}_{L}=\mathcal{\bar{C}}_{U}$. }

\emph{Moreover, if $\hat{T}\left(\boldsymbol{r}\right)$ is continuous function of $\boldsymbol{r}$
then, 
\begin{equation}
\mathcal{C}=\left\{ \boldsymbol{r}:\ \hat{T}\left(\boldsymbol{r}\right)\leq1\right\} .\label{eq:capacity2user-1}
\end{equation}
} 
\begin{IEEEproof}
Clearly $\bar{\mathcal{C}_{L}}\subseteq\bar{\mathcal{C}}_{U}$, hence it suffices to show that
$\bar{\mathcal{C}_{U}}\subseteq\bar{\mathcal{C}}_{L}$. For this, it suffices to show that
for any convergent sequence $\boldsymbol{r}_{k}\ k=1,2...$ in $\mathcal{C}_{U}$, if $\lim_{k\rightarrow\infty}\boldsymbol{r}_{k}=\boldsymbol{r},$
then $\boldsymbol{r}\in\bar{\mathcal{C}}_{L}$. Consider such a sequence and construct the
sequence $\tilde{\boldsymbol{r}}_{k}\in\bar{\mathcal{C}_{L}},$ as follows. If $\hat{T}\left(\boldsymbol{r}_{k}\right)<1,$
then $\tilde{\boldsymbol{r}}_{k}=\boldsymbol{r}_{k}$. If $\hat{T}\left(\boldsymbol{r}_{k}\right)=1,$
then pick a point $\tilde{\boldsymbol{r}}_{k}$ such that $\left|\boldsymbol{r}_{k}-\tilde{\boldsymbol{r}_{k}}\right|\leq1/k$
and $\hat{T}\left(\tilde{\boldsymbol{r}}_{k}\right)<1$. Such an $\tilde{\boldsymbol{r}}_{k}$
exists by the assumption in the corollary. Since $\hat{T}\left(\tilde{\boldsymbol{r}}_{k}\right)<1$
for all $k,$ we conclude that $\tilde{\boldsymbol{r}}_{k}\in\mathcal{C}_{L}$. Also, by construction
$\left|\boldsymbol{r}_{k}-\tilde{\boldsymbol{r}_{k}}\right|\leq1/k,$ which implies that 
\begin{eqnarray*}
\left|\tilde{\boldsymbol{r}}_{k}-\boldsymbol{r}\right| & \leq & \left|\tilde{\boldsymbol{r}}_{k}-\boldsymbol{r}_{k}\right|+\left|\boldsymbol{r}_{k}-\boldsymbol{r}\right|\\
 & \leq & \frac{1}{k}+\left|\boldsymbol{r}_{k}-\boldsymbol{r}\right|.
\end{eqnarray*}
Taking limits we have $\lim_{k\rightarrow\infty}\left|\tilde{\boldsymbol{r}}_{k}-\boldsymbol{r}\right|=0$.
Hence $\lim_{k\rightarrow\infty}\tilde{\boldsymbol{r}}=\boldsymbol{r}$ and we conclude that
$\boldsymbol{r}\in\bar{\mathcal{C}}_{L}.$

If in addition $\hat{T}\left(\boldsymbol{r}\right)$ is a continuous function of $\boldsymbol{r}$
then $\mathcal{C}_{U}$ is closed, hence $\mathcal{C}_{U}=\bar{\mathcal{C}}_{U},$ which together
with the fact that $\mathcal{C}=\bar{\mathcal{C}}_{U}$ shows (\ref{eq:capacity2user-1}). 
\end{IEEEproof}

\section{\label{sec:TheoremEquality}Proof of Theorem \ref{thm:EaualityPacketized}}

\textbf{\textit{\emph{Theorem }}}\textbf{\ref{thm:EaualityPacketized} }\emph{For the packetized
system and All-or-Nothing side information it holds, 
\[
\mathcal{C}_{p}=\left\{ \boldsymbol{r}\geq\boldsymbol{0}:\ \hat{T}\left(\boldsymbol{r}\right)\leq1\right\} \triangleq\mathcal{R}.
\]
} 
\begin{IEEEproof}
We first show that $\mathcal{R}\subseteq\mathcal{C}.$ For this, taking into account that both
$\mathcal{C}_{p}$ and $\mathcal{R}$ are closed sets and $\hat{T}\left(\boldsymbol{r}\right)$
positively homogenous, it suffices to show that if for some $\boldsymbol{r}$ it holds $\hat{T}\left(\boldsymbol{r}\right)<1,$
then there is a sequence of codes $C_{n}=(n,2^{\left\lceil nr_{1}\right\rceil L},\ldots,2^{\left\lceil nr_{N}\right\rceil L})$
with $\lim_{n\rightarrow\infty}\lambda_{n}=0$. Select $\delta>0$ such that 
\begin{equation}
\hat{T}\left(\boldsymbol{r}\right)+3\delta<1.\label{eq:0i-1}
\end{equation}
For positive integers $n$ and $n_{0}$, expressing $n$ in term of the quotient and remainder
with respect to $n_{0}$ we can write for integers $l_{n},$ $\upsilon_{n}$, 
\[
n=l_{n}n_{0}+\upsilon_{n},\:0\leq\upsilon_{n}<n_{0}.
\]
Since $nr_{i}<(l_{n}+1)n_{0}r_{i}\leq(l_{n}+1)\left\lceil n_{0}r_{i}\right\rceil $, it holds,
\begin{equation}
\left\lceil nr_{i}\right\rceil \leq\left(l_{n}+1\right)\left\lceil n_{0}r_{i}\right\rceil .\label{eq:cap41-1}
\end{equation}
Using Theorem \ref{thm:subadditiveTheorem} we can select and fix $n_{0}$ large enough so
that 
\begin{equation}
\frac{\bar{T}^{*}\left(\left\lceil n_{0}\boldsymbol{r}\right\rceil \right)}{n_{0}}\leq\hat{T}\left(\boldsymbol{r}\right)+\delta.\label{eq:2-1i-1}
\end{equation}
By the definition of $\bar{T}^{*}\left(\left\lceil n_{0}\boldsymbol{r}\right\rceil \right)$,
we can select a code $\tilde{C}_{n_{0}}$ such that 
\begin{equation}
\bar{T}_{\tilde{C}_{n_{0}}}\left(\left\lceil n_{0}\boldsymbol{r}\right\rceil \right)\leq\bar{T}^{*}\left(\left\lceil n_{0}\boldsymbol{r}\right\rceil \right)+n_{0}\delta.\label{eq:3i-1}
\end{equation}

Consider the following sequence of codes\textbf{ $\ensuremath{C_{n}}$ }for transmitting $\left\lceil n\boldsymbol{r}\right\rceil $
packets in $n$ channel uses.

a) Use $\tilde{C}_{n_{0}}$ to transmit successively $l_{n}+1$ bunches of $\left\lceil n_{0}\boldsymbol{r}\right\rceil $
packets (if the last bunch contains fewer than $\left\lceil n_{0}\boldsymbol{r}\right\rceil $
packets, use extra ``dummy'' packets independent from all the ``real'' packets and with
uniformly distributed bits), until they are decoded by all receivers. Let $T_{\tilde{C}_{n_{0}}}^{m}\left(\left\lceil n_{0}\boldsymbol{r}\right\rceil \right)$
be the (random) time it takes to transmit the $m$th bunch, and 
\[
\tilde{T}^{n}\left(\left\lceil n_{0}\boldsymbol{r}\right\rceil \right)=\sum_{m=1}^{l_{n}+1}T_{\tilde{C}_{n_{0}}}^{m}\left(\left\lceil n_{0}\boldsymbol{r}\right\rceil \right),
\]
be the total time it takes to transmit all the packets.

b) If 
\[
\tilde{T}^{n}\left(\left\lceil n_{0}\boldsymbol{r}\right\rceil \right)\leq n,
\]
all packets are correctly decoded. Else declare error.

The probability of error of this sequence of codes is computed as follows. Observing that 
\[
\lim_{n\rightarrow\infty}l_{n}=\infty,\:\lim_{n\rightarrow\infty}\frac{\upsilon_{n}}{l_{n}}=0,
\]
and taking into account (\ref{eq:0i-1}), pick $\tilde{n}$ large enough so that for all $n\geq\tilde{n}$
it holds, 
\begin{align}
\frac{l_{n}}{l_{n}+1}\left(1+\frac{\upsilon_{n}}{n{}_{0}l_{n}}\right) & =\left(1-\frac{1}{l_{n}+1}\right)\left(1+\frac{\upsilon_{n}}{n{}_{0}l_{n}}\right)\nonumber \\
 & =1+\frac{\upsilon_{n}}{n_{0}l_{n}}-\frac{1}{l_{n}+1}\left(1+\frac{\upsilon_{n}}{n{}_{0}l_{n}}\right)\nonumber \\
 & \geq1-\frac{1}{l_{n}+1}\left(1+\frac{\upsilon_{n}}{n{}_{0}l_{n}}\right)\nonumber \\
 & \geq\hat{T}\left(\boldsymbol{r}\right)+3\delta.\label{eq:4i-1}
\end{align}
Then, for $n\geq\tilde{n}$ we have, 
\begin{align*}
\lambda_{n} & =\Pr\left\{ \tilde{T}^{n}\left(\left\lceil n_{0}\boldsymbol{r}\right\rceil \right)>n\right\} \\
 & =\Pr\left\{ \sum_{m=1}^{l_{n}+1}T_{\tilde{C}_{n_{0}}}^{m}\left(\left\lceil n_{0}\boldsymbol{r}\right\rceil \right)>l_{n}n_{0}+\upsilon_{n}\right\} \\
 & =\Pr\left\{ \frac{\sum_{m=1}^{l_{n}+1}\frac{T_{\tilde{C}_{n_{0}}}^{m}\left(\left\lceil n_{0}\boldsymbol{r}\right\rceil \right)}{n_{0}}}{l_{n}+1}>\frac{l_{n}}{l_{n}+1}\left(1+\frac{\upsilon_{n}}{n{}_{0}l_{n}}\right)\right\} \\
 & \leq\Pr\left\{ \frac{\sum_{m=1}^{l_{n}+1}\frac{T_{\tilde{C}_{n_{0}}}^{m}\left(\left\lceil n_{0}\boldsymbol{r}\right\rceil \right)}{n_{0}}}{l_{n}+1}>\hat{T}\left(\boldsymbol{r}\right)+3\delta\right\} {\rm \qquad by\quad(\ref{eq:4i-1})}\\
 & \leq\Pr\left\{ \left|\frac{\sum_{m=1}^{l_{n}+1}\frac{T_{\tilde{C}_{n_{0}}}^{m}\left(\left\lceil n_{0}\boldsymbol{r}\right\rceil \right)}{n_{0}}}{l_{n}+1}-\frac{\bar{T}_{\tilde{C}_{n_{0}}}\left(\left\lceil n_{0}\boldsymbol{r}\right\rceil \right)}{n_{0}}\right|>\hat{T}\left(\boldsymbol{r}\right)-\frac{\bar{T}_{\tilde{C}_{n_{0}}}\left(\left\lceil n_{0}\boldsymbol{r}\right\rceil \right)}{n_{0}}+3\delta\right\} \\
 & \leq\Pr\left\{ \left|\frac{\sum_{m=1}^{l_{n}+1}\frac{T_{\tilde{C}_{n_{0}}}^{m}\left(\left\lceil n_{0}\boldsymbol{r}\right\rceil \right)}{n_{0}}}{l_{n}+1}-\frac{\bar{T}_{\tilde{C}_{n_{0}}}\left(\left\lceil n_{0}\boldsymbol{r}\right\rceil \right)}{n_{0}}\right|>\hat{T}\left(\boldsymbol{r}\right)-\frac{\bar{T}^{*}\left(\left\lceil n_{0}\boldsymbol{r}\right\rceil \right)}{n_{0}}+2\delta\right\} {\rm \quad{\rm by}\:}(\ref{eq:3i-1})\\
 & \leq\Pr\left\{ \left|\frac{\sum_{m=1}^{l_{n}+1}\frac{T_{\tilde{C}_{n_{0}}}^{m}\left(\left\lceil n_{0}\boldsymbol{r}\right\rceil \right)}{n_{0}}}{l_{n}+1}-\frac{\bar{T}_{\tilde{C}_{n_{0}}}\left(\left\lceil n_{0}\boldsymbol{r}\right\rceil \right)}{n_{0}}\right|>\delta\right\} \quad{\rm by\;}(\ref{eq:2-1i-1}).
\end{align*}
Due to the memorylessness of the channel and the fact that the bits in the packet contents
are i.i.d, the random variables $T_{\tilde{C}_{n_{0}}}^{m}\left(\left\lceil n_{0}\boldsymbol{r}\right\rceil \right),\: m=1,2...$
are i.i.d. Using the fact that $\lim_{n\rightarrow\infty}l_{n}=\infty$, we conclude from the
law of large numbers that 
\[
\lim_{t\rightarrow\infty}\frac{\sum_{m=1}^{l_{n}+1}\frac{T_{\tilde{C}_{n_{0}}}^{m}\left(\left\lceil n_{0}\boldsymbol{r}\right\rceil \right)}{n_{0}}}{l_{n}+1}=\frac{\bar{T}_{\tilde{C}_{n_{0}}}\left(\left\lceil n_{0}\boldsymbol{r}\right\rceil \right)}{n_{0}},\ a.e.,
\]
which implies that 
\begin{align*}
\lim_{n\rightarrow\infty}\lambda_{n} & \leq\lim_{t\rightarrow\infty}\Pr\left\{ \left|\frac{\sum_{m=1}^{l_{n}+1}\frac{T_{\tilde{C}_{n_{0}}}^{m}\left(\left\lceil n_{0}\boldsymbol{r}\right\rceil \right)}{n_{0}}}{l_{n}+1}-\frac{\bar{T}_{\tilde{C}_{n_{0}}}\left(\left\lceil n_{0}\boldsymbol{r}\right\rceil \right)}{n_{0}}\right|>\delta\right\} =0.
\end{align*}

Next we show that $\mathcal{C}\subseteq\mathcal{R}.$ Since $\mathcal{C},\ \mathcal{R}$ are
closed, it suffices to show that if $\boldsymbol{r}$ is achievable then $\hat{T}\left(\boldsymbol{r}\right)\leq1.$
Assume then that $\boldsymbol{r}$ is achievable, so that there is a sequence of coding algorithms
$C_{n}$ where $C_{n}$ operates for $n$ channel uses, with rate $\boldsymbol{r}$ and $\lim_{n\rightarrow\infty}\lambda_{n}=0$.

Let $\tilde{C}_{0}$ be the code that (re)transmits each packet until it is received correctly
by the corresponding destination. For this code it holds, 
\[
\bar{T}_{\tilde{C}_{0}}\left(\boldsymbol{k}\right)=\sum_{i=1}^{N}\frac{k_{i}}{1-\epsilon_{i}}.
\]
We construct a code $\tilde{C}$ for transmitting the $\left\lceil n\boldsymbol{r}\right\rceil $
packets as follows.

a) For and $\delta>0,$ select $n$ so that $\lambda_{n}<\delta$.

b) Implement code $C_{n}$ up to time $n$.

c) If all receivers decode correctly after $n$ channel uses, then stop.

e) Else resend all the $\left\lceil n\boldsymbol{r}\right\rceil $ packets using the one-by-one
code $\tilde{C}_{0}.$

We compute the average time to transmit the $\left\lceil n\boldsymbol{r}\right\rceil $ packets
under code $\tilde{C}$. Let $\mathcal{E}$ be the event that all destinations have decoded
correctly the packets by time $n$. Then, since on $\mathcal{E}^{c}$ it holds 
\[
T_{\tilde{C}}\left(\left\lceil n\boldsymbol{r}\right\rceil \right)=n+T_{\tilde{C}_{0}}\left(\left\lceil n\boldsymbol{r}\right\rceil \right),
\]
$T_{\tilde{C}_{0}}\left(\left\lceil n\boldsymbol{r}\right\rceil \right)$ is independent of
$\mathcal{E}^{c}$, and by choice $\Pr\left\{ \mathcal{E}^{c}\right\} =\lambda_{n}<\delta,$
we have 
\begin{eqnarray*}
\mathbb{E}\left[T_{\tilde{C}}\left(\left\lceil n\boldsymbol{r}\right\rceil \right)1_{\mathcal{E}^{c}}\right] & = & \Pr\left\{ \mathcal{E}^{c}\right\} \mathbb{E}\left[T_{\tilde{C}}\left(\left\lceil n\boldsymbol{r}\right\rceil \right)\right]\\
 & = & n\Pr\left\{ \mathcal{E}^{c}\right\} +\Pr\left\{ \mathcal{E}^{c}\right\} \bar{T}_{\tilde{C}_{0}}\left(\left\lceil n\boldsymbol{r}\right\rceil \right)\\
 & \leq & n\delta+\delta\left(\sum_{i=1}^{N}\frac{\left\lceil nr_{i}\right\rceil }{1-\epsilon_{i}}\right).
\end{eqnarray*}
Taking into account that $T_{\tilde{C}}\left(\left\lceil n\boldsymbol{r}\right\rceil \right)\leq n$
on $\mathcal{E},$ we conclude, 
\begin{align*}
\bar{T}_{\tilde{C}}\left(\left\lceil n\boldsymbol{r}\right\rceil \right) & =\mathbb{E}\left[T_{\tilde{C}}\left(\left\lceil n\boldsymbol{r}\right\rceil \right)1_{\mathcal{E}}\right]+\mathbb{E}\left[T_{\tilde{C}}\left(\left\lceil n\boldsymbol{r}\right\rceil \right)1_{\mathcal{E}^{c}}\right]\\
 & \leq n+\delta\left(n+\sum_{i=1}^{N}\frac{\left\lceil nr_{i}\right\rceil }{1-\epsilon_{i}}\right).
\end{align*}
Hence, 
\begin{align*}
\frac{\bar{T}^{*}\left(\left\lceil n\boldsymbol{r}\right\rceil \right)}{n} & \leq\frac{\bar{T}_{\tilde{C}}\left(\left\lceil n\boldsymbol{r}\right\rceil \right)}{n}\\
 & \leq1+\delta\frac{n+\sum_{i=1}^{N}\frac{\left\lceil nr_{i}\right\rceil }{1-\epsilon_{i}}}{n}.
\end{align*}
Taking limits, we obtain, 
\[
\hat{T}\left(\boldsymbol{r}\right)\leq1+\delta\left(\sum_{i=1}^{N}\frac{\left\lceil r_{i}\right\rceil }{1-\epsilon_{i}}+1\right),
\]
and since $\delta$ is arbitrary we conclude 
\[
\hat{T}\left(\boldsymbol{r}\right)\leq1.
\]

\end{IEEEproof}

\section{\label{sec:Proof-of-Lemmas}Proofs of Lemmas, Corollaries and Propositions in Section \ref{sub:Tightness-of-bound}}

\textbf{Corollary \ref{cor:CaseDirected}. }\emph{1. If $G$ is acyclic, then for any $\boldsymbol{r}\geq\boldsymbol{0},$
$W_{\boldsymbol{r}}^{*}=W_{\boldsymbol{r}}^{*}=\sum_{i=1}^{N}r_{i}$ and this bound can be
achieved by transmitting each packet separately, hence no benefit can be obtained by coding. }

\emph{2. If $G$ is a simple (directed) cycle then for any $\boldsymbol{r}\geq\boldsymbol{0},$
\[
W_{\boldsymbol{r}}^{*}=\sum_{i=1}^{N}r_{i}-\min_{i\in[N]}r_{i.}
\]
This bound can be achieved by pairwise XOR coding operations. } 
\begin{IEEEproof}
Part 1 follows directly by the definition of MWAIS. Transmitting without coding we get the
completion time $T\left(\boldsymbol{k}\right)=\sum_{i=1}^{N}k_{i}$. Hence, 
\begin{equation}
\lim_{n\rightarrow\infty}\frac{T\left(\left\lceil n\boldsymbol{r}\right\rceil \right)}{n}=\sum_{i=1}^{N}r_{i}.\label{eq:dircycle-1}
\end{equation}

Part 2 (\ref{eq:dircycle-1}) also follows directly from the definition of MWAIS. To describe
the code, assume without loss of generality that $k_{N}=\min_{i\in[N]}k_{i}$. Consider the
following code.

a) $k\leftarrow0$

b) $k\leftarrow k+1$

c) Transmit $p_{i}^{k}\oplus p_{i+1}^{k}$, $i\in[N-1]$.

d) If $Q_{N}$ empties go to d). Else got to b)

d) Transmit remaining packets in each of the other queues uncoded.

As observed in \cite{bar2006index} each time Part c) is implemented, (with $N-1$ transmissions)
one packet from each of the queue $Q_{i},\ i\in[N]$ is decoded by the corresponding receiver;
this is so since every receiver $i\in[N-1]$ can decode $p_{i}^{k}=\left(p_{i}^{k}\oplus p_{i+1}^{k}\right)\oplus p_{i+1}^{k}$
while receiver $N$ can decode $p_{i}^{k}=\left(\oplus_{i=1}^{N-1}\left(p_{i}^{k}\oplus p_{i+1}^{k}\right)\right)\oplus p_{1}^{k}$.
When $Q_{N}$ empties, (part d)), there are $k_{i}-k_{N}$ packets left in each of the queues,
hence since each of the latter packets is transmitted uncoded, 
\[
T\left(\boldsymbol{k}\right)=k_{N}\left(N-1\right)+\sum_{i=1}^{N-1}\left(k_{i}-k_{N}\right)=\sum_{i=1}^{N-1}k_{i},
\]
and 
\[
\lim_{n\rightarrow\infty}\frac{T\left(\left\lceil n\boldsymbol{r}\right\rceil \right)}{n}=\sum_{i=1}^{N-1}r_{i}.
\]

\end{IEEEproof}
In the arguments below concerning undirected graphs we will make use of the following simple
lemma which allows us to consider only integer weights. 
\begin{lem}
\label{lem:IntToReal}It holds for any real vector $\boldsymbol{r}\geq\boldsymbol{0},$ 
\[
\lim_{n\rightarrow\infty}\frac{W_{\left\lceil n\boldsymbol{r}\right\rceil }^{*}}{n}=W_{\boldsymbol{r}}^{*}.
\]
\end{lem}
\begin{IEEEproof}
Observe that for any set $\mathcal{S}$ it holds, 
\begin{eqnarray*}
\lim_{n\rightarrow\infty}\frac{W_{\left\lceil n\boldsymbol{r}\right\rceil }\left(\mathcal{S}\right)}{n} & = & \lim_{n\rightarrow\infty}\frac{\sum_{i\in\mathcal{S}}\left\lceil nr_{i}\right\rceil }{n}\\
 & = & \lim_{n\rightarrow\infty}\sum_{i\in\mathcal{S}}r_{i}\\
 & = & W_{\boldsymbol{r}}\left(\mathcal{S}\right).
\end{eqnarray*}
and hence since $\left|\mathcal{I}\right|<\infty$, by continuity of $\max\left\{ \cdot\right\} $,
\[
\lim_{n\rightarrow\infty}\frac{W_{\left\lceil n\boldsymbol{r}\right\rceil }^{*}}{n}=\lim_{n\rightarrow\infty}\max_{\mathcal{S}\in\mathcal{I}}\left\{ \frac{W_{\left\lceil n\boldsymbol{r}\right\rceil }\left(\mathcal{S}\right)}{n}\right\} =\max_{\mathcal{S}\in\mathcal{I}}\left\{ W_{\boldsymbol{r}}\left(\mathcal{S}\right)\right\} =W_{\boldsymbol{r}}^{*}.
\]

\end{IEEEproof}
A direct consequence of Lemma \ref{lem:IntToReal} is Corollary \ref{cor:Achieve}.

\textbf{\textit{\emph{Proposition \ref{lem:10-tree}. }}}\emph{If the information graph is
a tree then for any $\boldsymbol{k}\geq\boldsymbol{0}$ there is a code using pairwise XOR
operations and $T\left(\boldsymbol{k}\right)=W_{\boldsymbol{k}}^{*}$. The same holds if the
information graph is a forest.} 
\begin{IEEEproof}
Let the information graph be a tree. We prove by induction the following.

\emph{Induction hypothesis}:\emph{ }For any $l\geq0,$ if $W_{\boldsymbol{k}}^{*}\leq l$,
there is a code using only XOR combinations of packets with completion time $T\left(\boldsymbol{k}\right)=W_{\boldsymbol{k}}^{*}$.

If $l$=0 then no packets need to be transmitted and the statement holds by definition. Assume
now that the inductive hypothesis is true up to $l\geq0$, and let $W_{\boldsymbol{k}}^{*}=l+1\geq1.$
Then there is a least one node $i$ with $k_{i}\geq1$. Also among the nodes with positive
weight ($k_{i}\geq1)$ there is at least one, $i_{0}$, for which the following holds.

a) Either all neighbors of $i_{0}$ have zero weight,

b) or $i_{0}$ has degree (i.e. number of links which have node $i_{0}$ as one end) $d\left(i_{0}\right)=1$
and its neighbor, $i_{1},$ has $k_{i_{1}}\geq1$. This is so, since otherwise, i.e., if any
node $i$ with positive weight had some neighbor with positive weight and $i$ had degree at
least 2, then the tree would contain a cycle consisting of nodes with positive weight.

Consider first case a). Observe that if from any independent set $\mathcal{S}$ we remove any
neighbors of $i_{0}$ the set may contain, the resulting set $\mathcal{S}_{1}$ is independent
with weight $W(\mathcal{S}_{1})=W(\mathcal{S})$. Also, if $\mathcal{S}_{1}$ does not contain
$i_{0}$ the set $\mathcal{S}_{2}=\mathcal{S}_{1}\cup\{i_{0}\}$ is independent and $W(\mathcal{S}_{2})=W(\mathcal{S})+k_{i_{0}}\geq W(\mathcal{S})+1$.
It follows that,

\emph{Fact 1}:\emph{ }all independent sets with (maximum) weight $W_{\boldsymbol{k}}^{*}$
must contain $i_{0}$ and any independent set not containing $i_{0}$ has weight at most $W_{\boldsymbol{k}}^{*}-1$.

Apply now the simple code: transmit a packet from queue $Q_{i_{0}}$. After this transmission,
the new weights of the nodes are, $\tilde{k}_{i_{0}}=k_{i_{0}}-1$ and $\tilde{k}_{i}=k_{i}$
if $i\neq i_{0}.$ Taking into account the above mentioned Fact 1, we conclude that with the
new weights, $W_{\tilde{\boldsymbol{k}}}^{*}=W_{\boldsymbol{k}}^{*}-1=l.$ We can use now the
code implied by the inductive hypothesis to transmit the packets $\tilde{\boldsymbol{k}}$
in time $W_{\tilde{\boldsymbol{k}}}^{*}$ and the total length of the code is, 
\[
T\left(\boldsymbol{k}\right)=W_{\tilde{\boldsymbol{k}}}^{*}+1=W_{\boldsymbol{k}}^{*}.
\]

Consider next case b), and let $i_{1}\ $be the single neighbor of $i_{0}$ with $k_{i_{1}}\geq1.$
Then,

\emph{Fact 2}: any independent set with weight $W_{\boldsymbol{k}}^{*}$ must contain exactly
one of $i_{0},$ $i_{1}$ and any independent set not containing either of them has weight
at most $W_{\boldsymbol{k}}^{*}-1.$

Apply now the simple code: transmit an XOR combination of a pair of packets $p_{i_{0}}^{1}\oplus p_{i_{1}}^{1}.$
After this transmission the new weights of the nodes are, $\tilde{k}_{i_{0}}=k_{i_{0}}-1$,
$\tilde{k}_{i_{1}}=k_{i_{1}}-1$ and $\tilde{k}_{i}=k_{i},\ i\notin\left\{ i_{0},i_{1}\right\} $.
Taking into account Fact 2, we conclude again the with the new weights $W_{\tilde{\boldsymbol{k}}}^{*}=W_{\boldsymbol{k}}^{*}-1=l,$
and we can apply the inductive hypothesis as in case a).

For forests, the lemma follows by the fact that the weight of any independent set $\mathcal{S}$
in a forest is the sum of the weights of the components of $\mathcal{S}$ in each of the forest
trees. 
\end{IEEEproof}
\textbf{\textit{\emph{Proposition \ref{prop:CyclesGeneral}. }}}\emph{Let the information graph
be a cycle $\mathcal{C}=(1,2,...,N,1)$. If either one of the following conditions holds,} 
\begin{enumerate}
\item \emph{\label{enu:CyclesGeneralCase1-1}There is at least one node with zero weight;} 
\item \emph{\label{enu:CyclesGeneralCase2-1}$k_{i}>0$ for all $i\in\{1,...,N\}$ and there is an
independent set $\mathcal{S}=\left\{ i_{1},...,i_{\left|\mathcal{S}\right|}\right\} \in\mathcal{I}^{*},$
where $i_{1}<i_{2}<...<i_{\left|\mathcal{S}\right|}$, such that $d(i_{m},i_{m+1})=3$ and
$d(i_{k},i_{k+1})=3$ for some $i_{m},i_{k}\in\mathcal{S},\ i_{k}\neq i_{m};$ } 
\end{enumerate}
\emph{then there is a code using pairwise XOR operations with $T\left(\boldsymbol{k}\right)=W_{\boldsymbol{k}}^{*}.$} 
\begin{IEEEproof}
\emph{Case 1}. Assume without loss of generality that $k_{1}=0.$ In this case, for the path
$\mathcal{P}=\left(1,2,....,N\right)$, we have $W_{\mathcal{P},\boldsymbol{k}}^{*}$=$W_{\mathcal{C},\boldsymbol{k}}^{*}$.
To see this note first that since any independent set for $\mathcal{C}$ is an independent
set for $\mathcal{P},$ it holds $W_{\mathcal{P},\boldsymbol{k}}^{*}\geq$$W_{\mathcal{C},\boldsymbol{k}}^{*}$.
On the other hand, for any independent set $\mathcal{S}$ for $\mathcal{P}$ , since node $1$
has zero weight, we may assume that $1\notin\mathcal{S}$ (otherwise we may remove node $1$
from $\mathcal{S}$ without affecting its weight.). But then, $\mathcal{S}$ is an independent
set for $\mathcal{C}$ which implies that $W_{\mathcal{P},\boldsymbol{k}}^{*}\leq$$W_{\mathcal{C},\boldsymbol{k}}^{*}$.
The proposition now follows by applying Proposition \ref{lem:10-tree} to the path $\mathcal{P}.$

\emph{Case 2. }Note that since $k_{i}>0,\ i\in\{1,...,N\},$ for any $\mathcal{S}=\left\{ i_{1},...,i_{\left|\mathcal{S}\right|}\right\} \in\mathcal{I}_{\boldsymbol{k}}^{*},$
$i_{l}<i_{l+1},\ l\in\{1,...,\left|\mathcal{S}\right|-1\}$, any node of the graph not in $\mathcal{S}$
must be connected to a node in $\mathcal{S}$, otherwise $\mathcal{S}$ would not have maximum
weight. Hence for any node $i_{l}\in\mathcal{S},$ $d(i_{l},i_{l+1})\leq3.$

\begin{figure}
\begin{centering}
\textcolor{black}{\includegraphics[scale=0.2]{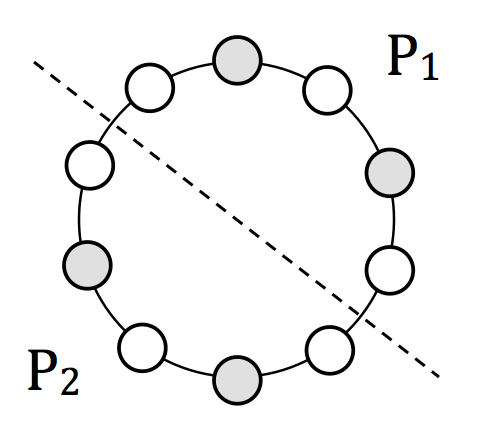}}
\par\end{centering}

\caption{\label{fig:The-separation-of}The separation of the cycle in two paths $P_{1}$ and $P_{2}$}
\end{figure}
Let $\mathcal{S}_{1},\ \mathcal{S}_{2}$ be the nodes of $\mathcal{S}$ in the paths $\mathcal{P}_{1}=\left(i_{k+1}-1,i_{m}+1\right)$,
$\mathcal{P}_{2}=\left(i_{m+1}-1,\ i_{k}+1\right)$ respectively, see Figure \ref{fig:The-separation-of}.
Since $\mathcal{S}=\mathcal{S}_{1}\cup\mathcal{S}_{1}$ and $\mathcal{S}$ has maximum weight,
$\mathcal{S}_{l},\ l\in\left\{ 1,2\right\} $ has maximum weight in the path $\mathcal{P}_{l}$;
otherwise if, say in $\mathcal{P}_{1}$, there is an independent set $\mathcal{G}_{1}$ with
larger weight than $\mathcal{S}_{1}$, the independent set $\mathcal{\hat{S}}=\mathcal{G}_{1}\cup\mathcal{S}_{2}$
would have larger weight than $\mathcal{S}.$

We now employ the following algorithm to transmit the packets: transmit the packets in path
$\mathcal{P}_{1}$ and then the packets in path $\mathcal{P}_{2},$ according to the algorithm
implied in Lemma \ref{lem:10-tree}. Then the total length is 
\begin{eqnarray*}
T\left(\boldsymbol{k}\right) & = & W_{\boldsymbol{k}}\left(\mathcal{S}_{1}\right)+W_{\boldsymbol{k}}\left(\mathcal{S}_{2}\right)\ \mbox{by Lemma \ref{lem:10-tree}}\\
 & = & W_{\boldsymbol{k}}\left(\mathcal{S}\right)\ \mbox{ since \ensuremath{\mathcal{S}=\mathcal{S}_{1}\cup\mathcal{S}_{2}}}\mbox{and \ensuremath{\mathcal{S}_{1}\cap\mathcal{S}_{2}=\emptyset}}\\
 & = & W_{\boldsymbol{k}}^{*}.
\end{eqnarray*}

\end{IEEEproof}
\textbf{\textit{\emph{Proposition \ref{lem:11}}}}\textbf{\emph{.}}\emph{ If the information
graph is an even cycle $\mathcal{C}=(1,2,...,N,1)$ with $N=2l,\ l\geq1$, then for any $\boldsymbol{k}\geq\boldsymbol{0}$
there is a code using pairwise XOR operations with $T\left(\boldsymbol{k}\right)=W_{\boldsymbol{k}}^{*}.$} 
\begin{IEEEproof}
Consider the following cases.

\emph{Case 1. }There is at least one node with zero weight; This case is handled in Proposition
\ref{prop:CyclesGeneral}.

\emph{Case 2. }All nodes have nonzero weight. We distinguish the following sub-cases.

\emph{Case 2.1 }The cycle has that following property $\mathbb{P}:$ there is at least one
independent set $\mathcal{S}=\left\{ i_{1},...,i_{\left|\mathcal{S}\right|}\right\} \in\mathcal{I}^{*},$
such that $d(i_{m},i_{m+1})=3$ for some $i_{m}\in\mathcal{S}.$

In this case there must be another node $i_{k}\in\mathcal{S}$ such that $d\left(i_{k},i_{k+1}\right)=3$
(it is possible that $k=m+1$); otherwise the cycle would have odd length. The proposition
then follows from Proposition \ref{prop:CyclesGeneral} case 2.

\emph{Case 2.2. }For any set $\mathcal{S}=\left\{ i_{1},...,i_{\left|\mathcal{S}\right|}\right\} \in\mathcal{I}^{*},$
$d(i_{m},i_{m+1})=2.$ Hence $\left|\mathcal{S}\right|=N/2$, and any such set (there are at
most two) contains exactly one of the nodes $i,i+1$ for all $i\in[N]$.

Select arbitrarily any pair of consecutive nodes $(i,i+1)$ and transmit an XOR combination
of a pair of packets, one from each of the queues $Q_{i},\ Q_{i+1}$. Note that (due to the
fact any $\mathcal{S}\in\mathcal{I}_{\boldsymbol{k}}^{*}$ contains exactly one of the nodes
$i,$ $i+1$), if $\boldsymbol{k}_{1}$ are the new node weights after the first transmission,
we have $W_{\boldsymbol{k}_{1}}^{*}=W_{\boldsymbol{k}}^{*}-1$. In general, send XOR combinations
of pairs of packets from queues $Q_{i},\ Q_{i+1}$ until after $l\geq1$ transmissions, either
with the resulting weights, $\boldsymbol{k}_{l}$, the cycle $\mathcal{C}$ has property $\mathbb{P},$
or the weight of one of the nodes $i,\ i+1$ becomes zero. We then have, 
\[
W_{\boldsymbol{k}_{l}}^{*}=W_{\boldsymbol{k}}^{*}-l.
\]
If after the $l$ transmissions the cycle $\mathcal{C}$ (with weights $\boldsymbol{k}_{l}$)
has property $\mathbb{P},$ then apply the coding in Case 2.1 to transmit the $\boldsymbol{k}_{l}$
packets in time $W_{\boldsymbol{k}_{l}}^{*}$. The resulting completion time is 
\[
T\left(\boldsymbol{k}\right)=T\left(\boldsymbol{\boldsymbol{k}_{l}}\right)+l=W_{\boldsymbol{k}_{l}}^{*}+l=W_{\boldsymbol{k}}^{*}.
\]
If on the other hand after the $l$ transmissions the weight of one of the nodes $i,\ i+1$
becomes zero, then follow Case 1 to derive the same conclusion. 
\end{IEEEproof}
\textbf{\textit{\emph{Proposition \ref{lem:OddCycle}. }}}\emph{Let the information graph be
an odd cycle $\mathcal{C}=(1,2,...,N,1)$ with $N=2g+1,\ g\geq2$. Then there is a code using
pairwise XOR operations with 
\[
T\left(\boldsymbol{k}\right)\leq W_{\boldsymbol{k}}^{*}+\left\lceil \frac{\min_{i}k_{i}}{2}\right\rceil .
\]
Hence it holds for any $\boldsymbol{r}\geq\boldsymbol{0},$ 
\[
\hat{T}\left(\boldsymbol{r}\right)\leq W_{\boldsymbol{r}}^{*}+\frac{\min_{i}r_{i}}{2}.
\]
} 
\begin{IEEEproof}
If Condition \ref{enu:CyclesGeneralCase1} of Proposition \ref{prop:CyclesGeneral} holds,
the result follows. Consider next that $k_{i}\geq1$ for all $i\in[N].$

Assume without loss of generality that $k_{1}=\min_{i\in[N]}k_{i}$ and consider the following
algorithm.

\emph{Algorithm I} 
\begin{enumerate}
\item $s\leftarrow1$; $\boldsymbol{k}^{1}\leftarrow\boldsymbol{k};$ 
\item \label{enu:loop1}while $s\leq\left\lfloor k_{1}/2\right\rfloor $;

\begin{enumerate}
\item Transmit two packets $p_{2s-1}^{1}\oplus p_{s}^{2}$ and $p_{2s}^{1}\oplus p_{s}^{N}$; 
\item \label{enu:oddcycleStep3-2}Set (remaining packets at each queue) $k_{1}^{s+1}\leftarrow k_{1}^{s}-2$;
$k_{2}^{s+1}\leftarrow k_{2}^{s}-1;$ $k_{N}^{s+1}\leftarrow k_{N}^{s}-1;$ $k_{i}^{s+1}\leftarrow k_{i}^{s},\ i\in[N]-\{N,1,2\};$ 
\item $s\leftarrow s+1$;
\end{enumerate}
\item If $k_{1}^{s}=1$ (at this point $s=\left\lfloor k_{1}/2\right\rfloor +1$ and $k_{1}=2s-1=\left\lfloor k_{1}/2\right\rfloor 2+1$)

\begin{enumerate}
\item Transmit packet $p_{k_{1}}^{1}\oplus p_{\left\lfloor k_{1}/2\right\rfloor +1}^{2}$
\item \label{enu:oddcycleStep2-2}Set (remaining packets at each queue) $k_{1}^{s+1}\leftarrow0;$
$k_{2}^{s+1}\leftarrow k_{2}^{s}-1;$ $k_{i}^{s+1}\leftarrow k_{i}^{s},\ i\in[N]-\{1,2\};$ 
\item $s\leftarrow s+1$
\end{enumerate}
\item \label{enu:oddcyclesStep4-1}(At this point, $k_{1}^{s}=0$). Use the algorithm implied by
Case \ref{enu:CyclesGeneralCase1}) of Proposition\emph{ \ref{prop:CyclesGeneral}} to transmit
the $\boldsymbol{k}^{s}$ packets; end; 
\end{enumerate}
We evaluate the completion time of this algorithm. Notice first the following. 
\begin{itemize}
\item When Step \ref{enu:oddcycleStep3-2} is executed, $k_{1}^{s}$ is reduced by 2 while $k_{i}^{s}$
is reduced by at most 1 for $i\neq1.$ 
\item When Step \ref{enu:oddcycleStep2-2} is executed, $k_{1}^{s}$ becomes $0,$ while $k_{i}^{s}$
is reduced by at most 1 for $i\neq1$. 
\end{itemize}
It follows from the above observations that the algorithm always ends, and (since $k_{1}=\min_{i}k_{i}$)
that when Steps \ref{enu:oddcycleStep3-2}, \ref{enu:oddcycleStep2-2} are executed, $k_{i}^{s}\geq k_{1}^{s}\geq1$
for all $i\in[N].$ 

We claim that if $k_{1}\geq2,$ at the $s$th execution of Step \ref{enu:oddcycleStep3-2}
it holds, 
\begin{equation}
W_{\boldsymbol{k}^{s+1}}^{*}\leq W_{\boldsymbol{k}^{s}}^{*}-1.\label{eq:BasicReccursion-1-1}
\end{equation}
To see this, notice that since $k_{i}^{s}\geq k_{1}^{s}\geq1,\ i\in[N]$, for any $\mathcal{S}=\left\{ i_{1},...,i_{\left|\mathcal{S}\right|}\right\} \in\mathcal{I}_{\boldsymbol{k}^{s}}^{*}$
where $i_{1}<i_{2}<...<i_{\left|{\cal S}\right|}$, it holds $d(i_{m},i_{m+1})\leq3,$ i.e.,
between any two successive nodes of ${\cal S}$ there are at most two nodes of the information
graph. Hence every ${\cal S}\in{\cal I}_{\boldsymbol{k}^{s}}^{*}$ must contain at least one
of the nodes $i_{N},i_{1},i_{2}.$ The latter observation implies that after the $s$th execution
of Step \ref{enu:oddcycleStep3-2}, the weights of all ${\cal S}\in{\cal I}_{\boldsymbol{k}^{s}}^{*}$
are reduced by at least 1. Since for all other independent sets $\ {\cal S}\in{\cal I}-{\cal I}_{\boldsymbol{k}^{s}}^{*},$
it also holds 
\[
W_{\boldsymbol{k}^{s+1}}\left({\cal S}\right)\leq W_{\boldsymbol{k}^{s}}\left({\cal S}\right)\leq W_{\boldsymbol{k}^{s}}^{*}-1,
\]
we conclude (\ref{eq:BasicReccursion-1-1}).

Let $\hat{s}$ be the times Loop \ref{enu:loop1} is executed, i.e., $\hat{s}=\left\lfloor k_{1}/2\right\rfloor $.
From (\ref{eq:BasicReccursion-1-1}) and since Loop \ref{enu:loop1} is executed $\hat{s}$
times, we have. 
\begin{equation}
W_{\boldsymbol{k}^{\hat{s}+1}}^{*}\leq W_{\boldsymbol{k}}^{*}-\hat{s}.\label{eq:MaxWeight-1}
\end{equation}
Consider two cases

\emph{Case 1. }\uline{$k_{1}$ is even.} Since it takes $W_{\boldsymbol{k}^{\hat{s}+1}}^{*}$
transmissions for the algorithm to complete once Step \ref{enu:oddcyclesStep4-1} is reached,
and 2 transmissions take place at each execution of the Loop \ref{enu:loop1}, the completion
time of the algorithm is in this case, 
\begin{eqnarray*}
T(\boldsymbol{k}) & = & 2\hat{s}+W_{\boldsymbol{k}^{\hat{s}+1}}^{*}\\
 & \leq & W_{\boldsymbol{k}}^{*}+\hat{s}\\
 & = & W_{\boldsymbol{k}}^{*}+\left\lfloor k_{1}/2\right\rfloor .
\end{eqnarray*}

\emph{Case 2. }\uline{$k_{1}$ is odd.} In this case, after Step \ref{enu:oddcycleStep2-2}
is executed it holds, 
\[
W_{\boldsymbol{k}^{\hat{s}+2}}^{*}\leq W_{\boldsymbol{k}^{\hat{s}+1}}^{*},
\]
 and taking into account (\ref{eq:MaxWeight-1}) we conclude, 
\[
W_{\boldsymbol{k}^{\hat{s}+2}}^{*}\leq W_{\boldsymbol{k}}^{*}-\hat{s}.
\]
Since it takes $2\hat{s}$ transmissions to complete Loop \ref{enu:loop1}, 1 transmission
to execute Step \ref{enu:oddcycleStep2-2} and $W_{\boldsymbol{k}^{\hat{s}+2}}^{*}$ transmissions
at step \ref{enu:oddcyclesStep4-1}, we have,

\begin{eqnarray*}
T\left(\boldsymbol{k}\right) & = & 2\hat{s}+1+W_{\boldsymbol{k}^{\hat{s}+2}}^{*}\\
 & \leq & 2\hat{s}+1+W_{\boldsymbol{k}}^{*}-\hat{s}\\
 & = & W_{\boldsymbol{k}}^{*}+\hat{s}+1\\
 & = & W_{\boldsymbol{k}}^{*}+\left\lceil \frac{k_{1}}{2}\right\rceil .
\end{eqnarray*}

\end{IEEEproof}
\textbf{\textit{\emph{Lemma \ref{lem:AntiholeMaxWeight}. }}}\emph{For an antihole $\mathcal{A}_{N}$,
for any $\boldsymbol{k}\geq\boldsymbol{0}$ it holds 
\[
W_{\boldsymbol{k}}^{*}=\max_{i\in\mathcal{H}_{N}}\left(k_{i}+k_{i+1}\right).
\]
} 
\begin{IEEEproof}
By Lemma \ref{lem:EvenHoleLemma} and the fact that by the definition of antihole there is
no link between consecutive pairs of nodes $i,\ i+1$, it follows that the independent sets
of $\mathcal{A}_{N}$ consist either of singletons or of consecutive pairs of nodes, $\left\{ i,i+1\right\} $.
Observing that if a singleton, say $\{i\},$ has maximum weight then both nodes $i-1$ and
$i+1$ must have zero weight (otherwise, if say $i+1$ had nonzero weight the independent set
$\{i,i+1\}$ would have larger weight than the weight of $\{i\}$), the lemma follows. 
\end{IEEEproof}
\textbf{\textit{\emph{Proposition \ref{prop:AntiholeGeneral}. }}}\emph{If for an antihole
${\cal A}_{N}$ it holds $k_{i}=0$ for some node, then there is a code using XOR operations
(not necessarily pairwise) with $T(\boldsymbol{k})=W_{\boldsymbol{k}}^{*}.$ } 
\begin{IEEEproof}
We prove the proposition by induction. Specifically we prove,

\emph{Induction Hypothesis}:\emph{ }For any $l\geq0,$ if $k_{i}=0$ for some $i\in{\cal A}_{N}$
and $W_{\boldsymbol{k}}^{*}=l$, there is a code using only XOR operations on packets with
length $T\left(\boldsymbol{k}\right)=$$l$.

The case $l=0$ is trivial. Assume now that the Induction Hypothesis holds for $l\geq0$ and
Consider an antihole for which $W_{\boldsymbol{k}}^{*}=l+1$ and, without loss of generality
assume that $k_{1}=0.$ We distinguish two cases.

\emph{Case 1.}\emph{\uline{ }}\uline{There is at least one $i$ with $k_{i}=W_{\boldsymbol{k}}^{*}$.}
Let $\mathcal{I}_{0}^{*}$ be the set of nodes $i$ with $k_{i}=W_{\boldsymbol{k}}^{*}.$ As
observed in Lemma \ref{lem:AntiholeMaxWeight}, for any $i\in\mathcal{I}_{0}^{*}$, both nodes
$i-1$ and $i+1$ must have zero weight. Hence for any $i,j$ in $\mathcal{I}_{0}^{*},\ i\neq j$
it holds $d(i,j)\geq2$. Moreover, for any $i\in$$\mathcal{I}_{0}^{*}$ and for any pair of
nodes $\left\{ j,j+1\right\} \in\mathcal{I}^{*}-\mathcal{I}_{0}^{*}$, i.e. pair with maximum
weight and $\min\left\{ k_{j},\ k_{j+1}\right\} >0,$ we also have $d(i,j)\geq2,$ and $d(j+1,i)\geq2$.
Therefore, we can construct a set ${\cal S}=\left\{ i_{1},i_{2},...,i_{m}\right\} $ consisting
of all nodes in $\mathcal{I}_{0}^{*}$ and one node from each of the pairs $\left\{ i,i+1\right\} $
with maximum weight and $\min\left\{ k_{j},\ k_{j+1}\right\} >0$, such that $d\left(i_{l},i_{l+1}\right)\geq2$
for all $l\in\left\{ 1,...,m\right\} $. According to Lemma \ref{lem:EvenHoleLemma}, ${\cal S}$
induces a complete subgraph. Transmit now an XOR combination of packets one from each of the
nodes of ${\cal S}$. Since the induced subgraph of ${\cal S}$ is complete all nodes in ${\cal S}$
are able to decode their corresponding messages. Hence the weights of all the pairs of nodes
in $\mathcal{I}^{*}$ is reduced by one. This means that with the new weights $\tilde{\boldsymbol{k}}$
we have $W_{\tilde{\boldsymbol{k}}}^{*}=W_{\boldsymbol{k}}^{*}-1=l$. Since again $\tilde{k}_{1}=0$
we can use the algorithm implied by the inductive hypothesis to transmit the remaining $\tilde{\boldsymbol{k}}$
packets. Hence, for the completion time of the algorithm we have, 
\[
T\left(\boldsymbol{k}\right)=1+T\left(\tilde{\boldsymbol{k}}\right)=1+W_{\tilde{\boldsymbol{k}}}^{*}=l+1.
\]

\emph{Case 2. }\uline{There is no $i\in{\cal A}_{N}$ with $k_{i}=W_{\boldsymbol{k}}^{*}$.}
Then, since $k_{1}=0$ sets $\{N,1\}$ and $\{1,2\}$ do not belong to ${\cal I}_{\boldsymbol{k}}^{*}$.
Consider now the set ${\cal S}=\left\{ 3,5,7,...,N-J_{N}\right\} $ where $J_{N}=1$ if $N$
is even and $J_{N}=2$ if $N$ is odd. Again, this set induces a complete subgraph. We transmit
as before an XOR combination of packets one from each of the nodes of ${\cal S}$ so that all
these nodes decode the packet destined to them. Also, since $\{i_{N},i_{1}\}$ and $\{i_{1},i_{2}\}$
do not belong to ${\cal I}_{\boldsymbol{k}}^{*}$ and by the choice of ${\cal S}$ the weight
of all sets in ${\cal I}_{\boldsymbol{k}}^{*}$ is reduced by one, we can repeat the previous
argument to complete the proof. 
\end{IEEEproof}
\textbf{\textit{\emph{Proposition \ref{prop:EvenHoleProposition}. }}}\emph{For an even antihole
$\mathcal{A}_{N},\ N=2g$ and any $\boldsymbol{k}\geq\boldsymbol{0}$ there is a code using
XOR operations with $T\left(\boldsymbol{k}\right)=W_{\boldsymbol{k}}^{*}$}. 
\begin{IEEEproof}
We prove the lemma by induction. Specifically we prove the following.

\emph{Induction hypothesis}:\emph{ }For any $l\geq0,$ where $W_{\boldsymbol{k}}^{*}=l$, there
is a code using only XOR operations with length $T\left(\boldsymbol{k}\right)=$$l$.

For $l=0$ there is nothing to prove. Assume now that the hypothesis is true up to $l\geq0$.
We will show that the hypothesis holds for $W_{\boldsymbol{k}}^{*}=l+1.$

If there is at least one $i\in{\cal A}_{N}$ with $k_{i}=W_{\boldsymbol{k}}^{*},$ then necessarily
either $k_{i-1}=0$ or $k_{i+1}=0$ and the Induction Hypothesis holds by Proposition \ref{prop:AntiholeGeneral}.
Assume next that there is no $i$ with $k_{i}=W_{\boldsymbol{k}}^{*}$, hence all sets $\left\{ i,i+1\right\} $
with maximum weight have $\min\left\{ k_{i},\ k_{i+1}\right\} >0.$ Transmit an XOR combination
of packets from nodes in $\left\{ 1,3,5,...,2g-1\right\} $ that have nonzero weight. Since
the graph has even number of nodes the conditions in Lemma \ref{lem:EvenHoleLemma} hold and
we conclude that the induce subgraph of these nodes is complete; hence all these nodes are
able to decode their corresponding messages. Hence after the first transmission the number
of packets at each of these nodes is reduced by one. But since for each pair of nodes $\left\{ i,i+1\right\} $
with maximum weight either $i$ or $i+1$ is odd number and both weights $k_{i},\ k_{i+1}$,
are positive, the weights of all these pairs is reduced by one. This means that with the new
weights we have $W_{\tilde{\boldsymbol{k}}}^{*}=W_{\boldsymbol{k}}^{*}-1=l$. We can therefore
use the algorithm implied by the inductive hypothesis to transmit the remaining $\tilde{\boldsymbol{k}}$
packets. Hence, 
\[
T\left(\boldsymbol{k}\right)=1+T\left(\tilde{\boldsymbol{k}}\right)=1+W_{\tilde{\boldsymbol{k}}}^{*}=l+1.
\]

\end{IEEEproof}
\textbf{\textit{\emph{Proposition \ref{prop:OddAntihole}. }}}\emph{Let the information graph
be an odd antihole ${\cal A}_{N},\ N=2g+1$. Then there is a code using XOR operations such
that, 
\[
T\left(\boldsymbol{k}\right)\leq W_{\boldsymbol{k}}^{*}+\left\lceil \frac{\min_{i\in[N]}k_{i}}{\left\lfloor N/2\right\rfloor }\right\rceil .
\]
Hence, 
\[
\hat{T}\left(\boldsymbol{r}\right)\leq W_{\boldsymbol{r}}^{*}+\frac{\min_{i\in[N]}r_{i}}{\left\lfloor N/2\right\rfloor }.
\]
} 
\begin{IEEEproof}
If $k_{i}=0$ for some $i\in{\cal A}_{N}$, the result follows from Proposition \ref{prop:AntiholeGeneral}.
Assume now that $k_{1}=\min_{i\in[N]}k_{i}>0$ and consider the following algorithm.

\emph{Algorithm } 
\begin{enumerate}
\item $\boldsymbol{k}^{1}\leftarrow\boldsymbol{k};$ $\hat{s}=\left\lfloor k_{1}/g\right\rfloor $;
\item $\upsilon\leftarrow k_{1}-\hat{s}g;$
\item \label{enu:oddcycleStep2-1-1}For $s=1$ to $\hat{s}$

\begin{enumerate}
\item \label{enu:substepOddCycl-1}For $m=1,...,g,$ transmit $g$ packets of the form 
\[
\oplus_{l=1}^{g}p^{2m+2l},\ p^{2m+2l}\in Q_{2m+2l}.
\]
Due to Lemma \ref{lem:EvenHoleLemma} all packet in the XOR combination can be decoded by the
corresponding receivers. For example, for $N=9,$ i.e., $g=4$, the $4$ transmitted packets
are of the form, 
\[
p^{4}\oplus p^{6}\oplus p^{8}\oplus p^{1},
\]
\[
p^{6}\oplus p^{8}\oplus p^{1}\oplus p^{3},
\]
\[
p^{8}\oplus p^{1}\oplus p^{3}\oplus p^{5},
\]
\[
p^{1}\oplus p^{3}\oplus p^{5}\oplus p^{7}.
\]

\item \label{enu:Substep2OddCylce-1}Set (remaining packets at each queue) 
\begin{equation}
k_{2h+1}^{s+1}\leftarrow k_{2h+1}^{s}-(g-h),\ 0\leq h\leq g,\label{eq:Odd-step2}
\end{equation}
\begin{equation}
k_{2h}^{s+1}\leftarrow k_{2h}^{s}-(h-1),\ 1\leq h\leq g.\label{eq:even-step2}
\end{equation}

\end{enumerate}
\item \label{enu:oddcycleStep3-1-1}For $m=1$ to $\upsilon$

\begin{enumerate}
\item \label{enu:ODDCycleEnd-1}Transmit $\upsilon$ packets of the form,
\[
\oplus_{l=1}^{g}p^{2m+2l},\ p^{2m+2l}\in Q_{2m+2l}.
\]
Due to Lemma \ref{lem:EvenHoleLemma} all packet in the XOR combination can be decoded by the
corresponding receivers. For example for $N=9,$ and $k_{1}^{s}=2$, the $2$ transmitted packets
are of the form,
\[
p^{4}\oplus p^{6}\oplus p^{8}\oplus p^{1},
\]
\[
p^{6}\oplus p^{8}\oplus p^{1}\oplus p^{3}.
\]

\item Set (remaining packets at each queue) 
\begin{equation}
k_{2h+1}^{s+1}\leftarrow k_{2h+1}^{s}-\max\left\{ \upsilon-h,0\right\} ,\ 0\leq h\leq g.\label{eq:odd-step3}
\end{equation}
\begin{equation}
k_{2h}^{s+1}\leftarrow k_{2h}^{s}-\min\{\upsilon,(h-1)\},\ 1\leq h\leq g.\label{eq:even-step3}
\end{equation}

\item $s\leftarrow s+1;$ 
\end{enumerate}
\item \label{enu:OddHolsFinalStep-1}At this step, $k_{1}^{s}=0$ and $k_{i}^{s}\geq0$ for $i\in[N]$.
Therefore, use the algorithm implied by Proposition \ref{prop:AntiholeGeneral} to transmit
the $\boldsymbol{k}^{s}$ packets; 
\item end; 
\end{enumerate}
Formulas (\ref{eq:odd-step3}), (\ref{eq:even-step3}) can be shown by induction on $\upsilon$
to hold for $0\leq\upsilon\leq g$. The special case $\upsilon=g$ gives formulas (\ref{eq:Odd-step2}),
(\ref{eq:even-step2}). 

Next we compute the number of transmissions needed for this algorithm to complete, and its
relation to $W_{\boldsymbol{k}}^{*}$. From (\ref{eq:Odd-step2}), (\ref{eq:even-step2}) we
derive, 
\[
k_{2h+1}^{s+1}+k_{2h}^{s+1}=k_{2h+1}^{s}+k_{2h}^{s}-g+1,\ 1\leq h\leq g
\]
\[
k_{2(h-1)+1}^{s+1}+k_{2h}^{s+1}=k_{2(h-1)+1}^{s}+k_{2h}^{s}-g,\ 1\leq h\leq g
\]
\[
k_{2g+1}^{s+1}+k_{1}^{s+1}=k_{2g+1}^{s}+k_{1}^{s}-g
\]
Using Lemma \ref{lem:AntiholeMaxWeight} and the equalities above we have, 
\begin{align}
W_{\boldsymbol{k}^{s+1}}^{*} & =\max_{i\in\mathcal{H}_{N}}\left(k_{i}^{s+1}+k_{i+1}^{s+1}\right)\nonumber \\
 & \leq\max_{i\in\mathcal{H}_{N}}\left(k_{i}^{s}+k_{i+1}^{s}\right)-g+1\nonumber \\
 & =W_{\boldsymbol{k}^{s}}^{*}-g+1\label{eq:WeightIneq}
\end{align}
 Since the loop in Step \ref{enu:oddcycleStep2-1-1} is executed $\hat{s}$ times and $W_{\boldsymbol{k}^{1}}^{*}=W_{\boldsymbol{k}}^{*}$,
we conclude from (\ref{eq:WeightIneq}) that, 
\begin{equation}
W_{\boldsymbol{k}^{\hat{s}+1}}^{*}\leq W_{\boldsymbol{k}}^{*}-\left\lfloor \frac{k_{1}}{g}\right\rfloor g+\left\lfloor \frac{k_{1}}{g}\right\rfloor .\label{eq:MaxWIneq}
\end{equation}
 Since each time this loop is executed $g$ packets are transmitted, the total number of transmissions
that take place in this loop is $\left\lfloor k_{1}/g\right\rfloor g$. 

Consider now the following two cases. 

\emph{Case 1.}\uline{ $k_{1}$ is a multiple of $g$.} In this case, $k_{1}^{\hat{s}+1}=\upsilon=0$
and by Proposition \ref{prop:AntiholeGeneral} the completion time of step $\ref{enu:OddHolsFinalStep-1}$
is $W_{\boldsymbol{k}^{\hat{s}+1}}^{*}$. Hence the completion time of the algorithm is 
\begin{eqnarray*}
T(\boldsymbol{k}) & = & \left\lfloor \frac{k_{1}}{g}\right\rfloor g+W_{\boldsymbol{k}^{\hat{s}+1}}^{*}\\
 & \leq & \left\lfloor \frac{k_{1}}{g}\right\rfloor g+W_{\boldsymbol{k}}^{*}-\left\lfloor \frac{k_{1}}{g}\right\rfloor g+\left\lfloor \frac{k_{1}}{g}\right\rfloor \\
 & = & W_{\boldsymbol{k}}^{*}+\left\lfloor \frac{k_{1}}{g}\right\rfloor .
\end{eqnarray*}

\emph{Case 2. }\uline{$k_{1}$ is not a multiple of $g$.} When Loop \ref{enu:oddcycleStep3-1-1}
is entered, $s=\hat{s}+1$, $k_{1}^{s}=k_{1}^{\hat{s}+1}=\upsilon$ and hence $\upsilon$ packet
transmissions take place in Loop \ref{enu:oddcycleStep3-1-1}. Also, from (\ref{eq:odd-step3}),
(\ref{eq:even-step3}) we have,
\begin{equation}
k_{2h+1}^{s+1}+k_{2h}^{s+1}=k_{2h+1}^{s}+k_{2h}^{s}-\max\left\{ \upsilon-h,0\right\} -\min\{\upsilon,(h-1)\},\ 1\leq h\leq g,\label{eq:fin1}
\end{equation}
\begin{equation}
k_{2(h-1)+1}^{s+1}+k_{2h}^{s+1}=k_{2(h-1)+1}^{s}+k_{2h}^{s}-\max\left\{ \upsilon-h+1,0\right\} -\min\{\upsilon,(h-1)\},\ 1\leq h\leq g,\label{eq:fin2}
\end{equation}
\begin{equation}
k_{2g+1}^{s+1}+k_{1}^{s+1}=k_{2g+1}^{s}+k_{1}^{s}-\upsilon.\label{eq:fin3}
\end{equation}
It is easy to see that 
\begin{align}
\max\left\{ \upsilon-h,0\right\} +\min\{\upsilon,(h-1)\} & =\left\{ \begin{array}{ccc}
\upsilon-1 & {\rm if} & \upsilon\geq h\\
\upsilon & {\rm if} & \upsilon\leq h-1
\end{array},\right.\nonumber \\
 & \geq\upsilon-1\label{eq:fin4}
\end{align}
and 
\begin{equation}
\max\left\{ \upsilon-h+1,0\right\} +\min\{\upsilon,(h-1)\}=\upsilon.\label{eq:fin5}
\end{equation}
From (\ref{eq:fin1})-(\ref{eq:fin5}), and using Lemma \ref{lem:AntiholeMaxWeight} we have,
\begin{align*}
W_{\boldsymbol{k}^{\hat{s}+2}}^{*} & =\max_{i\in\mathcal{H}_{N}}\left(k_{i}^{\hat{s}+2}+k_{i+1}^{\hat{s}+2}\right)\\
 & \leq\max_{i\in\mathcal{H}_{N}}\left(k_{i}^{\hat{s}+1}+k_{i+1}^{\hat{s}+1}\right)-\upsilon+1\\
 & =W_{\boldsymbol{k}^{\hat{s}+1}}^{*}-\upsilon+1\\
 & \leq W_{\boldsymbol{k}}^{*}-\left\lfloor \frac{k_{1}}{g}\right\rfloor g+\left\lfloor \frac{k_{1}}{g}\right\rfloor -\upsilon+1\ {\rm by\ (\ref{eq:MaxWIneq})}\\
 & =W_{\boldsymbol{k}}^{*}-k_{1}+\left\lfloor \frac{k_{1}}{g}\right\rfloor +1.
\end{align*}
Since in this case it takes $\left\lfloor \frac{k_{1}}{g}\right\rfloor g+\upsilon=k_{1}$ transmissions
to reach Step \ref{enu:OddHolsFinalStep-1} and $W_{\boldsymbol{k}^{\hat{s}+2}}^{*}$ transmissions
to transmit the remaining packets, we conclude, 
\begin{eqnarray*}
T(\boldsymbol{k}) & = & k_{1}+W_{\boldsymbol{k}^{\hat{s}+2}}^{*}\\
 & \leq & k_{1}+W_{\boldsymbol{k}}^{*}-k_{1}+\left\lfloor \frac{k_{1}}{g}\right\rfloor +1\\
 & = & W_{\boldsymbol{k}}^{*}+\left\lfloor \frac{k_{1}}{g}\right\rfloor +1\\
 & = & W_{\boldsymbol{k}}^{*}+\left\lceil \frac{k_{1}}{g}\right\rceil .
\end{eqnarray*}
Since $g=\left\lfloor N/2\right\rfloor ,$ the proposition follows.\end{IEEEproof}

\end{document}